\documentclass[11pt,a4paper]{article}
\pdfoutput=1
\usepackage{cancel}
\usepackage{enumitem}
\usepackage{etex}
\usepackage{amsthm}
\usepackage{geometry}
\usepackage{dcolumn}
\usepackage{epsf}
\usepackage{mathrsfs}
\usepackage{multirow}
\usepackage{booktabs}
\usepackage{tabularx}
\usepackage{array}
\usepackage{slashed}
\usepackage{float}
\usepackage{leftidx}
\usepackage{setspace}
\usepackage{verbatim}
\usepackage{adjustbox}
\usepackage{tikz}
\usepackage[caption=false]{subfig}
\usepackage{arydshln}
\usepackage{jheppub}
\usepackage{shuffle}
\usepackage{amsmath}
\usepackage{cases}

\numberwithin{equation}{section}
\usetikzlibrary{arrows,decorations.markings,shapes.arrows,patterns,positioning}

\newcommand{\bea}{\begin{eqnarray}}
\newcommand{\eea}{\end{eqnarray}}
\newcommand{\bean}{\begin{eqnarray*}}
\newcommand{\eean}{\end{eqnarray*}}
\newcommand{\nn}{\nonumber\\}
\newcommand{\Sl}{\sum\limits}

\def\W #1{\widetilde{#1}}

\def\Label#1{\label{#1}%
  \smash{\hbox to0pt{\raise1ex\hbox{\tiny[#1]}\hss}}}
\def\Label#1{\label{#1}}
\renewcommand{\eqref}[1]{eq.~(\ref{#1})}
\newcommand{\figref}[1]{Fig.~\ref{#1}}
\newcommand{\tabref}[1]{table~\ref{#1}}
\newcommand{\secref}[1]{section~\ref{#1}}

\def\Sl{\sum\limits}

\newcommand{\ctobedelete}[1]{}
\allowdisplaybreaks

\title{ Off-shell extended graphic rule and the expansion of Berends-Giele currents in Yang-Mills theory}

\author[a]{Konglong Wu} \author[a,b,c]{Yi-Jian Du\footnote{Corresponding author}}

\affiliation[a]{Department of Physics, School of Physics and Technology,
Wuhan University, \\
No.299 Bayi Road, Wuhan 430072, China}
\affiliation[b]{ Hubei Key Laboratory of Nuclear Solid Physics, School of Physics and Technology, Wuhan University,\\
No.299 Bayi Road, Wuhan 430072, China}
\affiliation[c]{Suzhou Institute of Wuhan University,\\
No.377 Linquan Street, Suzhou, 215123, China}

\emailAdd{wukonglong@whu.edu.cn,yijian.du@whu.edu.cn}

\date{\today}
\abstract{Tree-level color-ordered Yang-Mills (YM) amplitudes can be decomposed in terms of $(n-2)!$ bi-scalar (BS) amplitudes, whose expansion coefficients form a basis of Bern-Carrasco-Johansson (BCJ) numerators. By the help of the recursive expansion of Einstein-Yang-Mills (EYM) amplitudes, the BCJ numerators are given by  polynomial functions of Lorentz contractions which are conveniently described by graphic rule. In this work, we extend the expansion of YM amplitudes to off-shell level. We define  different types of off-shell extended numerators that can be generated by graphs. By the use of these extended numerators, we propose a general decomposition formula of off-shell Berends-Giele currents in YM.  This formula consists of three terms: (i). an effective current which is expanded as a combination of the Berends-Giele currents in BS theory (The expansion coefficients are one type of off-shell extended numerators) (ii). a term proportional to the total momentum  of on-shell lines and (iii). a term expressed by the sum of lower point  Berends-Giele currents in which some polarizations and momenta are replaced by vectors proportional to off-shell momenta appropriately. In the on-shell limit, the last two terms vanish while the decomposition of effective current precisely reproduces the decomposition of on-shell YM amplitudes with the the expected coefficients (BCJ numerators in DDM basis). We further symmetrize these coefficients such that the Lie symmetries are satisfied. These symmetric BCJ numerators simultaneously satisfy the relabeling property of external lines and the algebraic properties (antisymmetry and Jacobi identity). }

\begin{document}
\maketitle \flushbottom

\section{Introduction}

The recursive expansion relation of tree level Einstein-Yang-Mills (EYM) amplitudes \cite{Fu:2017uzt,Chiodaroli:2017ngp,Teng:2017tbo,Du:2017kpo,Du:2017gnh} serves as a bridge between Einstein gravity (GR) and Yang-Mills theory (YM). By this expansion, any tree-level EYM amplitude can be given by a combination of amplitudes with fewer gravitons and/or gluon traces. The expansion coefficients are polynomial functions of Lorentz contractions between external polarizations and/or momenta. When the recursive expansion is applied repeatedly, an EYM amplitude is finally written in terms of color-ordered pure YM amplitudes, which was  earlier studied in \cite{Stieberger:2016lng,Nandan:2016pya,delaCruz:2016gnm,Schlotterer:2016cxa} by evaluating explicit examples. By the help of the recursive expansion as well as the relationship between GR and EYM amplitudes \cite{Fu:2017uzt}, one can finally express a tree-level GR amplitude as a combination of $(n-2)!$ color-ordered YM amplitudes. The polynomial coefficients in this expansion can be considered as the Bern-Carrasco-Johansson (BCJ) \cite{Bern:2008qj,Bern:2010ue} numerators (which are characterized by cubic diagrams and  satisfy antisymmetry and Jacobi identity) in Del~Duca-Dixon-Maltoni (DDM) basis \cite{DelDuca:1999rs}.

It was shown in \cite{Fu:2017uzt,Teng:2017tbo,Du:2017kpo,Gao:2017dek,Du:2017gnh,He:2021lro} that the EYM recursive expansion and the resulted pure-YM expansion of GR amplitudes could be understood from the framework of Cachazo-He-Yuan (CHY) \cite{Cachazo:2013hca,Cachazo:2013gna,Cachazo:2013iea, Cachazo:2014nsa,Cachazo:2014xea} formula. This fact allows one to generalize the expansions to various theories. For example, a color-ordered YM amplitude can be expanded in terms of color ordered bi-adjoint scalar (BS) amplitudes\footnote{The supersymmetric version of this expansion can be found in \cite{Mafra:2016ltu, Broedel:2013tta}.}, with the same expansion coefficients (i.e. BCJ numerators in DDM form) in the pure-YM expansion of GR amplitudes. Systematic study of the recursive expansions in different theories can be found in \cite{Feng:2019tvb,Zhou:2020mvz}.

When the Lorentz contractions between external polarizations and/or momenta are expressed by graphs \cite{Huang:2017ydz,Teng:2017tbo,Du:2017kpo,Du:2017gnh,Lam:2018tgm,Hou:2018bwm,Du:2019vzf}, the coefficients (BCJ numerators) for the pure-YM (BS) expansion of GR (YM) amplitudes can be given by a sum over proper graphs. This method provides a convenient approach to the study of related properties of GR, EYM and YM amplitudes. In particular, the relationship between symmetry induced identities and BCJ relations in YM was founded through the graphs \cite{Hou:2018bwm,Du:2019vzf}. In addition, sectors of EYM (YM) amplitudes were naturally constructed through graphic rule, which further inspired a symmetric formula \cite{Tian:2021dzf} of double trace maximally-helicity-violeting (MHV) amplitude in EYM theory.

Although the graphic expansion has already been applied to investigate many properties of on-shell tree amplitudes, it is still lack of a study at the off-shell level, which may be helpful for  understanding loop amplitudes. In this paper, we generalize the graphic rule to off-shell level by studying Berends-Giele currents \cite{Berends:1987me} in YM theory. We first generalize the graphs to three types of off-shell extended graphs and then define the corresponding off-shell extended numerators. Using Berends-Giele recursion \cite{Berends:1987me}, we prove that a Berends-Giele current in YM can be decomposed into the following three terms: \textbf{(i).}  a combination of the Berends-Giele currents in BS theory \cite{Mafra:2016ltu} accompanied by off-shell extended polynomial coefficients (numerators), \textbf{(ii).} a term which is proportional to the total momentum as well as \textbf{(iii).} a combination of lower point Berends-Giele currents in which the polarizations and momenta of some external lines are proportional to off-shell momenta in a proper way. The latter two terms in fact vanish under on-shell limit while the expansion coefficients in the first term have the same graphic interpretation with those for on-shell amplitudes \cite{Du:2017kpo}. Thus this decomposition of Berends-Giele currents precisely reproduces the graphic expansion of on-shell YM amplitudes. Since the Berends-Giele recursion is essentially a proper way for collecting Feynman diagrams and the expansion of on-shell YM amplitudes has been derived from CHY formula  \cite{Du:2017kpo}, this work establishes a  connection between  CHY formula and Feynman diagram approach, via graphs.

The expansion coefficients of BS currents can be considered as the BCJ numerators in DDM basis. Other BCJ numerators at tree-level are generated by Jacobi identity and antisymmetry straightforwardly. Nevertheless, these numerators do not satisfy relabeling properties, i.e., distinct numerators for a given topology of cubic graphs can not be related to one another by relabeling the external lines. To remedy this disadvantage, we symmetrize the off-shell extended numerators so that they satisfy Lie symmetries. As pointed in \cite{Mafra:2015vca,Lee:2016tbd,Bridges:2019siz}, numerators with Lie symmetries naturally satisfy the relabeling property and the algebraic properties (Jacobi identity and antisymmetry).

The structure of this paper is following. In \secref{Sec:Review}, we review the Berends-Giele recursion for YM and BS amplitudes as well as the graphic rule for the expansion of YM amplitudes. Three types of off-shell extended graphs and numerators are introduced in \secref{sec:ExtededNumerators}.  We show the expansion of Berends-Giele currents in YM by explicit examples in \secref{Sec:Examples}. The general expansion formula and its proof are presented in \secref{sec:genExpansion}.  Off-shell extended BCJ numerators with Lie symmetries are constructed in \secref{sec:NumLieSym}. We summarize this work in \secref{sec:Conclusions} and provide the proofs of some  helpful formulas in the appendix.



\section{Preparations: Berends-Giele recursions and the graphic expansion}\label{Sec:Review}
In this section, we review the Berends-Giele recursions in YM theory and BS theory. The graphic expansion of color-ordered YM amplitudes is also reviewed.

\subsection{Berends-Giele recursion in YM}

Color-ordered YM amplitudes can be reconstructed by Berends-Giele recursion which is essentially the sum of all Feynman diagrams. Specifically, an $(n-1)$-point Berends-Giele current $J^{\rho}(1,...,n-1)$ in Feynman gauge is expressed by lower-point currents as follows \cite{Berends:1987me}
\bea
J^{\rho}(1,...,n-1)&=&\frac{1}{s_{1...n-1}}\bigg[\Sl_{1\leq i<n-1}V_3^{\mu\nu\rho}J_{\mu}(1,...,i)J_{\nu}(i+1,...,n-1)\nn
&&+\Sl_{1\leq i<j<n-1}V_4^{\mu\nu\tau\rho}J_{\mu}(1,...,i)J_{\nu}(i+1,...,j)J_{\tau}(j+1,...,n-1)\bigg].\Label{Eq:BerendsGiele}
\eea
In the above equation, $s_{1...n-1}\equiv (k_1+k_2+...+k_{n-1})^2$ where $k^{\mu}_i$ ($i=1,...,n-1$) denotes the momentum of external gluon $i$.
The 3-point vertex $V_3^{\mu\nu\rho}$ and 4-point vertex $V_4^{\mu\nu\tau\rho}$ are respectively given by\footnote{In standard textbook, 3-point vertex, 4-point vertex and propagator are correspondingly dressed by factors $\frac{-i}{\sqrt{2}}$, $\frac{i}{2}$ and $-i$.  In this paper, an overall factor $\left(\frac{1}{\sqrt{2}}\right)^{n-2}(-i)^{n}$, which comes from these factors, has been absorbed into the normalization factor for convenience. This does not affect our discussions.}
 \bea
  V_3^{\mu\nu\rho}&=&\eta^{\mu\nu}(k_A-k_B)^\rho+\eta^{\nu\rho}(k_B-k_C)^\mu +\eta^{\rho\mu}(k_C-k_A)^\nu\Label{Eq:CubicVertex}\\
  V_4^{\mu\nu\tau\rho}&=&2\eta^{\mu\rho}\eta^{\nu\lambda}-\eta^{\mu\nu}\eta^{\rho \lambda}-\eta^{\mu\lambda}\eta^{\nu\rho}.\Label{Eq:QuarticVertex}
  \eea
  In the 3-point vertex, $k_A$, $k_B$ and $k_C$ correspondingly denote the momenta of lines $A$, $B$ and $C$ (dressed by the Lorentz indices $\mu$, $\nu$ and $\rho$) that are attached to the vertex.  The $\eta^{\mu\nu}$ stands for Minkowskian metric.
  In this paper, we define so-called \emph{effective 3-point vertex} $\W V_3^{\mu\nu\rho}$ by
  \bea
  \W V_3^{\mu\nu\rho}\equiv \left[\big(2k_B^{\mu}\eta^{\nu\rho}-\eta^{\mu\nu}2k_B^{\rho}\big)-\eta^{\mu\rho}2k_A^{\nu}\right].\Label{Eq:EffCubicVertex}
  \eea
  When momentum conservation $k^{\mu}_A+k^{\mu}_B+k^{\mu}_C=0$ is applied, the full 3-point vertex (\ref{Eq:CubicVertex}) is rewritten as
  \bea
  V_3^{\mu\nu\rho}&=&\W V_3^{\mu\nu\rho}+\eta^{\nu\rho}k_A^{\mu}-\eta^{\mu\rho}k_B^{\nu}+\eta^{\mu\nu}(k_A+k_B)^{\rho}.\Label{Eq:CubicVertex1}
  \eea
The starting point of the Berends-Giele  recursion is the one-point current $J^{\rho}(l)=\epsilon^{\rho}_l$, where $\epsilon^{\rho}_l$ is the polarization vector of the external gluon $l$. The color ordered on-shell amplitude $A(1,...,n)$ is obtained through the following on-shell limit
\bea
A(1,...,n)=\Big[s_{1...n-1}\,\epsilon_n\cdot J(1,...,n-1)\Big]\Big|_{k_n^2=s_{1...n-1}\to 0}.\Label{Eq:On-shellLimit}
\eea

An important identity satisfied by the Berends-Giele current $J^{\rho}(1,...,n-1)$ in YM theory is the `conservation condition'
\bea
k_{1,n-1}\cdot J(1,...,n-1)=0, \Label{Eq:YMBGProperty-1}
\eea
where the total momentum $k_{1,n-1}^{\rho}\equiv\sum_{l}k_{l}^{\rho}$ of on-shell lines $1,...,n-1$ is not necessarily on-shell. This identity can be further extended to more generic cases:
\bea
0&=& k_{1,n-1}\cdot J\bigl(1,...,a_1-1,k_{a_1,b_1},b_1+1,...,a_2-1,k_{a_2,b_2},...,k_{a_I,b_I},b_I+1...,n-1\bigr),\Label{Eq:YMBGProperty-2}\\
0&=&\epsilon_n\cdot J\bigl(1,...,a_1-1,k_{a_1,b_1},b_1+1,...,a_2-1,k_{a_2,b_2},...,k_{a_I,b_I},b_I+1...,n-1\bigr),\Label{Eq:YMBGProperty-3}
\eea
where $a_i$ and  $b_i$ (for a given $i$) are two arbitrary elements s.t. $a_i\leq b_i$. The $k^{\rho}_{a_i,b_i}$ ($i=1,...,I$) denotes the sum of momenta of all elements in the consecutive sequence  $\{a_i, a_{i}+1,...,b_i\}$. The   $J^{\rho}(1,...,a_1-1,k_{a_1,b_1},b_1+1,...,a_2-1,k_{a_2,b_2},...,k_{a_I,b_I},b_I+1...,n-1)$
stands for the `Berends-Giele current' where $k^{\rho}_{a_i,b_i}$ is considered as both the momentum and the polarization of the line between $a_i-1$ and $b_i+1$. Explicit examples are displayed as follows
\bea
J^{\rho}(1,k_{2,3})&=&\frac{1}{s_{1...3}}V^{\mu\nu\rho}_{3}\,J_{\mu}(1)\,\big(k_{2,3}\big)_{\nu},\nn
J^{\rho}(k_{1},k_{2,3})&=&\frac{1}{s_{1...3}}V^{\mu\nu\rho}_{3}\,{(k_1)}_{\mu}\,\big(k_{2,3}\big)_{\nu},\nn
J^{\rho}(1,k_{2,3},4)&=&\frac{1}{s_{1...4}}\Bigl[V^{\mu\nu\rho}_{3}\,J_{\mu}(1,k_{2,3})\,J_{\nu}(4)+V^{\mu\nu\rho}_{3}\,J_{\mu}(1)\,J_{\nu}(k_{2,3},4) +V^{\mu\nu\tau\rho}_4\,J_{\mu}(1)\big(k_{2,3}\big)_{\nu}\,J_{\tau}(4)\Bigr],\nn
J^{\rho}\bigl(1,k_{2,3},4,k_{5,7}\bigr)&=&\frac{1}{s_{1...7}}\Big[V^{\mu\nu\rho}_{3}\,J_{\mu}(1)\,J_{\nu}\bigl(k_{2,3},4,k_{5,7}\bigr)+V^{\mu\nu\rho}_{3}\,J_{\mu}(1,k_{2,3})\,J_{\nu}\bigl(4,k_{5,7}\bigr)\nn
&&~~~~~~~~~~~~~\,+V^{\mu\nu\rho}_{3}\,J_{\mu}(1,k_{2,3},4)\,\big(k_{5,7}\big)_{\nu}+V^{\mu\nu\tau\rho}_4\,J_{\mu}(1,k_{2,3})\,J_{\nu}(4)(k_{5,7})_{\tau}\nn
&&~~~~~~~~~~~~~~~~~~~+V^{\mu\nu\tau\rho}_4\,J_{\mu}(1)\,J_{\nu}(k_{2,3},4)\,\big(k_{5,7}\big)_{\tau}+V^{\mu\nu\tau\rho}_4\,J_{\mu}(1)\,\big(k_{2,3}\big)_{\nu}\,J_{\tau}(4,k_{5,7})\Big].\nonumber
\eea
In the property (\ref{Eq:YMBGProperty-3}), the momentum of $n$ is considered as $k^{\mu}_n=-(k^{\mu}_1+...+k^{\mu}_{n-1})$. The on-shell condition $k_n^2=0$ and the physical condition $\epsilon_n\cdot k_n=0$ are implied in (\ref{Eq:YMBGProperty-3}).

\subsection{Berends-Giele recursion for BS amplitudes}

The Berends-Giele current  $\phi(1,...,n-1\big|\sigma_{1},...,\sigma_{n-1})$ in BS theory is defined by \cite{Mafra:2016ltu}
\bea \phi\big(1,...,n-1\big|\sigma_{1},...,\sigma_{n-1}\big)&=&\frac{1}{s_{1...n-1}}\Sl_{i=1}^{n-2}\Big[\phi\big(1,...,i\big|\sigma_{1},...,\sigma_{i}\big)\phi\big(i+1,...,n-1\big|\sigma_{i+1},...,\sigma_{n-1}\big)\nn
&&~~~~~~~~~~-\phi\big(1,...,i\big|\sigma_{n-i},...,\sigma_{n-1}\big)\phi\big(i+1,...,n-1\big|\sigma_{1},...,\sigma_{n-i}\big)\Big],\Label{Eq:BerendsGieleBS}
\eea
where $\pmb{\sigma}=\{\sigma_{1},...,\sigma_{n-1}\}$ is a permutation of external lines $1,...,n-1$. The starting point of the recursion is
\bea
 \phi(l|l')=\left\{
              \begin{array}{cc}
                1& (\,l'=l) \\
                0 & (\,l'\neq l) \\
              \end{array}\right ..
\eea
 As a result of (\ref{Eq:BerendsGieleBS}), the BS current $\phi\big(a_1,...,a_i\big|b_1,...,b_i\big)$ has to vanish when  $\{a_1,...,a_i\}\setminus\{b_1,...,b_i\}\neq\emptyset$. The on-shell BS amplitude $\mathcal{A}(1,...,n|\sigma_{1},...,\sigma_n)$ is then obtained by taking the following   limit
\bea
\mathcal{A}(1,2,...,n|\sigma_{1},...,\sigma_n)=\Big[s_{1...n-1}\phi\big(1,2,...,n-1\big|\sigma_{1},...,\sigma_{n-1}\big)\Bigr]\Big|_{s_{1...n-1}=k_n^2=0}.
\eea
The BS current (\ref{Eq:BerendsGieleBS}) satisfies many important relations which were first founded in YM theory:

\begin{itemize}
\item Reflection relation
\bea
\phi\big(1,...,n-1\big|\sigma_{1},...,\sigma_{n-1}\big)=(-1)^{n}\phi\big(1,...,n-1\big|\sigma_{n-1},...,\sigma_{1}\big).\Label{Eq:BSBGProperty-1}
\eea
\item Kleiss-Kuijf (KK) relation \cite{Kleiss:1988ne}
\bea
\phi\big(1,2,...,n-1\big|\pmb{\beta},1,\pmb{\alpha}\big) =\Sl_{\shuffle}(-1)^{|\pmb{\beta}|}\phi\big(1,2,...,n-1\big|1,\pmb{\alpha}\shuffle \pmb{\beta}^T\big).\Label{Eq:BSBGProperty-2}
\eea
\item Two generalized KK relations
\bea
\Sl_{\shuffle}\phi\big(1,2,...,n-1\big|\pmb{\alpha}\shuffle\pmb{\beta}\big)&=&0,\Label{Eq:BSBGProperty-4}\\
\Sl_{\shuffle}\phi\big(1,2,...,n-1\big|\pmb{\beta}\shuffle\pmb{\gamma}^T,1,\pmb{\alpha}\big)&=&\Sl_{\shuffle}(-1)^{|\pmb{\gamma}|}\phi\big(1,2,...,n-1\big|\pmb{\beta},1,\pmb{\alpha}\shuffle\pmb{\gamma}\big).\Label{Eq:BSBGProperty-3}
\eea

\end{itemize}
In these relations $\pmb{\alpha}$, $\pmb{\beta}$ and $\pmb{\gamma}$ stand for ordered sets. The $|\pmb{\beta}|$, $|\pmb{\gamma}|$  denote the number of elements in $\pmb{\beta}$ and $\pmb{\gamma}$, while $\pmb{\beta}^T$ is the inverse permutation of $\pmb{\beta}$. The shuffling permutations $\pmb{A}\shuffle \pmb{B}$ of two ordered sets $\pmb{A}$ and $\pmb{B}$ are defined by all those permutations obtained by merging $\pmb{A}$ and $\pmb{B}$ together such that the relative order of elements in each set is preserved. The reflection relation (\ref{Eq:BSBGProperty-1}) is apparently the KK relation (\ref{Eq:BSBGProperty-2}) in the special case $\pmb{\alpha}=\emptyset$, the relation (\ref{Eq:BSBGProperty-4}) was proved by KK relation(as pointed in \cite{Du:2011js}), while the relation (\ref{Eq:BSBGProperty-3}) can also be proven by the KK relation (\ref{Eq:BSBGProperty-2}) straightforwardly.

\begin{figure}
\centering
\includegraphics[width=1\textwidth]{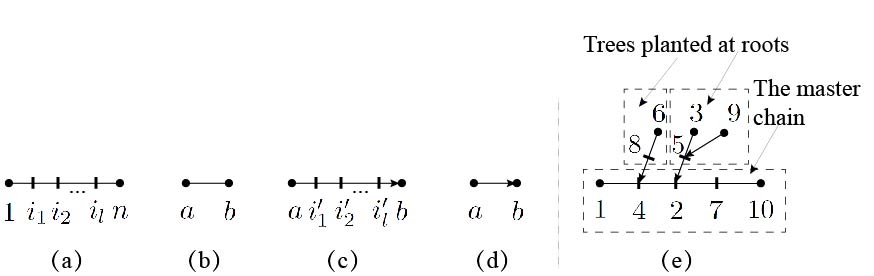}
\caption{  The master chain of the form (a) corresponds to a factor $(-1)^{n-l}\epsilon_1\cdot F_{i_1}\cdot...\cdot F_{i_l}\cdot\epsilon_n$. If $\{i_1,...,i_l\}=\emptyset$, the master chain becomes (b) which contributes a factor $(-1)^{n}\epsilon_1\cdot\epsilon_n$. A chain defined by step-3 and -4 has the general form (c), which stands for  a factor $\epsilon_a\cdot F_{i_1'}...\cdot F_{i_l'}\cdot 2k_b$. The special case with $\{i_1,...,i_l\}=\emptyset$ is given by (d), corresponding to the factor $\epsilon_a\cdot 2k_b$. A full graph is shown by (e). }
 \label{Fig:ChainsAndGraph}
\end{figure}

\subsection{Graphic expansion of color-ordered YM amplitudes}\label{sec:ExpansionYM}

Tree level color-ordered YM amplitude $A(1,\dots,n)$  can be expanded in terms of tree level BS amplitudes $\mathcal{A}\left(1,...,n|1,\pmb{\sigma},n\right)$:
\bea
A(1,\dots,n)=\Sl_{\pmb{\sigma}\in P(2,n-1)}N (1,\pmb{\sigma},n )\mathcal{A}\left(1,...,n|1,\pmb{\sigma},n\right),\Label{Eq:OnShellDec}
\eea
where the expansion coefficients $N\big(1,\pmb{\sigma},n\big)$ are known as BCJ numerators in DDM basis \cite{DelDuca:1999rs} and $P(2,n-1)$ denotes the set of all $(n-2)!$ permutations of elements $2$, $3$, ..., $n-1$. Through the recursive expansion of Yang-Mills-scalar amplitudes \cite{Fu:2017uzt}, the BCJ numerators $N\big(1,\pmb{\sigma},n\big)$ are constructed as polynomial functions of Lorentz contractions between external momenta and polarizations according to the following rule \cite{Du:2017kpo}.
\newline
\newline
$~~~~~~~~~~~~~~~~~~~~~~~~~~~~~~~~~~~~~~~~~~~~~~~~${\bf \emph {Graphic rule for $N\left(1,\pmb{\sigma},n\right)$}}
\begin{itemize}
\item {\bf\emph{Step-1}}~~Define a reference order $\mathsf{R}=\{1,\pmb{\gamma},n\}$ where $\pmb{\gamma}\equiv\{\gamma_1,...,\gamma_{n-2}\}\in P(2,n-1)$. In sections \ref{sec:ExtededNumerators}, \ref{Sec:Examples}, \ref{sec:genExpansion}, $\mathsf{R}$ is chosen as the the inverse of the normal order, i.e.,  $\mathsf{R}=\{n,n-1,...,2,1\}$. Other reference orders will be used in the symmetrization in \secref{sec:NumLieSym}. The position of each element in $\mathsf{R}$ is called its \emph{weight}.
\item {\bf \emph{Step-2}}~~Pick out $1$, $n$ as well as some nodes $i_1,i_2,....,i_l$ such that $\sigma^{-1}(i_1)<\sigma^{-1}(i_2)<...<\sigma^{-1}(i_l)$\footnote{In this paper, $\sigma^{-1}(a)$ denotes the position of element $a$ in the permutation $\pmb{\sigma}$. This can be understood as follows, if ${\sigma}_x\equiv{\sigma}(x)=a$, then $x=\sigma^{-1}(a)$.}, then construct a chain of the following form
    \bea
    \epsilon_1\cdot F_{i_1}\cdot  F_{i_2}\cdot...\cdot F_{i_l}\cdot \epsilon_n, \Label{Eq:MainChain}
    \eea
    which is accompanied by a factor\footnote{The factor and the following definition of  $F^{\mu\nu}_{i}$ differ from  those in the original paper \cite{Fu:2017uzt} by a $(-1)$ and a factor $2$ respectively. This difference can be absorbed into the normalization factor and does not affect our discussions. } $(-1)^{n-l}$.  The strength tensor $F^{\mu\nu}_{i}$ is defined by $F_{i}^{\mu\nu}\equiv 2 k_i^{\mu}\epsilon_i^{\nu}-2 k_i^{\nu}\epsilon_i^{\mu}$. The chain (\ref{Eq:MainChain}) in this paper is mentioned as a \emph{master chain} and presented by the graphs \figref{Fig:ChainsAndGraph} (a) or (b) (here we follow the graphs used in \cite{Lam:2018tgm}).
    Redefine the reference order $\mathsf{R}$ by removing $1,i_1,...,i_l,n$:
    \bea
    \mathsf{R}\to\mathsf{R}\setminus\{1,i_1,...,i_l,n\}.
    \eea
    Define the root set $ \mathcal{R}$ by
    \bea
    \mathcal{R}=\{1,i_1,...,i_l\},
    \eea
where the element $n$ is not involved.

\item {\bf \emph{Step-3}}~~Pick out the highest-weight element (i.e. the last element in the reference order that was redefined in the previous step) say, $a$  as well as elements $i'_1,...,i'_{l'}$, which satisfy $\sigma^{-1}\big(i'_1\big)<...<\sigma^{-1}\big(i'_{l'}\big)<\sigma^{-1}(a)$, from the new defined $\mathsf{R}$ and construct a chain towards an element $b\in \mathcal{R}$ satisfying $\sigma^{-1}(b)<\sigma^{-1}(i'_1)$
    \bea
    \epsilon_{a}\cdot F_{i'_{l'}}\cdot...\cdot F_{i'_{1}}\cdot 2k_{b}.
    \eea
   The above chain is conveniently presented by \figref{Fig:ChainsAndGraph} (c) and (d), where the nodes $a$, $i'_1,...,i'_{l'}$ and the node $b$ are respectively mentioned as the starting node, the internal nodes and the ending node of this chain.  Redefine the reference order and the root set respectively as follows
    \bea
    \mathsf{R}\to\mathsf{R}\setminus\{i'_1,...,i'_{l'},a\},~~~~\mathcal{R}\to \mathcal{R}\cup\{i'_1,...,i'_{l'},a\}.
    \eea

\item {\bf \emph{Step-4}}~~ Repeat step-3 until the ordered set $\mathsf{R}$ becomes empty, then we obtain a connected tree graph $\mathcal{F}$ (as shown by \figref{Fig:ChainsAndGraph} (e)) which stands for a term $C_{\mathcal{F}}(\pmb{\sigma})$ in the BCJ numerator $N(1,\pmb{\sigma},n)$. When all graphs $\mathcal{F}\in \mathcal{G}[\pmb{\sigma}]$ corresponding to the permutation $\pmb{\sigma}$ are collected together, we get the full BCJ numerator
    \bea
    N(1,\pmb{\sigma},n)=\Sl_{\mathcal{F}\in \mathcal{G}[\pmb{\sigma}]}C_{\mathcal{F}},\Label{Eq:BCJOnShell}
    \eea
    which is a polynomial function of the Lorentz contractions $\epsilon\cdot\epsilon$, $\epsilon\cdot 2k$ and $2k\cdot 2k$.
\end{itemize}
 Briefly speaking, \emph{a graph is described by {\bf(i)} a given reference order $\mathsf{R}$, {\bf(ii)} a master chain (\ref{Eq:MainChain}) and {\bf(iii)} trees planted at roots in the set $\mathcal{R}=\{1,i_1,...,i_l\}$, which are constructed according to step-3 and step-4. A numerator $N(1,\pmb{\sigma},n)$ is given by the sum of all graphs that are consistent\footnote{The `consistent' means that the graphs can be generated from the permutation $\pmb{\sigma}$ according to the graphic rule.} with the permutation $\pmb{\sigma}$.} It is worth pointing out that the decomposition formula (\ref{Eq:OnShellDec}) can be rearranged as \cite{Hou:2018bwm,Du:2019vzf}
\bea
A(1,\dots,n)=\Sl_{\mathcal{F}}C_{\mathcal{F}}\left[\Sl_{\pmb{\sigma}^{\mathcal{F}}}A(1,...,n|1,\pmb{\sigma}^{\mathcal{F}},n)\right],\Label{Eq:OnShellDec1}
\eea
where we have summed over all possible graphs and all permutations $\pmb{\sigma}^{\mathcal{F}}$ allowed by a given graph. The permutations $\pmb{\sigma}^{\mathcal{F}}$  are determined as follows (see \cite{Hou:2018bwm,Du:2019vzf}): (i). the element $1$ is the leftmost one, (ii) for any two adjacent nodes $i$ and $j$, if $i$ is nearer to $1$ than $j$, we have $i\prec j$\footnote{Here $i\prec j$ denotes $({\sigma}^{\mathcal{F}})^{-1}(i)\prec ({\sigma}^{\mathcal{F}})^{-1}(j)$ for short.} in $\pmb{\sigma}^{\mathcal{F}}$, (iii). if there are more than one branch attached to a node, we should shuffle the permutations established by these branches together.
An explicit example is that the permutations $\pmb{\sigma}^{\mathcal{F}}$ established by the graph \figref{Fig:ChainsAndGraph} (e) are given by
\bea
\pmb{\sigma}^{\mathcal{F}}\in\Biggl\{1,4,\{8,6\}\shuffle\bigg\{2,\Big\{5,\{3\}\shuffle\{9\}\Big\}\shuffle\{7\}\bigg\},10\Biggr\}.
\eea

In the coming sections, graphs are naturally generalized to off-shell level, based on which, off-shell numerators are defined. We further provide an expansion formula for off-shell Berends-Giele currents in YM with these off-shell extended numerators. In the on-shell limit, this expansion reproduces the formula (\ref{Eq:OnShellDec}).

\section{Off-shell extended graphs and numerators}\label{sec:ExtededNumerators}

 In the previous section, we have introduced a graphic rule to express the polynomial expansion coefficients (i.e., BCJ numerators in DDM basis) for on-shell color-ordered YM amplitudes. In this section, we generalize these graphs and numerators to three-types of off-shell extended graphs and numerators, by extracting the polarization vector $(\epsilon_{n})_{\rho}$ or both  $(\epsilon_{1})_{\rho}$ and $(\epsilon_{n})_{\rho}$ out from the master chain in a proper way.

\subsection{Type-A graphs and numerators}

\begin{figure}
\centering
\includegraphics[width=0.3\textwidth]{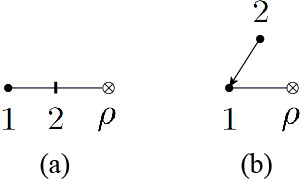}
\caption{Type-A graphs with two on-shell nodes}\label{Fig:2ptGraphsA}
\end{figure}
 \begin{figure}
\centering
\includegraphics[width=0.9\textwidth]{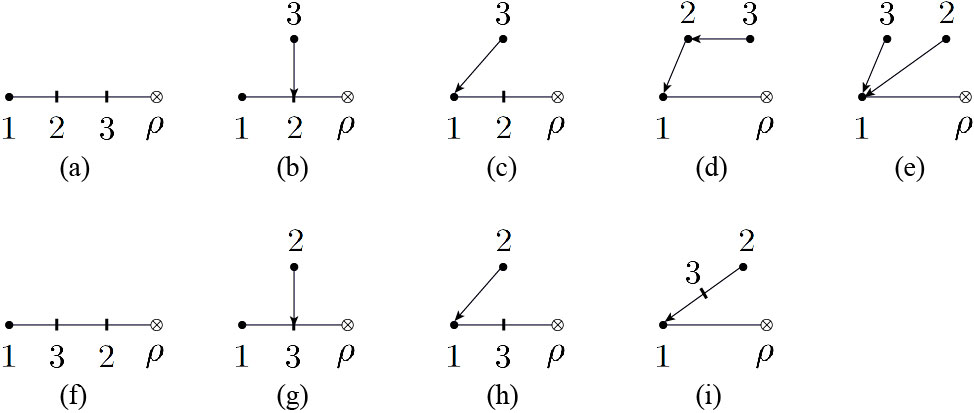}
\caption{Type-A graphs with three on-shell nodes}\label{Fig:3ptGraphsA}
\end{figure}

A type-A graph is a graph where the $(\epsilon_{n})_{\rho}$ on the master chain (\ref{Eq:MainChain}) is extracted out. In a type-A graph, the outer Lorentz index $\rho$ (expressed by a cross in this paper) plays as the rightmost node and the reference order is given by $\mathsf{R}=\{\rho,n-1,...,2,1\}$. The master chain has the form $(-1)^{n-l}\left(\epsilon_1\cdot F_{i_1}\cdot...\cdot F_{i_l}\right)^{\rho}$ and the root set of a graph has the form $\mathcal{R}=\{1,i_1,...,i_l\}$ which does not involve the the outer Lorentz index $\rho$. A type-A numerator $N^{\rho}_{A}(1,\pmb{\sigma})$ is defined by collecting all type-A graphs corresponding to the permutation $\pmb{\sigma}\in P(2,n-1)$, i.e.,
\bea
N^{\rho}_{A}(1,\pmb{\sigma})\equiv\Sl_{\mathcal{F}\in \mathcal{G}^A[\pmb{\sigma}]} C_{\mathcal{F}}^{\rho},\Label{Eq:Type-A}
\eea
where $\mathcal{G}^A[\pmb{\sigma}]$ denotes the set of all type-A off-shell extended graphs that correspond to the permutation $1,\pmb{\sigma}$ ($\pmb{\sigma}\in P(2,n-1)$). In the special case  $n=2$, the permutation $\pmb{\sigma}=\emptyset$ and $N^{\rho}_{A}=\epsilon_1^{\rho}$.

{\bf\emph{Example-1}}~~All possible type-A graphs with on-shell nodes $1$, $2$ and Lorentz index $\rho$ are displayed by \figref{Fig:2ptGraphsA}. Thus the corresponding type-A numerator $N^{\rho}_A(1,2)$ is given by
\bea
N^{\rho}_A(1,2)=[\epsilon_1\cdot F_2-\epsilon_1(\epsilon_2\cdot 2k_1)]^{\rho},  \Label{Eq:2ptOffShellNum}
\eea
 where the first term in the square bracket refers to $\epsilon_{1\nu} F_2^{\nu\rho}$ while the second means $(\epsilon_2\cdot 2k_1)\epsilon_{1\nu}\eta^{\nu\rho}$ (Such convention will be widely used in the coming discussions).

{\bf\emph{Example-2}}~~All type-A graphs with on-shell nodes $1$, $2$, $3$ and Lorentz index $\rho$ are displayed by \figref{Fig:3ptGraphsA}. The type-A numerators $N^{\rho}_{A}(1,\pmb{\sigma})$ ($\pmb{\sigma}=\{2,3\} \text{and} \{3,2\}$) are then explicitly written as
\bea
N_A^{\rho}(1,2,3)&=&\Big[\epsilon_1\cdot (F_2-\epsilon_2\cdot 2k_1)\cdot(F_3-\epsilon_3\cdot (2k_2+2k_1))\Big]^{\rho}, \nn
N_A^{\rho}(1,3,2)&=&\Big[\epsilon_1\cdot F_3\cdot(F_2-\epsilon_2\cdot 2k_3)-(\epsilon_2\cdot 2k_1) \epsilon_1\cdot F_3 \nn
&&-(\epsilon_3\cdot 2k_1) \epsilon_1\cdot F_2+(\epsilon_2\cdot 2k_1)(\epsilon_3\cdot 2k_1)\epsilon_1+(\epsilon_2\cdot F_3\cdot 2k_1)\epsilon_1\Big]^{\rho},\Label{Eq:3ptOffShellNum}
\eea
which are respectively depicted by  \figref{Fig:3ptGraphsA} (a), (b), (c), (d), (e), (h) and (f), (g), (h), (c), (e), (i).

When contracted with the polarization vector $\epsilon^{\rho}_n$, a type-A graph precisely has the same structure with a graph for on-shell amplitudes. Hence the type-A numerator $N^{\rho}_{A}(1,\pmb{\sigma})$ reproduces the polynomial numerator $N(1,\pmb{\sigma},n)$ in \eqref{Eq:BCJOnShell} by
    \bea
    \epsilon_{n}\cdot N_{A}(1,\pmb{\sigma})=N(1,\pmb{\sigma},n).
    \eea

\begin{figure}
\centering
\includegraphics[width=0.4\textwidth]{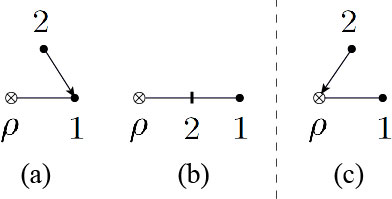}
\caption{Type-B graphs with two on-shell nodes}\label{Fig:2ptGraphsB}
\end{figure}
\begin{figure}
\centering
\includegraphics[width=0.95\textwidth]{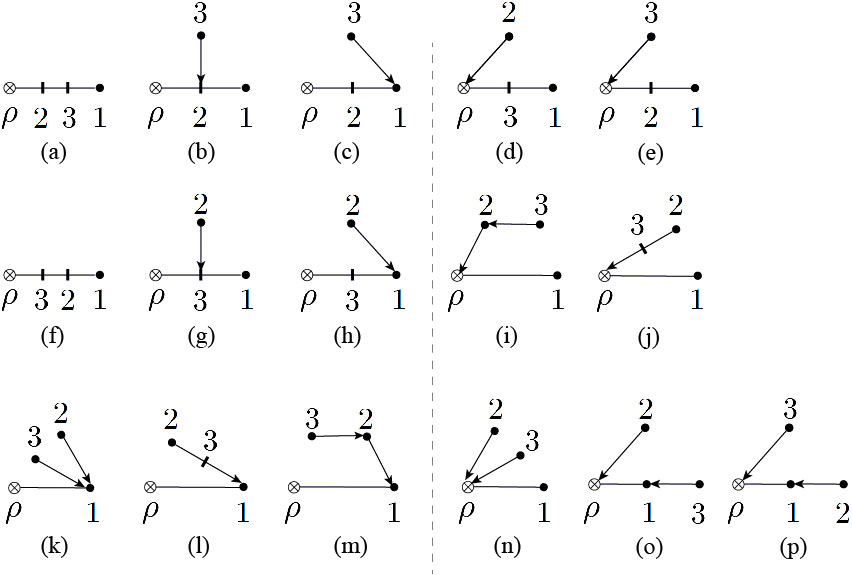}
\caption{Type-B graphs with three on-shell nodes }\label{Fig:3ptGraphsB}
\end{figure}

\subsection{Type-B graphs and numerators}

A type-B graph is obtained when we extract out the $\epsilon_n$ from the master chain and reverse the order of elements on the master chain in a proper way. Particularly, the reference order is defined by $\mathsf{R}=\{\rho,n-1,...,2,1\}$, while the master chain for each graph of type B has the form $(-1)^{n-l}\left(F_{i_l}\cdot ...\cdot F_{i_1}\cdot \epsilon_1\right)^{\rho}$ (note that the relative order of nodes is now $\{\rho, i_l,...,i_1\}$) and the corresponding root set is given by
    $\mathcal{R}=\{\rho,i_l,...,i_1,1\}$ which involves both the outer Lorentz index $\rho$ and the last element (i.e. the element $1$) on the master chain.
     If there is a tree planted at the root $\rho$, it always means that any node in the substructure which is contracted with $\rho$ can play as the root. Type-B numerator $N^{\rho}_{B}(\pmb{\sigma})$ is defined by
\bea
N^{\rho}_{B}(\pmb{\sigma})\equiv\Sl_{\mathcal{F}\in \mathcal{G}^B[\pmb{\sigma}]} C_{\mathcal{F}}^{\rho},\Label{Eq:Type-B}
\eea
where we have summed over all type-B graphs for the permutation $\pmb{\sigma}$ ($\pmb{\sigma}\in P(1,n-1)$). In the special case $n=2$, $\pmb{\sigma}=\{1\}$ the type-B numerator is defined by $N^{\rho}_{B}=\epsilon_1^{\rho}$. Type-B graphs/numerators have the following two features that are different from the type-A ones.  \textbf{(i).} A chain connected to the leftmost node $\rho$ on the master chain has the form $\epsilon_{a}\cdot F_{i'_{l'}}\cdot...\cdot F_{i'_{1}}\cdot 2k_{A}$ where $k_A$ is the total momentum of the substructure which is contracted with the type-B graph.
 \textbf{(ii).} Since the last element $1$ on the master chain may also be a root, it may  not play as the last element in the full permutation $\pmb{\sigma}$ (any node $x$ living on the tree planted at $1$ must satisfy $\sigma^{-1}(1)\prec \sigma^{(-1)}(a)$). As a result, $\pmb{\sigma}$ is an element in $P(1,n-1)$. Explicit examples are given as follows.
\begin{table}
\begin{tabular}{c|c|c|c|c|c|c}
  \hline
  Type-B numerators & $N_{B}^{\rho}(1,2,3)$ & $N_{B}^{\rho}(1,3,2)$  & $N_{B}^{\rho}(2,1,3)$  & $N_{B}^{\rho}(2,3,1)$  & $N_{B}^{\rho}(3,1,2)$  & $N_{B}^{\rho}(3,2,1)$  \\ \hline
    \begin{tabular}{c}
            Graphs in \figref{Fig:3ptGraphsB} \\
            with explicit labels \\
          \end{tabular}
          &\begin{tabular}{c}
           (i)(k)(m)\\
           (n)(o)(p)\\
          \end{tabular}
   & \begin{tabular}{c}
           (j)(k)(l)\\
           (n)(o)(p)\\
          \end{tabular}& \begin{tabular}{c}
           (b)(c)(e)\\
           (i)(n)(o)\\
          \end{tabular} & \begin{tabular}{c}
           (a)(b)(d)\\
           (e)(i)(n) \\
          \end{tabular} &\begin{tabular}{c}
           (d)(g)(h)\\
           (j)(n)(p) \\
          \end{tabular} & \begin{tabular}{c}
           (d)(e)(f)\\
           (g)(j)(n) \\
          \end{tabular} \\
  \hline
\end{tabular}
\caption{The type-B graphs in \figref{Fig:3ptGraphsB} contributing to each type-B numerator $N_{B}^{\rho}(\pmb{\sigma})$ ($\pmb{\sigma}\in P(1,n-1)$) }\label{Table:1}
\end{table}

{\bf\emph{Example-1}}~~All type-B graphs with on-shell nodes $1$, $2$ and Lorentz index $\rho$ are shown by \figref{Fig:2ptGraphsB}. The type-B numerators $N^{\rho}_{B}(1,2)$ and $N^{\rho}_{B}(2,1)$ are explicitly given by
\bea
N_{B}^{\rho}(1,2)&=&(-\epsilon_1^{\rho})\bigl[(\epsilon_2\cdot 2k_1)+(\epsilon_2\cdot 2k_A)\bigr],\nn
N_{B}^{\rho}(2,1)&=&( F_2\cdot\epsilon_1)^{\rho}+(-\epsilon_1^{\rho})(\epsilon_2\cdot 2k_A),\Label{Eq:TYPE-BEG2pt}
\eea
which get contributions from \figref{Fig:2ptGraphsB} (a), (c) and \figref{Fig:2ptGraphsB} (b), (c), respectively. \emph{The momentum $k_A$ denotes the total momentum of all nodes in the substructure that is contracted with this graph.}

{\bf\emph{Example-2}}~~All type-B graphs with on-shell nodes $1$, $2$, $3$ and the outer Lorentz index $\rho$ are shown by \figref{Fig:3ptGraphsB},  while the contributions of type-B graphs to the numerators with all $3!=6$ permutations are presented in \tabref{Table:1}. Therefore, these numerators are explicitly given by
\bea
N_{B}^{\rho}(1,2,3)&=& \epsilon_1^{\rho}(\epsilon_2\cdot 2k_1)(\epsilon_3\cdot 2k_1+\epsilon_3\cdot 2k_2)\nn
&&+\epsilon_1^{\rho}[(\epsilon_2\cdot 2k_A)(\epsilon_3\cdot 2k_2+\epsilon_3\cdot 2k_A+\epsilon_3\cdot 2k_1)+(\epsilon_3\cdot 2k_A)(\epsilon_2\cdot 2k_1)]\nn
N_{B}^{\rho}(1,3,2)&=& \epsilon_1^{\rho}[(\epsilon_2\cdot 2k_1)(\epsilon_3\cdot 2k_1)+\epsilon_2\cdot F_3\cdot 2k_1],\nn
&&+\epsilon_1^{\rho}[\epsilon_2\cdot F_3\cdot 2k_A+(\epsilon_2\cdot 2k_A)(\epsilon_3\cdot 2k_A+\epsilon_3\cdot 2k_1)+(\epsilon_3\cdot 2k_A)(\epsilon_2\cdot 2k_1)],\nn
N_{B}^{\rho}(2,1,3)&=& -(F_2\cdot\epsilon_1)^{\rho}(\epsilon_3\cdot 2k_1+\epsilon_3\cdot 2k_2)\nn
&&+[-(\epsilon_3\cdot 2k_A)(F_2\cdot\epsilon_1)^{\rho}+\epsilon_1^{\rho}(\epsilon_2\cdot 2k_A)(\epsilon_3\cdot 2k_A+\epsilon_3\cdot 2k_2)+\epsilon_1^{\rho}(\epsilon_3\cdot 2k_A)(\epsilon_2\cdot 2k_1)],\nn
N_{B}^{\rho}(2,3,1)&=& [( F_2\cdot F_3\cdot\epsilon_1)^{\rho}-(F_2\cdot\epsilon_1)^{\rho}(\epsilon_3\cdot 2k_2)]\nn
&&+[-(\epsilon_2\cdot 2k_A)(F_3\cdot\epsilon_1)^{\rho}-(\epsilon_3\cdot 2k_A)(F_2\cdot\epsilon_1)^{\rho}+\epsilon_1^{\rho}(\epsilon_2\cdot 2k_A)(\epsilon_3\cdot 2k_2+\epsilon_3\cdot 2k_A)],\nn
N_{B}^{\rho}(3,1,2)&=& -(F_3\cdot\epsilon_1)^{\rho}\big(\epsilon_2\cdot 2k_1+\epsilon_2\cdot 2k_3\big)\nn
&&+[-(\epsilon_2\cdot 2k_A)(F_3\cdot\epsilon_1)^{\rho}+\epsilon_1^{\rho}(-(F_3\cdot\epsilon_2)^{\rho}+(\epsilon_3\cdot 2k_A)(\epsilon_2\cdot 2k_A+\epsilon_2\cdot 2k_1))],\nn
N_{B}^{\rho}(3,2,1)&=& \bigl[-(F_3\cdot\epsilon_1)^{\rho}(\epsilon_2\cdot 2k_3)+(F_3\cdot F_2\cdot\epsilon_1)^{\rho}\bigr]\nn
&&+[-(\epsilon_2\cdot 2k_A)(F_3\cdot\epsilon_1)^{\rho}-(\epsilon_3\cdot 2k_A)(F_2\cdot\epsilon_1)^{\rho}+\epsilon_1^{\rho}(\epsilon_2\cdot F_3\cdot 2k_A+(\epsilon_2\cdot 2k_A)(\epsilon_3\cdot 2k_A))].\Label{Eq:Type-BEG}
\eea

 \begin{figure}
\centering
\includegraphics[width=0.75\textwidth]{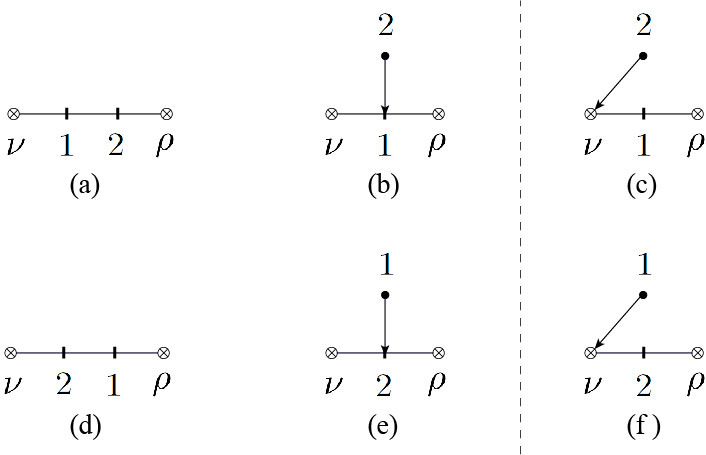}
\caption{Type-C graphs with two on-shell nodes}\label{Fig:2ptGraphsC}
\end{figure}

\begin{figure}
\centering
\includegraphics[width=0.95\textwidth]{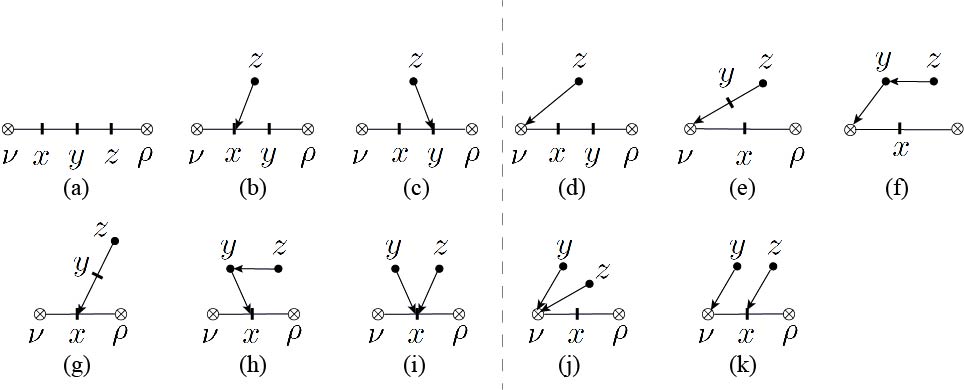}
\caption{ Possible structures of type-C graphs with three on-shell nodes }\label{Fig:3ptGraphsC}
\end{figure}

\begin{table}
\begin{tabular}{c|c|c|c|c|c|c}
  \hline
  Type-C numerators & $N_{C}^{\nu\rho}(1,2,3)$ & $N_{C}^{\nu\rho}(1,3,2)$  & $N_{C}^{\nu\rho}(2,1,3)$  & $N_{C}^{\nu\rho}(2,3,1)$  & $N_{C}^{\nu\rho}(3,1,2)$  & $N_{C}^{\nu\rho}(3,2,1)$  \\ \hline
   \begin{tabular}{c}
            Graphs in \figref{Fig:3ptGraphsC} \\
            with explicit labels \\
          \end{tabular}  &\begin{tabular}{c}
            (a)[1,2,3]\\
            (b)[1,2,3]\\
            (b)[1,3,2]\\
            (c)[1,2,3]\\
            (d)[1,2,3]\\
            (d)[1,3,2]\\
            (d)[2,3,1]\\
            (f)[1,2,3]\\
            (f)[2,1,3]\\
            (f)[3,1,2]\\
            (h)[1,2,3]\\
            (i)[1,2,3]\\
            (j)[1,2,3]\\
            (j)[2,1,3]\\
            (j)[3,1,2]\\
            (k)[1,2,3]\\
            (k)[1,3,2]\\
            (k)[2,1,3]
          \end{tabular}
   & \begin{tabular}{c}
            (a)[1,3,2]\\
            (b)[1,2,3]\\
            (b)[1,3,2]\\
            (c)[1,3,2]\\
            (d)[1,2,3]\\
            (d)[1,3,2]\\
            (d)[3,2,1]\\
            (e)[1,3,2]\\
            (f)[2,1,3]\\
            (f)[3,1,2]\\
            (g)[1,3,2]\\
            (i)[1,2,3]\\
            (j)[1,2,3]\\
            (j)[2,1,3]\\
            (j)[3,1,2]\\
            (k)[1,2,3]\\
            (k)[1,3,2]\\
            (k)[3,1,2]
          \end{tabular}& \begin{tabular}{c}
            (a)[2,1,3]\\
            (b)[2,1,3]\\
            (b)[2,3,1]\\
            (c)[2,1,3]\\
            (d)[2,1,3]\\
            (d)[2,3,1]\\
            (d)[3,1,2]\\
            (e)[3,2,1]\\
            (f)[2,1,3]\\
            (f)[1,2,3]\\
            (h)[2,1,3]\\
            (i)[2,1,3]\\
            (j)[1,2,3]\\
            (j)[2,1,3]\\
            (j)[3,1,2]\\
            (k)[1,2,3]\\
            (k)[2,1,3]\\
            (k)[2,3,1]\\
          \end{tabular} & \begin{tabular}{c}
            (a)[2,3,1]\\
            (b)[2,3,1]\\
            (b)[2,1,3]\\
            (c)[2,3,1]\\
            (d)[2,3,1]\\
            (d)[3,1,2]\\
            (d)[2,1,3]\\
            (e)[2,3,1]\\
            (e)[3,2,1]\\
            (f)[1,2,3]\\
            (g)[2,3,1]\\
            (i)[2,1,3]\\
            (j)[1,2,3]\\
            (j)[2,1,3]\\
            (j)[3,1,2]\\
            (k)[2,3,1]\\
            (k)[2,1,3]\\
            (k)[3,2,1]
          \end{tabular} &\begin{tabular}{c}
            (a)[3,1,2]\\
            (b)[3,1,2]\\
            (b)[3,2,1]\\
            (c)[3,2,1]\\
            (d)[3,1,2]\\
            (d)[3,2,1]\\
            (d)[2,1,3]\\
            (e)[2,3,1]\\
            (e)[1,3,2]\\
            (f)[3,1,2]\\
            (h)[3,1,2]\\
            (i)[3,1,2]\\
            (j)[1,2,3]\\
            (j)[2,1,3]\\
            (j)[3,1,2]\\
            (k)[3,1,2]\\
            (k)[3,2,1]\\
            (k)[1,3,2]
          \end{tabular} & \begin{tabular}{c}
            (a)[3,2,1]\\
            (b)[3,2,1]\\
            (b)[3,1,2]\\
            (c)[3,2,1]\\
            (d)[3,2,1]\\
            (d)[3,1,2]\\
            (d)[2,1,3]\\
            (e)[1,3,2]\\
            (e)[2,3,1]\\
            (e)[3,2,1]\\
            (g)[3,2,1]\\
            (i)[3,1,2]\\
            (j)[1,2,3]\\
            (j)[2,1,3]\\
            (j)[3,1,2]\\
            (k)[3,1,2]\\
            (k)[3,2,1]\\
            (k)[2,3,1]
          \end{tabular} \\
  \hline
\end{tabular}
\caption{Contributions of graphs for all type-C numerators $N_{C}^{\nu\rho}(\pmb{\sigma})$ where $\pmb{\sigma}\in P(1,3)$. The notation, e.g., (a)[1,2,3] stands for the graph which is obtained from  \figref{Fig:3ptGraphsC} (a), via replacing $x$, $y$, and $z$ by nodes $1$, $2$ and $3$, respectively.}\label{table:2}
\end{table}

\subsection{Type-C graphs and numerators}

 In a graph of type-C, both ends of the master chain are outer Lorentz indices. The reference order is $\mathsf{R}=\{\rho,n-1,...,2,1,\nu\}$ where $\nu$, $\rho$ are the two outer indices and $1$, ..., $n-1$ are the on-shell nodes. The master chain for each graph of this type has the form $(-1)^{n+1-l}\left(F_{i_1}\cdot...\cdot F_{i_l}\right)^{\nu\rho}$ while the root set corresponding to this master chain is $\mathcal{R}=\{\nu,i_1,...,i_l\}$. Type-C numerators $N^{\nu\rho}_{C}(\pmb{\sigma})$ ($\pmb{\sigma}\in P(1,n-1)$) are defined by collecting all type-C graphs corresponding to the permutation $\pmb{\sigma}$, i.e.
\bea
N^{\nu\rho}_{C}(\pmb{\sigma})=\Sl_{\mathcal{F}\in \mathcal{G}^C[\pmb{\sigma}]} C_{\mathcal{F}}^{\nu\rho}, \Label{Eq:TypeC}
\eea
where $\pmb{\sigma}\in P(1,n-1)$ and the momentum $k_A$ of the leftmost node $\nu$ is the total momentum of all nodes in the subgraph which is connected to the index $\nu$. According to the definition, the case  $n=2$ is trivial, $N^{\nu\rho}_{C}(1)=F_{1}^{\nu\rho}$. The nontrivial cases with $n=3$ and $n=4$ are given as follows.

{\bf\emph{Example-1}}~~Type-C graphs with $n=3$ are given by \figref{Fig:2ptGraphsC}, while type-C numerators $N_{C}^{\nu\rho}(1,2)$ and $N_{C}^{\nu\rho}(2,1)$ are expressed as
\bea
N_{C}^{\nu\rho}(1,2)&=&\big[F_1\cdot(F_2-\epsilon_2\cdot 2k_1)-F_1(\epsilon_2\cdot 2k_A)-F_2(\epsilon_1\cdot 2k_A)\big]^{\nu\rho},\nn
N_{C}^{\nu\rho}(2,1)&=&\big[F_2\cdot(F_1-\epsilon_1\cdot 2k_2)-F_1(\epsilon_2\cdot 2k_A)-F_2(\epsilon_1\cdot 2k_A)\big]^{\nu\rho},\Label{Eq:TYPE-CEG2pt}
\eea
which are respectively characterized by the type-C graphs \figref{Fig:2ptGraphsC} (a), (b), (c), (f) and (d), (e), (c), (f).

{\bf\emph{Example-2}}~~All possible structures of type-C graphs for $n=4$ are shown by \figref{Fig:3ptGraphsC}. When collecting graphs corresponding to different permutations $\pmb{\sigma}\in P(1,3)$ as shown by \tabref{table:2}, we get all type-C numerators
\bea
N^{\nu\rho}_{C}(1,2,3)&=&\left(F_1\cdot F_2\cdot F_3\right)^{\nu\rho}-\left(F_1\cdot F_2\right)^{\nu\rho}\epsilon_3\cdot 2(k_1+k_2)-\left(F_1\cdot F_3\right)^{\nu\rho}(\epsilon_2\cdot 2k_1)\nn
&&+(\epsilon_3\cdot 2k_2)(\epsilon_2\cdot 2k_1)F_1^{\nu\rho}+(\epsilon_2\cdot 2k_1)(\epsilon_3\cdot 2k_1)F_1^{\nu\rho}+(...),\nn
N^{\nu\rho}_{C}(1,3,2)&=&\left(F_1\cdot F_3\cdot F_2\right)^{\nu\rho}-\left(F_1\cdot F_3\right)^{\nu\rho}\epsilon_2\cdot 2(k_1+k_3)-\left(F_1\cdot F_2\right)^{\nu\rho}(\epsilon_3\cdot 2k_1)\nn
&&+(\epsilon_2\cdot F_3\cdot 2k_1)F_1^{\nu\rho}+(\epsilon_2\cdot 2k_1)(\epsilon_3\cdot 2k_1)F_1^{\nu\rho}+(...),\nn
N^{\nu\rho}_{C}(2,1,3)&=&\left(F_2\cdot F_1\cdot F_3\right)^{\nu\rho}-\left(F_2\cdot F_1\right)^{\nu\rho}\epsilon_3 \cdot 2(k_1+k_2)-\left(F_2\cdot F_3\right)^{\nu\rho}(\epsilon_1\cdot 2k_2)\nn
&&+(\epsilon_3\cdot 2k_1)(\epsilon_1\cdot 2k_2)F_2^{\nu\rho}+(\epsilon_3\cdot 2k_2)(\epsilon_1\cdot 2k_2)F_2^{\nu\rho}+(...),\nn
N^{\nu\rho}_{C}(2,3,1)&=&\left(F_2\cdot F_3\cdot F_1\right)^{\nu\rho}-\left(F_2\cdot F_3\right)^{\nu\rho}\epsilon_1\cdot 2(k_2+k_3)-\left(F_2\cdot F_1\right)^{\nu\rho}(\epsilon_3\cdot 2k_2)\nn
&&+(\epsilon_1\cdot F_3\cdot 2k_2)F_2^{\nu\rho}+(\epsilon_3\cdot 2k_2)(\epsilon_1\cdot 2k_2)F_2^{\nu\rho}+(...),\nn
N^{\nu\rho}_{C}(3,1,2)&=&\left(F_3\cdot F_1\cdot F_2\right)^{\nu\rho}-\left(F_3\cdot F_1\right)^{\nu\rho}\epsilon_2\cdot2(k_1+k_3)-\left(F_3\cdot F_2\right)^{\nu\rho}(\epsilon_1\cdot 2k_3)\nn
&&+(\epsilon_2\cdot 2k_1)(\epsilon_1\cdot 2k_3)F_3^{\nu\rho}+(\epsilon_2\cdot 2k_3)(\epsilon_1\cdot 2k_3)F_3^{\nu\rho}+(...),\nn
N^{\nu\rho}_{C}(3,2,1)&=&\left(F_3\cdot F_2\cdot F_1\right)^{\nu\rho}-\left(F_3\cdot F_2\right)^{\nu\rho}\epsilon_1\cdot2(k_2+k_3)-\left(F_3\cdot F_1\right)^{\nu\rho}(\epsilon_2\cdot 2k_3)\nn
&&+(\epsilon_1\cdot F_2\cdot 2k_3)F_3^{\nu\rho}+(\epsilon_2\cdot 2k_3)(\epsilon_1\cdot 2k_3)F_3^{\nu\rho}+(...), \Label{Eq:TYPE-CEG3pt}
\eea
where we have only explicitly written down the contribution of graphs in which no tree is planted at $\nu$ (i.e. graphs (a), (b), (c), (g), (h), (i)).  The dots in each numerator denote the graphs involving tree(s) planted at $\nu$, whose expression contains the total momentum (say $k_A$) of all nodes in the substructure that is contracted with the leftmost Lorentz index $\nu$.

 \begin{figure}
\centering
\includegraphics[width=0.7\textwidth]{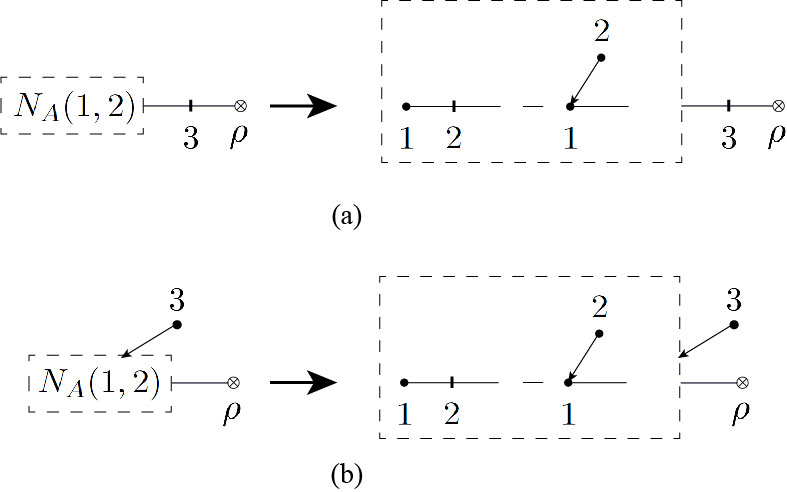}
\caption{Relations between the type-A graphs $\mathcal{F}\in \mathcal{G}^A[2,3]$ and the graphs with fewer points for the division $\pmb{\sigma}=\{2,3\}\to\pmb{\sigma}_L=\{2\},\pmb{\sigma}_R=\{3\}$}\label{Fig:3ptTypesRelations1}
\end{figure}
 \begin{figure}
\centering
\includegraphics[width=0.95\textwidth]{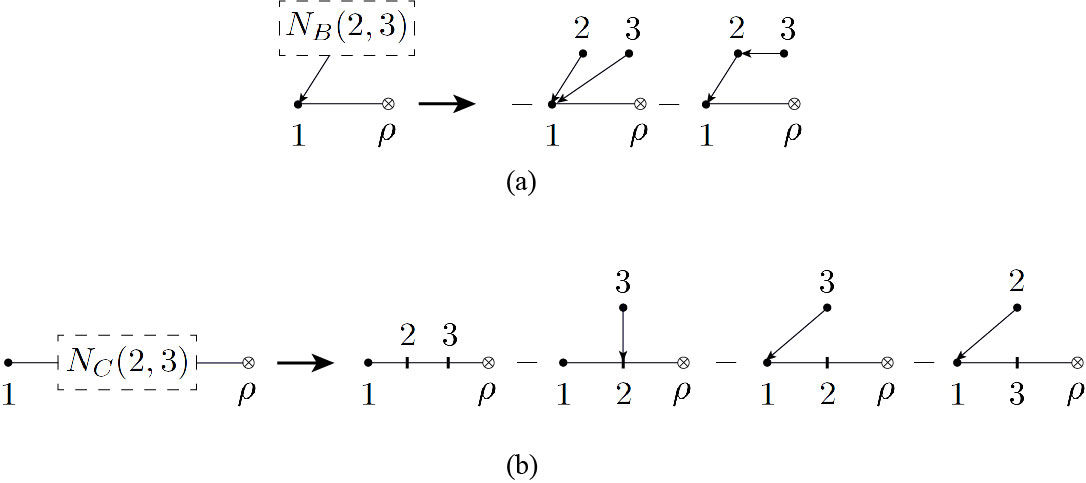}
\caption{Relations between the type-A graphs $\mathcal{F}\in \mathcal{G}^A[2,3]$ and the graphs with fewer points for the division $\pmb{\sigma}=\{2,3\}\to\pmb{\sigma}_L=\emptyset,\pmb{\sigma}_R=\{2,3\}$}\label{Fig:3ptTypesRelations2}
\end{figure}

\subsection{A relation between the three types of numerators}

 \begin{figure}
\centering
\includegraphics[width=0.95\textwidth]{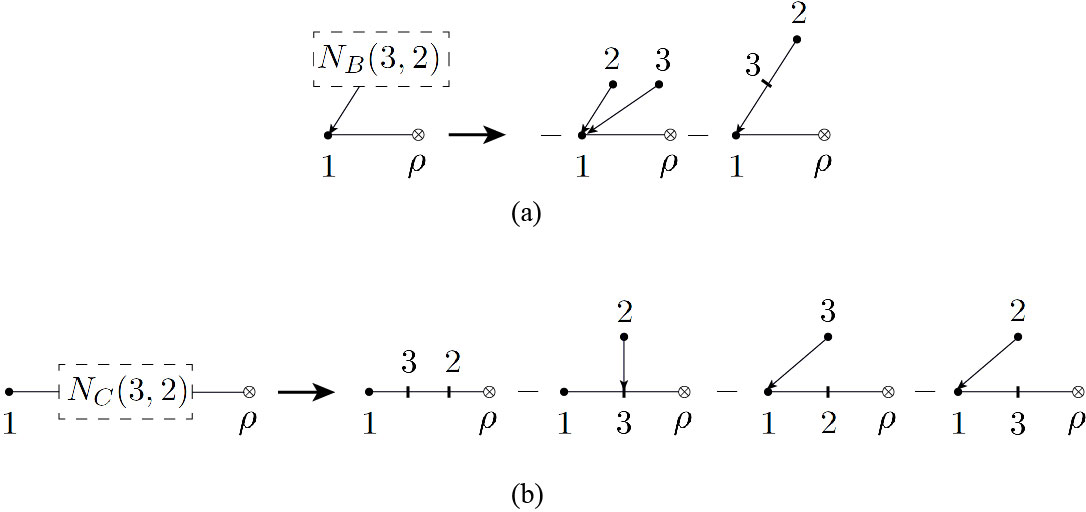}
\caption{Relations between the type-A graphs $\mathcal{F}\in \mathcal{G}^A[3,2]$ and the graphs with fewer points for the division $\pmb{\sigma}=\{2,3\}\to\pmb{\sigma}_L=\emptyset,\pmb{\sigma}_R=\{3,2\}$}\label{Fig:3ptTypesRelations3}
\end{figure}

Now we demonstrate an important property of the type-A numerator $N_A^{\rho}(1,\pmb{\sigma})$: if a permutation $\pmb{\sigma}=\{\sigma_2,...,\sigma_{n-1}\}\in P(2,n-1)$ can be divided into two parts $\pmb{\sigma}_L=\{\sigma_2,...,\sigma_{i-1}\}$ and $\pmb{\sigma}_R=\{\sigma_{i},...,\sigma_{n-1}\}$ which respectively satisfy $\pmb{\sigma}_L\in P(2,i-1)$ and $\pmb{\sigma}_R\in P(i,n-1)$ (in the special case $i=2$, $\pmb{\sigma}_L=\emptyset$),  the numerator $N_A^{\rho}(1,\pmb{\sigma})$ can be expressed by lower-point type-A, -B and -C numerators via the following relation

\bea
N_A^{\rho}(1,\pmb{\sigma})=\bigl[N_A(1,\pmb{\sigma}_L)\cdot N_C(\pmb{\sigma}_R)-N_A(1,\pmb{\sigma}_L)N_{B}(\pmb{\sigma}_R)\cdot 2k_{1,i-1}\bigr]^{\rho}.\Label{Eq:Num Relations}
\eea
 \begin{figure}
\centering
\includegraphics[width=1\textwidth]{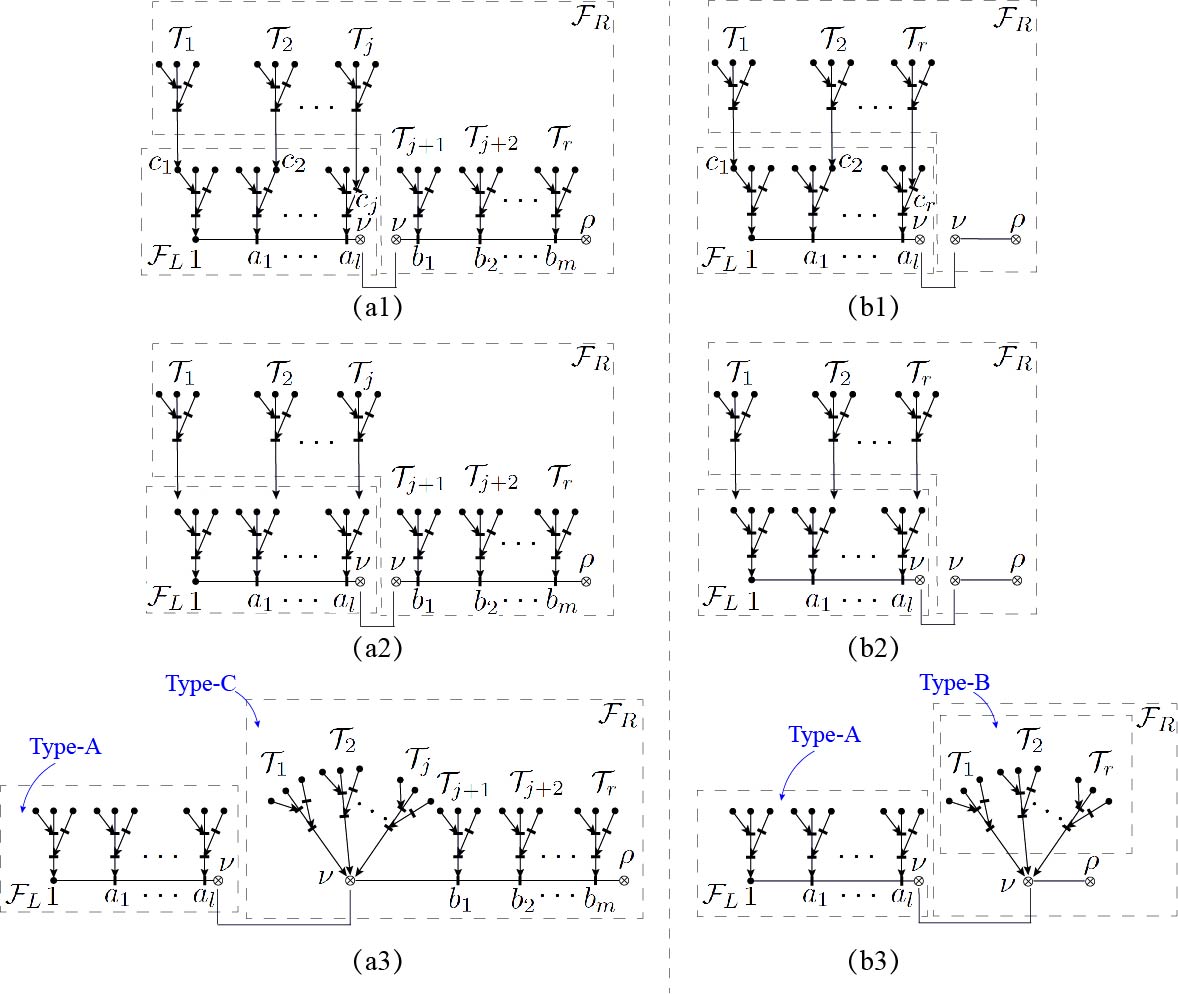}
\caption{If $\pmb{\sigma}=\{\sigma_1,...,\sigma_{n-1}\}$ can be divided into $\pmb{\sigma}_L=\{\sigma_1,...,\sigma_{i-1}\}\in P(1,i-1)$ and $\pmb{\sigma}_R=\{\sigma_i,...,\sigma_{n-1}\}\in P(i,n-1)$, a graph $\mathcal{F}\in \mathcal{G}^A[\pmb{\sigma}]$ is decomposed into $\mathcal{F}_L$ and $\mathcal{F}_R$, as shown by (a1) and (b1). If a graph $\mathcal{F}_R$ is consistent with the permutation $\pmb{\sigma}_R$, a change of the roots of trees $\mathcal{T}_s$ which are planted at $c_s\in \mathcal{F}_L$ must also result in a graph consistent with the permutation $\pmb{\sigma}_R$. Thus all choices of $c_s$ must be included in the summation over $\mathcal{F}_R$, as shown by (a2) and (b2). Here, an arrow line pointing to a boxed region means that we sum over all possible graphs where the arrow line ends at a node in this regions. A type-A graph is finally factorized into a type-A graph and a type-C graph (see (a3)), or a type-A graph and a type-B graph (see (b3)).}\label{Fig:TypesRelationsGen}
\end{figure}

To understand \eqref{Eq:Num Relations}, we first study the type-A numerator  $N_A^{\rho}(1,2,3)$. In this case $\pmb{\sigma}=\{2,3\}$ can be divided into $\pmb{\sigma}_L=\{2\},\pmb{\sigma}_R=\{3\}$ or $\pmb{\sigma}_L=\emptyset,\pmb{\sigma}_R=\{2,3\}$. Correspondingly, the relation (\ref{Eq:Num Relations}) gives
\bea
N_A^{\rho}(1,2,3)&=&\bigl[N_A(1,2)\cdot N_C(3)-N_A(1,2)N_{B}(3)\cdot 2k_{1,2}\bigr]^{\rho}\nn
&=&\bigl[N_A(1)\cdot N_C(2,3)-N_A(1)N_{B}(2,3)\cdot 2k_{1}\bigr]^{\rho},\Label{Eq:Num RelationsEG1}
\eea
where $N_A(1)=\epsilon_1$.  The above equation can be verified by contractions between graphs straightforwardly. Particularly, the two terms in the first equality are shown by \figref{Fig:3ptTypesRelations1} (a), (b) which respectively reproduce \figref{Fig:3ptGraphsA} (d), (e) and \figref{Fig:3ptGraphsA} (a), (h), (b), (c). The two terms in the second equality can be respectively understood by \figref{Fig:3ptTypesRelations2} (a) and (b) which reproduce \figref{Fig:3ptGraphsA} (a), (h) and  \figref{Fig:3ptGraphsA} (b), (c), (d), (e). Hence, both divisions give rise to the expected expression of $N_A^{\rho}(1,2,3)$ (see \eqref{Eq:3ptOffShellNum}).

Another example is given by $N_A^{\rho}(1,3,2)$. The division $\{3,2\}\to \pmb{\sigma}_L=\{3\},\pmb{\sigma}_R=\{2\}$ is invalid because $\{3\}\notin P\{2\}$. Thus there is only one possible division: $\{3,2\}\to \pmb{\sigma}_L=\emptyset,\pmb{\sigma}_R=\{3,2\}$ which corresponds to the following relation
\bea
N_A^{\rho}(1,3,2)&=&\bigl[N_A(1)\cdot N_C(3,2)-N_A(1)N_{B}(3,2)\cdot 2k_{1}\bigr]^{\rho}.\Label{Eq:Num RelationsEG2}
\eea
In the above expression, the first term on the RHS is shown by \figref{Fig:3ptTypesRelations3} (a) which reproduces \figref{Fig:3ptGraphsA} (e), (i), while the second term is shown by \figref{Fig:3ptTypesRelations3} (b) which reproduces \figref{Fig:3ptGraphsA} (f), (g), (h), (c). All together, we get the expected expression of $N_A^{\rho}(1,3,2)$  (see \eqref{Eq:3ptOffShellNum}).

\subsubsection{A general proof of \eqref{Eq:Num Relations}}

Now let us understand the general relation (\ref{Eq:Num Relations}) by manipulations of graphs. According to the definition (\ref{Eq:Type-A}), $N_A^{\rho}(1,\pmb{\sigma})$ is given by the sum of all graphs which are consistent with the permutation $\pmb{\sigma}$. Since $\pmb{\sigma}$ can be divided into a left part $\pmb{\sigma}_L=\{\sigma_2,...,\sigma_{i-1}\}\in P(2,i-1)$ and a right part $\pmb{\sigma}_R=\{\sigma_i,...,\sigma_{n-1}\}\in P(i,n-1)$, a graph $\mathcal{F}\in \mathcal{G}^A[\pmb{\sigma}]$ in general can be given by a contraction of two subgraphs $\mathcal{F}_L$ and $\mathcal{F}_R$ (as shown by \figref{Fig:TypesRelationsGen} (a1) and (b1)) which are respectively consistent with the permutations $\pmb{\sigma}_L (\in P(2,i-1))$ and $\pmb{\sigma}_R (\in P(i,n-1))$. These subgraphs have the following crucial properties:
\begin{itemize}

\item [{\bf{(i).}}] \emph{The structure of $\mathcal{F}_L$}~~The subgraph $\mathcal{F}_L$ (see \figref{Fig:TypesRelationsGen} (a1) and (b1)) is just a type-A graph whose reference order is  $\mathsf{R}_L=\{\nu, i-1,...,2,1\}$, the master chain has the form $(-1)^{i-l}(\epsilon_1\cdot F_{a_1}\cdot...\cdot F_{a_{l}})^{\nu}$ ($a_1,...,a_l\in \{2,...,n-1\}$), while the corresponding root set is $\mathcal{R}_L=\{1,a_1,...,a_l\}$.

\item [{\bf{(ii).}}] \emph{The structure of $\mathcal{F}_R$}~~The subgraph $\mathcal{F}_R$, is determined by the reference order $\mathsf{R}_R=\{\rho,n-1,...,i,\nu\}$. As displayed in \figref{Fig:TypesRelationsGen} (a1),  the master chain in $\mathcal{F}_R$  has the form $ (-1)^{(n-i+2)-m}(F_{b_1}\cdot...\cdot F_{b_{m}})^{\nu\rho}$, while the corresponding root set is given by $\mathcal{R}_R=\{1,2,...,i-1\}\cup\{b_1,...,b_m,\rho\}$. The special case $\{b_1,...,b_m\}=\emptyset$ is shown by \figref{Fig:TypesRelationsGen} (b1), in which the master chain has the form  $ (-1)^{n-i+2}\eta^{\nu\rho}$ and the root set is given by $\mathcal{R}_R=\{1,2,...,i-1\}$.

\item [{\bf{(iii).}}] \emph{Any node in $\mathcal{F}_L$ can not play as an internal node of a chain which starts from $\mathcal{F}_R$.}~~This can be understood as follows. If a node $x\in \mathcal{F}_L$ is an internal node of a chain whose starting node is $y\in\mathcal{F}_R$, $y$ must have a higher weight than the node $x$ in the reference order $\mathsf{R}=\{n-1,...,1\}$, i.e., $y<x$. Consequently, $\pmb{\sigma}_L$ cannot be a permutation of elements $1,...,i-1$, which is in conflict with our assumption $\pmb{\sigma}_L\in P(1,i-1)$.

\item [{\bf{(iv).}}] {\emph{If a graph $\mathcal{F}_R$, which involves trees $\mathcal{T}_1,...,\mathcal{T}_j$ correspondingly planted at given nodes $c_1,...,c_j\in \{1,...,i-1\}$ (see \figref{Fig:TypesRelationsGen} (a1) and (b1)), is consistent with the permutation $\pmb{\sigma}_R$, those graphs obtained by changing  $c_1,...,c_j\in \{1,...,i-1\}$ arbitrarily are also consistent with $\pmb{\sigma}_R$.}} This is because a change of roots $c_1,...,c_j\in \mathcal{F}_R$  does not affect the possible relative orders of nodes in $\mathcal{F}_R$.

\end{itemize}
With the above features, the type-A numerator $N_A^{\rho}(1,\pmb{\sigma})$ is decomposed as follows
\bea
N_A^{\rho}(1,\pmb{\sigma})=\Sl_{\mathcal{F}\in \mathcal{G}^A(\pmb{\sigma})}C_{\mathcal{F}}^{\,\rho}&=&\Biggl(\Sl_{\mathcal{F}_L\in \mathcal{G}^A(\pmb{\sigma}_L)}C_{\mathcal{F}_L}^{\,\nu}\Biggr)\, \Biggl(\Sl_{\mathcal{F}_R}C^{\,\nu\rho}_{\mathcal{F}_R}\Biggr),\Label{Eq:Num Relations2}
\eea
where the outer Lorentz indices in each factor are carried by the master chain. Besides, the upper indice $\nu$ in $C_{\mathcal{F}_L}^{\,\nu}$ and $C^{\,\nu\rho}_{\mathcal{F}_R}$ here refers to a contraction of the two parts.  As pointed in property (i), the expression in the first parenthesis is just the type-A numerator $N_A^{\nu}(1,\pmb{\sigma}_L)$. In the following, we study the expression in the second  parenthesis according to whether $\{b_1,...,b_m\}$ is empty.

{\bf\emph{The case $\{b_1,...,b_m\}\neq\emptyset$}}~~~Assuming there are trees $\mathcal{T}_1$,...,$\mathcal{T}_j$ in $\mathcal{F}_R$  rooted at $c_1,...,c_j\in\{1,2,...,i-1\}$ and trees $\mathcal{T}_{j+1}$,...,$\mathcal{T}_r$ rooted at $d_1,...,d_k\in\{b_1,...,b_m\}\neq \emptyset$, the second factor in \eqref{Eq:Num Relations2} is explicitly written as
\bea
\Sl_{\mathcal{F}_R}C^{\,\nu\rho}_{\mathcal{F}_R}&=&\Sl_{\text{master chains}}\,(-1)^{(n-i+2)-m}(F_{b_1}\cdot...\cdot F_{b_m})^{\nu\rho}\Label{Eq:Num Relations3}\\
&&\times\Sl_{\substack{\text{tree structures}\\ \mathcal{T}_{1},...,\mathcal{T}_r}}\Biggl[\,\Sl_{c_1,...,c_j}\,\prod\limits_{s=1}^j \Big(C_{\mathcal{T}_s}\cdot 2k_{c_s}\Big)\Biggr]\Biggl[\,\Sl_{d_1,...,d_k}\,\prod\limits_{t=j+1}^r \Big(C_{\mathcal{T}_t}\cdot 2k_{d_t}\Big)\Biggr],\nonumber
\eea
where $(C_{\mathcal{T}_s}\cdot 2k_{c_s})$ and $(C_{\mathcal{T}_t}\cdot 2k_{d_t})$ are respectively the contributions of trees planted at roots $c_s\in \mathcal{F}_L$ and $d_t\in \mathcal{F}_R$ (noting that $c_s$ must be an ending  node of a chain which starts from $\mathcal{F}_R$, as pointed in property (iii)).
According to (iv), the summation over $c_1,...,c_j$ should be taken over all $c_1,...,c_j\in \{1,...,i-1\}$ (see \figref{Fig:TypesRelationsGen} (a2)). It follows that the second factor in \eqref{Eq:Num Relations2} can be reexpressed by
\bea
\Sl_{c_1,...,c_j\in \{1,...,i-1\}}\,\prod\limits_{s=1}^j \Big(C_{\mathcal{T}_s}\cdot 2k_{c_s}\Big)=\prod\limits_{s=1}^j \Big(C_{\mathcal{T}_s}\cdot 2k_{1,i-1}\Big),
\eea
where $k_{1,i-1}=\Sl_{c_s\in\{1,...,i-1\}}k_{c_s}$ is the total momentum of on-shell nodes in $\mathcal{F}_L$. Then \eqref{Eq:Num Relations3} turns into
 \bea
&&~~\Sl_{\mathcal{F}_R}C^{\,\nu\rho}_{\mathcal{F}_R}\Label{Eq:Num Relations4}\\
&=&\Sl_{\substack{\text{master}\\ \text{chains}}}\,\Sl_{\substack{\text{tree structures}\\ \mathcal{T}_{1},...,\mathcal{T}_r}}\Sl_{d_1,...,d_k}(-1)^{(n-i+2)-m}(F_{b_1}\cdot...\cdot F_{b_m})^{\nu\rho}\Biggl[\prod\limits_{s=1}^j \Big(C_{\mathcal{T}_s}\cdot 2k_{1,i-1}\Big)\prod\limits_{t=j+1}^r \Big(C_{\mathcal{T}_t}\cdot 2k_{d_t}\Big)\Biggr],\nonumber
\eea
where the first factor in the square brackets can be considered as the contribution of trees planted at $\nu$ (whose momentum is defined as the total momentum of elements in the subgraph $\mathcal{F}_L$), the second factor is the contribution of trees planted at the roots in $\{b_1,...,b_k\}$. Therefore, each term in the above expression is just a contribution of a type-C graph (as shown by \figref{Fig:TypesRelationsGen} (a3)) with the reference order $\mathsf{R}_R=\{\rho,n-1,...,i,\nu\}$, the master chain  of the form $(-1)^{(n-i+2)-m}(F_{b_1}\cdot...\cdot F_{b_{m}})^{\nu\rho}$ and the root set $\mathcal{R}_R=\{\nu,b_1,...,b_m,\rho\}$.
When all possible configurations of the master chains, tree structures and all choices of roots $d_1,...,d_k\in \{i,...,n-1\}$, which are consistent with the permutation $\pmb{\sigma}_R$,  are summed over, \eqref{Eq:Num Relations4} becomes the type-C numerator
\bea
\Sl_{\mathcal{F}_R}C^{\,\nu\rho}_{\mathcal{F}_R}&=&N^{\,\nu\rho}_C (\pmb{\sigma}_R).\Label{Eq:Num Relations5}
\eea

{\bf\emph{The case $\{b_1,...,b_m\}=\emptyset$}}~~~The master chain of $\mathcal{F}_R$ in this special case is just the metric $\eta^{\nu\rho}$ and the summation over $\mathcal{F}_R$ in \eqref{Eq:Num Relations2} becomes
\bea
\Sl_{\mathcal{F}_R}C^{\,\nu\rho}_{\mathcal{F}_R}&=&(-1)^{n-i}\eta^{\nu\rho}\Sl_{\substack{\text{tree structures}\\ \mathcal{T}_{1},...,\mathcal{T}_r}}\Biggl[\,\Sl_{c_1,...,c_r}\,\prod\limits_{s=1}^r \Big(C_{\mathcal{T}_s}\cdot 2k_{c_s}\Big)\Biggr]\nn
&=&\eta^{\nu\rho}\Sl_{\substack{\text{tree structures}\\ \mathcal{T}_{1},...,\mathcal{T}_r}}\Biggl[\,(-1)^{n-i}\prod\limits_{s=1}^r \Big(C_{\mathcal{T}_s}\cdot 2k_{1,i-1}\Big)\Biggr],
\Label{Eq:Num Relations6}
\eea
where trees $\mathcal{T}_1,...,\mathcal{T}_r$ are planted at roots $c_1,...,c_r\in \mathcal{F}_L$. We have summed over all choices of roots $c_1,...,c_r\in\mathcal{F}_L$ as shown by \figref{Fig:TypesRelationsGen} (b2) and used $k_{1,i-1}=\sum_{c_s\in\{1,...,i-1\}}k_{c_s}$ on the second line as shown by \figref{Fig:TypesRelationsGen} (b3).
Now we illustrate that the expression in the square brackets is $\left[-N_{B}(\pmb{\sigma}_R)\cdot 2k_{1,i-1}\right]$. Without loss of generality, we assume that the tree $\mathcal{T}_1$ contains the chain that is led by the highest-weight node $i$ in $\mathcal{F}_R$. The expression in the square brackets in \eqref{Eq:Num Relations6} is thus written as
\bea
-\Bigl[(-1)^{n-i+1}C_{\mathcal{T}_1}^{\,\mu}\prod\limits_{s=2}^r \Big(C_{\mathcal{T}_s}\cdot 2k_{1,i-1}\Big)\Bigr]\cdot (k_{1,i-1})_{\mu}.
\eea
Then the trees $\mathcal{T}_2,...,\mathcal{T}_r$ can be considered as trees planted at the node $\mu$, while the $\mathcal{T}_1$ stands for the structure with trees planted at the on-shell nodes on the chain that is led by $i$. The expression in the square brackets is just the contribution of a type-B graph whose reference order is $\mathsf{R}_R=\{\mu,n-1,n-2,...,i\}$,  the master chain has the form $(-1)^{n-i+1}(\epsilon_i\cdot F_{x_1}\cdot ... \cdot F_{x_u})^{\mu}=(-1)^{(n-i+1)-u} (F_{x_u}\cdot ... \cdot F_{x_1}\cdot\epsilon_i)^{\mu}$ and the corresponding root set is $\{x_1,...,x_u,i\}$. Thus the above expression is just (upto a minus) the type-B graph contracted with $(k_{1,i-1})_{\mu}$ (as shown by \figref{Fig:TypesRelationsGen} (b3)) and \eqref{Eq:Num Relations6} can be written as
\bea
\Sl_{\mathcal{F}_R}C^{\,\nu\rho}_{\mathcal{F}_R}&=&\eta^{\nu\rho}\Sl_{\substack{\text{tree structures}\\ \mathcal{T}_{1},...,\mathcal{T}_r}}\Biggl[\,-\prod\limits_{s=1}^rN_{B}(\pmb{\sigma}_R)\cdot 2k_{1,i-1}\Biggr].
\Label{Eq:Num Relations7}
\eea

When contracted with the first factor in \eqref{Eq:Num Relations2}, the expressions (\ref{Eq:Num Relations5}) and (\ref{Eq:Num Relations7}) reproduce the first term and the second term in \eqref{Eq:Num Relations}, respectively. Hence \eqref{Eq:Num Relations} has been proven.

\section{Decompositions of two- and three-point Berends-Giele currents}\label{Sec:Examples}

In this section, we investigate an off-shell extension of the decomposition formula (\ref{Eq:OnShellDec}), through direct evaluations of two-point and three-point Berends-Giele currents in YM theory. The starting point of the recursion is the expansion of one-point current
\bea
\W J^{\rho}(l)=J^{\rho}(l)=\epsilon_l^{\rho}\phi(l|l)=N_A^{\rho}(l)\phi(l|l),
\eea
where\emph{ the effective current} $\W J^{\rho}(l)$ is introduced as the part of the current $J^{\rho}(l)$ that reproduces the same graphic expansion formula  (\ref{Eq:OnShellDec}) for on-shell amplitudes. In this special case, $\W J^{\rho}(l)$ is defined by $J^{\rho}(l)$.

\subsection{Two-point current}

To express the two-point current $J^{\rho}(1,2)$ appropriately, we substitute the $3$-point vertex (\ref{Eq:CubicVertex1}) into the Berends-Giele recursion (\ref{Eq:BerendsGiele}). Then the current $J^{\rho}(1,2)$ is given by
\bea
J^{\rho}(1,2)&=&\frac{1}{s_{12}}V_3^{\mu\nu\rho}J_{\mu}(1)J_{\nu}(2)\nn
&=&\frac{1}{s_{12}}\bigl\{\left[J(1)\cdot F_2-J(1)(J(2)\cdot 2k_1)\right]^{\rho}+\left[J(1)\cdot J(2)\right] k_{1,2}^{\rho}\bigr\}\Label{Eq:2ptCurrentForm1}\nn
&=&[\epsilon_1\cdot F_2-\epsilon_1(\epsilon_2\cdot 2k_1)]^{\rho}\mathcal{\phi}(12|12)+\frac{1}{s_{12}}[\W J(1)\cdot \W J(2)] k_{1,2}^{\rho},
\eea
where the form of $3$-point vertex (\ref{Eq:CubicVertex1}), the fact $J(1)\cdot k_1=J(2)\cdot k_2=0$ and the Berends-Giele recursion (\ref{Eq:BerendsGieleBS}) for BS amplitude have been applied. The first term on the last line is the \emph{effective current} that satisfies the following expansion formula
\bea
\W J^{\rho}(1,2)\equiv[\epsilon_1\cdot F_2-\epsilon_1(\epsilon_2\cdot 2k_1)]^{\rho}\mathcal{\phi}(12|12)=N^{\rho}_A(1,2)\mathcal{\phi}(12|12),\Label{Eq:2ptEffectiveCurrent1}
\eea
in which, the coefficient of $\phi(12|12)$ is just the type-A numerator (\ref{Eq:2ptOffShellNum}) which gives rise to the expected BCJ numerator $N(1,2,3)$ when contracted with $\epsilon_3^{\rho}$.
The second term on the last line of \eqref{Eq:2ptCurrentForm1}, i.e., $K^{\rho}(1,2)\equiv\frac{1}{s_{12}}[\W J(1)\cdot \W J(2)] k_{1,2}^{\rho}$ is proportional to the total momentum of on-shell lines and has to vanish under the on-shell limit since $\epsilon_3\cdot k_{1,2}=-\epsilon_3\cdot k_3=0$.

Noting that $\mathcal{\phi}(12|12)=-\mathcal{\phi}(12|21)$, the effective current (\ref{Eq:2ptEffectiveCurrent1}) can be rearranged into a sum over two permutations:
\bea
\W J^{\rho}(1,2)&=&-\epsilon_1^{\rho}(\epsilon_2\cdot 2k_1)\mathcal{\phi}(12|12)-(\epsilon_1\cdot F_2)^{\rho}\mathcal{\phi}(12|21)\Label{Eq:2ptCurrentForm2}\nn
&=&(-\epsilon_1^{\rho})\bigl[(\epsilon_2\cdot 2k_1)+(\epsilon_2\cdot 2k_A)\bigr]\mathcal{\phi}(12|12)+\bigl[( F_2\cdot\epsilon_1)^{\rho}+(-\epsilon_1^{\rho})(\epsilon_2\cdot 2k_A)\big]\mathcal{\phi}(12|21)\nn
&=& N_{B}^{\rho}(1,2)\mathcal{\phi}(12|12)+N_{B}^{\rho}(2,1)\mathcal{\phi}(12|21),
\eea
where the antisymmetry of the strength tensor has been applied. The $k_A$ on the second line is an arbitrarily chosen momentum, because the terms with $k_A$ are finally cancelled with each other due to the antisymmetry of  $\mathcal{\phi}(12|12)$. When we choose $k_A$ as the total momentum of the subgraph which are contracted with this part, the expansion coefficients are just the type-B numerators $N_{B}^{\rho}(1,2)$ and $N_{B}^{\rho}(2,1)$ which are given by \eqref{Eq:TYPE-BEG2pt}.

Having defined the effective current $\W J^{\rho}(1,2)$, we further generalize the strength tensor $F_{i}^{\mu\nu}$ to an off-shell extended one
\bea
\W F^{\nu\rho}(1,2)\equiv 2 k_{1,2}^{\nu}\W J^{\rho}(1,2)-2 k_{1,2}^{\rho}\W J^{\nu}(1,2), \Label{Eq:2ptGeneralizedF}
\eea
where $k_{1,2}^{\nu}\equiv k_1^{\nu}+k_2^{\nu}$. When the explicit expression (\ref{Eq:2ptCurrentForm2}) is inserted into  \eqref{Eq:2ptGeneralizedF}, $\W F^{\nu\rho}(1,2)$ is expanded as
\bea
\W F^{\nu\rho}(1,2)&=&\big[F_1\cdot(F_2-\epsilon_2\cdot 2k_1)\big]^{\nu\rho}\phi(12|12)+\big[F_2\cdot(F_1-\epsilon_1\cdot 2k_2)\big]^{\nu\rho}\phi(12|21)+2(\epsilon_1^{\nu}\epsilon_2^{\rho}-\epsilon_2^{\nu}\epsilon_1^{\rho})\nn
&=&\big[F_1\cdot(F_2-\epsilon_2\cdot 2k_1)-F_1(\epsilon_2\cdot 2k_A)-F_2(\epsilon_1\cdot 2k_A)\big]^{\nu\rho}\phi(12|12)\nn
&&+\big[F_2\cdot(F_1-\epsilon_1\cdot 2k_2)-F_1(\epsilon_2\cdot 2k_A)-F_2(\epsilon_1\cdot 2k_A)\big]^{\nu\rho}\phi(12|21)+2(\epsilon_1^{\nu}\epsilon_2^{\rho}-\epsilon_2^{\nu}\epsilon_1^{\rho})\nn
&=&N_{C}^{\nu\rho}(1,2)\phi(12|12)+N_{C}^{\nu\rho}(2,1)\phi(12|21)+2(\epsilon_1^{\nu}\epsilon_2^{\rho}-\epsilon_2^{\nu}\epsilon_1^{\rho}),\Label{Eq:2ptGeneralizedF2}
\eea
in which, the first equality can be straightforwardly verified, by expanding $F_1$ and $F_2$ according to the definition and considering the antisymmetry of $\phi(12|12)$.
An arbitrarily chosen momentum $k_A$ is introduced in the second equality. It is easy to see that the terms containing $k_A$ cancel with each other, due to the antisymmetry of $\phi(12|12)$. When we choose $k_A$ as the total momentum of nodes in the subgraph which is contracted with the $\nu$ index, the expansion coefficients for $\phi(12|12)$ and $\phi(12|21)$ in \eqref{Eq:2ptGeneralizedF2} are just the type-C numerators $N_{C}^{\nu\rho}(1,2)$ and $N_{C}^{\nu\rho}(2,1)$ which are presented by \eqref{Eq:TYPE-CEG2pt}.

\subsection{Three-point current}

According to Berends-Giele recursion (\ref{Eq:BerendsGiele}), the three-point current $J^{\rho}(1,2,3)$ can be expressed by one- and two-point currents
\bea
J^{\rho}(1,2,3)={1\over s_{123}}\left[V_{3}^{\mu\nu\rho}J_{\mu}(1,2)J_{\nu}(3)+V_{3}^{\mu\nu\rho}J_{\mu}(1)J_{\nu}(2,3)+V_4^{\mu\nu\tau\rho}J_{\mu}(1)J_{\nu}(2)J_{\tau}(3)\right].\Label{Eq:3pt}
\eea
In the following, we evaluate these three terms separately.
\noindent\begin{itemize}
\item {\bf(i).} When the $3$-point vertex (\ref{Eq:CubicVertex1}) and the expression (\ref{Eq:2ptCurrentForm1}) of two-point current are considered, we write the first term of \eqref{Eq:3pt} as
\bea
J^{\rho}((1,2)(3))\equiv{1\over s_{123}}\W V_{3}^{\mu\nu\rho}\left[\W J_{\mu}(1,2)+K_{\mu}(1,2)\right]J_{\nu}(3)+\frac{1}{s_{123}}(J(1,2)\cdot J(3))k_{1,3}^{\rho},\Label{Eq:3pt2}
\eea
where we have applied $J(1,2)\cdot k_{1,2}=J(3)\cdot k_3=0$. The first term in the square brackets in \eqref{Eq:3pt2} contributes
\bea
{1\over s_{123}}\W V_{3}^{\mu\nu\rho}\W J_{\mu}(1,2)J_{\nu}(3)&=&{1\over s_{123}}\left\{\W J_{\mu}(1,2)\cdot\left[F_3-\epsilon_3\cdot(2k_1+2k_2)\right]\right\}^{\rho}\phi(3|3)\Label{Eq:3pt2-1}\\
&=&{1\over s_{123}}\Bigl\{N_A(1,2)\cdot\left[F_3-\epsilon_3\cdot(2k_1+2k_2)\right]\Bigr\}^{\rho}\phi(12|12)\phi(3|3)\nn
&=&{1\over s_{123}}N^{\rho}_A(1,2,3)\phi(12|12)\phi(3|3).\nonumber
\eea
In the above expression, the explicit definition of $\W V_{3}^{\mu\nu\rho}$ and the fact  $\phi(3|3)=1$ were considered, while the decomposition (\ref{Eq:2ptEffectiveCurrent1}) for $\W J_{\mu}(1,2)$ and the relation (\ref{Eq:Num RelationsEG1})  were applied.
The second term in the square brackets in \eqref{Eq:3pt2} contributes
\bea
&&{1\over s_{123}}\W V_{3}^{\mu\nu\rho}K_{\mu}(1,2)J_{\nu}(3)\nn
&=&{1\over s_{123}}\left[V_{3}^{\mu\nu\rho}K_{\mu}(1,2)J_{\nu}(3)-\W J^{\rho}(3)K(1,2)\cdot (k_1+k_2)-(K(1,2)\cdot J(3))k^{\rho}_{1,3}\right]\nn
&=&{1\over s_{123}}\left[V_{3}^{\mu\nu\rho}K_{\mu}(1,2)J_{\nu}(3)-\W J^{\rho}(3) \left(\W J(1)\cdot\W J(2)\right)-(K(1,2)\cdot \W J(3))k^{\rho}_{1,3}\right],\Label{Eq:3pt2-2}
\eea
where the effective $3$-point vertex $\W V_{3}^{\mu\nu\rho}$ was expressed by the usual vertex $V_{3}^{\mu\nu\rho}$ via \eqref{Eq:CubicVertex1}, the definition of $K^{\mu}(1,2)$ and the fact $J^{\rho}(3)=\W J^{\rho}(3)=\epsilon^{\rho}_3$ were considered. Summing  \eqref{Eq:3pt2-1} and  \eqref{Eq:3pt2-2} together, we get
\bea
J^{\rho}((1,2)(3))&=&{1\over s_{123}}N^{\rho}_A(1,2,3)\phi(12|12)\phi(3|3)\Label{Eq:3pt(12)(3)}\\
&&+{1\over s_{123}}\Bigl[V_{3}^{\mu\nu\rho}K_{\mu}(1,2)J_{\nu}(3)-\W J^{\rho}(3) \left(\W J(1)\cdot\W J(2)\right)\Bigr]+\frac{1}{s_{123}}\left(\W J(1,2)\cdot \W J(3)\right)k_{1,3}^{\rho}.\nonumber
\eea
\item {\bf(ii).} The second term in \eqref{Eq:3pt} reads
\bea
J^{\rho}((1)(2,3))&\equiv&\frac{1}{s_{123}}\W V_{3}^{\mu\nu\rho}\W J_{\mu}(1)\Bigl[\W J_{\nu}(2,3)+K_{\nu}(2,3)\Bigr]+\frac{1}{s_{123}}(J(1)\cdot J(2,3)) k_{1,3}^{\rho}.\Label{Eq:3pt3}
\eea
When the explicit expression of the reduced $3$-point vertex $\W V_{3}^{\mu\nu\rho}$ is considered,  the first term in the square brackets of \eqref{Eq:3pt3} becomes
\bea
&&\frac{1}{s_{123}}\W V_{3}^{\mu\nu\rho}\W J_{\mu}(1)\W J_{\nu}(2,3)\nn
&=&\frac{1}{s_{123}}\Bigl[\epsilon_1\cdot\W F(2,3)-\epsilon_1(\W J(2,3)\cdot 2k_1)\Bigr]^{\rho},\Label{Eq:3pt4}\nn
&=&\frac{1}{s_{123}}\biggl\{\Sl_{\pmb{\sigma}\in P\{2,3\}}\Bigl[\epsilon_1\cdot N_{C}(\pmb{\sigma})-\epsilon_1(N_{B}(\pmb{\sigma})\cdot 2k_1)\Bigr]^{\rho}\phi(23|\pmb{\sigma})+\Bigl[2(\epsilon_1\cdot\epsilon_2)\epsilon_3^{\rho}-2(\epsilon_1\cdot\epsilon_3)\epsilon_2^{\rho}\Bigr]\biggr\}\nn
&=&\frac{1}{s_{123}}\biggl\{N^{\rho}_A(1,2,3)\phi(23|23)+N^{\rho}_A(1,3,2)\phi(23|32)+\Bigl[2(\epsilon_1\cdot\epsilon_2)\epsilon_3^{\rho}-2(\epsilon_1\cdot\epsilon_3)\epsilon_2^{\rho}\Bigr]\biggr\}.
\eea
where the decomposition formulas  (\ref{Eq:2ptCurrentForm2}), (\ref{Eq:2ptGeneralizedF2}) (with the replacement $1,2\to 2,3$) and the relation (\ref{Eq:Num RelationsEG1}) were applied. The second term in the square brackets of \eqref{Eq:3pt3} is given by
\bea
&&~\frac{1}{s_{123}}\,\W V_{3}^{\mu\nu\rho}J_{\mu}(1)K_{\nu}(2,3)\nn
&=&{1\over s_{123}}\Bigl[ V_{3}^{\mu\nu\rho}J_{\mu}(1)K_{\nu}(2,3)+J^{\rho}(1) K(2,3)\cdot (2k_{2,3})-(J(1)\cdot K(2,3)) k_{1,3}^{\rho}\Bigr]\nn
&=&{1\over s_{123}}\Bigl[V_{3}^{\mu\nu\rho}J_{\mu}(1)K_{\nu}(2,3)+\W J^{\rho}(1) \left(\W J(2)\cdot\W J(3)\right)-(J(1)\cdot K(2,3)) k_{1,3}^{\rho}\Bigr].\Label{Eq:3pt5}
\eea
Here, the $\W V_{3}^{\mu\nu\rho}$ was expressed via \eqref{Eq:CubicVertex1} and the fact $J_{\mu}(1)=\W J_{\mu}(1)$ was considered. Substituting \eqref{Eq:3pt4} and \eqref{Eq:3pt5} into \eqref{Eq:3pt3} and considering the explicit expressions of $N^{\nu}_{B}(\pmb{\sigma})$ and $N^{\nu\rho}_{C}(\pmb{\sigma})$ (see \eqref{Eq:2ptCurrentForm2} and \eqref{Eq:2ptGeneralizedF}), we obtain
\bea
J^{\rho}((1)(2,3))&=&\frac{1}{s_{123}}\Big(N^{\rho}_A(1,2,3)\phi(1|1)\phi(23|23)+N^{\rho}_A(1,3,2)\phi(1|1)\phi(23|32)\Big)\nn
&&~~+\frac{1}{s_{123}}\Big[2\W J^{\rho}(3)\left(\W J(1)\cdot \W J(2)\right) -2\W J^{\rho}(2)\left(\W J(1)\cdot \W J(3)\right)+\W J^{\rho}(1)\left(\W J(2)\cdot\W J(3)\right)\Big]\nn
&&~~~~+\frac{1}{s_{123}}\left(\W J(1)\cdot \W J(23)\right) k_{1,3}^{\rho}+{1\over s_{123}} V_{3}^{\mu\nu\rho}J_{\mu}(1)K_{\nu}(2,3), \Label{Eq:3pt(1)(23)}
\eea
where $\W J^{\rho}(l)=\epsilon^{\rho}_l$ has been used.

\item {\bf(iii).} The third term in \eqref{Eq:3pt} reads
\bea
J^{\rho}((1)(2)(3))\equiv\frac{1}{s_{123}}\Bigl[2\W J^{\rho}(2)\left(\W J(1)\cdot \W J(3)\right)-\W J^{\rho}(3)\left(\W J(1)\cdot \W J(2)\right)-\W J^{\rho}(1)\left(\W J(2)\cdot\W J(3)\right)\Bigr].\Label{Eq:3pt(1)(2)(3)}
\eea
\end{itemize}
Summing \eqref{Eq:3pt(12)(3)}, \eqref{Eq:3pt(1)(23)} and \eqref{Eq:3pt(1)(2)(3)} together and considering the Berends-Giele recursion expression of BS currents
\bea
\phi(123|123)&=&\frac{1}{s_{123}}[\phi(1|1)\phi(23|23)+\phi(12|12)\phi(3|3)],~~\phi(123|132)=\frac{1}{s_{123}}\phi(1|1)\phi(23|32),
\eea
we find that the terms of the form $\W J^{\rho}(\W J\cdot \W J)$ all cancel out and the current $J^{\rho}(1,2,3)$ can be decomposed as
\bea
J^{\rho}(1,2,3)=\W J^{\rho}(1,2,3)+K^{\rho}(1,2,3)+L^{\rho}(1,2,3).\Label{Eq:3ptTildeJKL}
\eea
The $\W J^{\rho}(1,2,3)$ term in \eqref{Eq:3ptTildeJKL} is explicitly written as
\bea
\W J^{\rho}(1,2,3)=N_A^{\rho}(1,2,3)\phi(123|123)+N_A^{\rho}(1,3,2)\phi(123|132),\Label{Eq:3ptEFFJ}
\eea
where $N_A^{\rho}(1,2,3)$ and $N_A^{\rho}(1,3,2)$ are just the type-A numerators which were already given by \eqref{Eq:3ptOffShellNum}.
The $K^{\rho}(1,2,3)$ term in \eqref{Eq:3ptTildeJKL} is proportional to the total momentum of the on-shell lines
\bea
K^{\rho}(1,2,3)=\frac{1}{s_{123}}\Big[\W J(1,2)\cdot \W J(3)+\W J(1)\cdot \W J(2,3)\Big]k_{1,3}^{\rho}.\Label{Eq:3ptKterm}
\eea
The $L^{\rho}(1,2,3)$ term is given by
\bea
L^{\rho}(1,2,3)={1\over s_{123}}\Bigl[V_{3}^{\mu\nu\rho} K_{\mu}(1,2)J_{\nu}(3)+V_{3}^{\mu\nu\rho} J_{\mu}(1)K_{\nu}(2,3)\Bigr]=J^{\rho}(K(1,2),3)+J^{\rho}(1,K(2,3)),\Label{Eq:3ptLterm}
\eea
where $J^{\rho}(K(1,2),3)$ is obtained by replacing the polarization and the momentum of an external line in a two-point current by $K^{\mu}(1,2)$ and $k^{\mu}_{1,2}$, respectively.

In on-shell limit, the expansion formula (\ref{Eq:3ptEFFJ}) for the effective current precisely reproduce the expansion of on-shell amplitude $A(1,2,3)$. The $K^{\rho}(1,2,3)$ term and $L^{\rho}(1,2,3)$ term have to vanish due to $\epsilon_4\cdot k_{1,3}=-\epsilon_4\cdot k_{4}=0$ and the property (\ref{Eq:YMBGProperty-3}), respectively.

\section{The general expansion formula of Berends-Giele currents}\label{sec:genExpansion}

From the three-point example (\ref{Eq:3ptTildeJKL}), we learn that the Berends-Giele current in YM can be decomposed into three terms: (i). an effective current $\W J^{\rho}$ which is written in terms of BS currents whose coefficients are the type-A numerators in DDM form, (ii). a term $K^{\rho}$  which is proportional to the total momentum and (iii). an $L^{\rho}$ term that is a sum of currents where the polarization vector and momenta of some external lines are replaced by lower-point $K^{\rho}$ terms and the corresponding momenta, respectively. This pattern also holds for the two-point current $J^{\rho}(1,2)$ if we define $L^{\rho}(1,2)\equiv 0$ and for the one-point current if we define $K^{\rho}(l)=L^{\rho}(l)\equiv0$. In this section, we generalize the above observation to an arbitrary-point current  $J^{\,\rho}(1,2,...,n-1)$ in YM theory:
\bea
J^{\,\rho}(1,2,...,n-1)=\W J^{\,\rho}(1,2,...,n-1)+K^{\,\rho}(1,2,...,n-1)+L^{\,\rho}(1,2,...,n-1),\Label{Eq:GenForm1}
\eea
where the three terms are respectively demonstrated as follows.

{\bf(i).} The $\W J^{\rho}(1,2,...,n-1)$ term in \eqref{Eq:GenForm1} is mentioned as the \emph{effective current} which can be decomposed in terms of BS currents $\phi(1,2,...,n-1 | 1,\pmb{\sigma})$  accompanied by type-A numerators $N_A^{\rho}(1,\pmb{\sigma})$:
\bea
\W J^{\rho}(1,2,\dots,n-1)=\Sl_{\pmb{\sigma}\in P(2,n-1)}\,N_A^{\rho}(1,\pmb{\sigma})\phi(1,2,\dots,n-1 | 1,\pmb{\sigma}).\Label{Eq:GenForm1-1}
\eea
In the above expression, the summation is taken over all possible permutations $\pmb{\sigma}\in P(2,n-1)$ of elements $2,3,...n-1$. This decomposition reproduces \eqref{Eq:OnShellDec} via $s_{1...n-1}\big[\epsilon_n\cdot\W J(1,2,\dots,n-1)\big]$.

{\bf(ii).} The $K^{\rho}(1,2,...,n-1)$ term in \eqref{Eq:GenForm1} is proportional to the total momentum $k^{\,\rho}_{1,n-1}\equiv \sum_{j=1}^{n-1}k_j^{\,\rho}$ of the on-shell lines $1,...,n-1$ and is expressed via Lorentz contraction of lower-point effective currents
\bea
K^{\,\rho}\big(1,2,...,n-1\big)=\frac{1}{s_{12\dots n-1}}\,k^{\,\rho}_{1,n-1}\,\Sl_{i=1}^{n-2}\W J(1,...,i)\cdot \W J(i+1,...,n-1).\Label{Eq:GenK}
\eea

{\bf(iii).} The $L^{\rho}\big(1,2,...,n-1\big)$ term in \eqref{Eq:GenForm1} is presented as
\bea
    &&~~L^{\rho}\big(1,2,...,n-1\big)\nonumber \\
    &=&\Sl_{\{a_i,b_i\}\subset\{1,...,n-1\}}(-1)^{I+1}J^{\rho}\bigl(S_{1,a_1-1},K_{(a_1,b_1)},S_{b_1+1,a_2-1},K_{(a_2,b_2)},...,K_{(a_I,b_I)},S_{b_I+1,n-1}\bigr).\Label{Eq:GenL}
\eea
In the above expression, we use $S_{1,a_1-1}$ to denote the sequence $1,2,...,a_1-1$ and $K_{(a_i,b_i)}$ to denote $K(a_i,...,b_i)$ for short. The $J^{\rho}(S_{1,a_1-1},K_{(a_1,b_1)},...,K_{(a_I,b_I)},S_{b_I+1,n-1})$ stands for the Berends-Giele current when $\{a_i,...,b_i\}$ ($i=1,...,I$) is considered as a single external line with the polarization vector $K^{\mu}_{(a_i,b_i)}\equiv K^{\mu}(a_i,...,b_i)$ and momentum $k^{\mu}_{a_i,b_i}$.  We have summed over all possible choices of ordered pairs $\{a_i,b_i\}$  which satisfy the following conditions: (i). $1\leq a_1<b_1<a_2<b_2<...<a_I<b_I\leq n-1$, (ii). if $I=1$, i.e. there is only  $K_{(a_1,b_1)}$, $\{a_1,b_1\}$ cannot be chosen as $\{1,n-1\}$. To clarify this summation, we take the  $n=5$ case as an example, in which we can pick out $a_i,b_i$ pairs from the ordered set $\{1,2,3,4\}$. If $I=1$, the $\{a_1,b_1\}$ can be either one of
\bea
\{1,2\},\{2,3\},\{3,4\},\{1,3\},\{2,4\}.
\eea
If $I=2$, the  $\{a_1,b_1\}$ and $\{a_2,b_2\}$ can only be given by
\bea
\{a_1,b_1\}=\{1,2\},\{a_2,b_2\}=\{3,4\}.
\eea
Therefore, $L^{\rho}(1,2,3,4)$ is explicitly expressed as
\bea
L^{\rho}(1,2,3,4)&=&J^{\rho}\big(K_{(1,2)},3,4\big)+J^{\rho}\big(1,K_{(2,3)},4\big)+J^{\rho}\big(1,2,K_{(3,4)}\big)\nn
&&+J^{\rho}\big(K_{(1,2,3)},4\big)+J^{\rho}\big(1,K_{(2,3,4)}\big)-J^{\rho}\big(K_{(1,2)},K_{(3,4)}\big),
\eea
where the first five terms correspond to contributions with $I=1$, while the last term corresponds to the $I=2$ case.
The following properties of $L^{\rho}\big(1,2,...,n-1\big)$ term will be helpful in the coming discussions:
\begin{itemize}
\item Since $K^{\nu}_{(a_i,b_i)}$ is proportional to the total momentum $k^{\mu}_{a_i,b_i}$ of gluons in the sector $\{a_i,a_i+1,...,b_i\}$, each term in \eqref{Eq:GenL} is proportional to $J^{\rho}(S_{1,a_1-1},k_{a_1,b_1},...,k_{a_I,b_I},S_{b_I+1,n-1})$. Consequently, $L^{\rho}(1,2,...,n-1)$  satisfies the identity
\bea
k_{1,n-1}\cdot L(1,2,...,n-1)=0 \Label{Eq:LID1}
\eea
and the following identity under on-shell limit
\bea
\epsilon_{n}\cdot L(1,2,...,n-1)=0,\Label{Eq:LID2}
\eea
which correspond to the properties (\ref{Eq:YMBGProperty-2}), (\ref{Eq:YMBGProperty-3}) satisfied by $J^{\rho}\bigl(S_{1,a_1-1},k_{a_1,b_1},...,k_{a_I,b_I},S_{b_I+1,n-1}\bigr)$.

\item If we write each  $J^{\rho}(S_{1,a_1-1},K_{a_1,b_1},...,K_{a_I,b_I},S_{b_I+1,n-1})$ in \eqref{Eq:GenL} according to Berends-Giele recursion and reorganize the terms by collecting all contributions with the same partition of momenta connecting to the $3$- or $4$-point vertex, the $L^{\rho} (1,2,...,n-1 )$ can be expressed by lower-point $J$, $K$ and $L$. Roughly speaking
    \bea
  L&\sim& \frac{1}{s_{1...n-1}}\Sl\Bigl[V_3 (L+K)J+V_3 J(L+K)-V_3 (L+K)(L+K)\Bigr]\nn
    &&+\frac{1}{s_{1...n-1}}\Sl\Bigl[V_4 (L+K)JJ+V_4 J(L+K)J+V_4 (L+K)JJ\nn
    &&~~~~~~~~~~~~~~~\,-V_4 (L+K)(L+K)J-V_4 J(L+K)(L+K)-V_4 (L+K)J(L+K)\nn
    &&~~~~~~~~~~~~~~~~~~~~+V_4 (L+K)(L+K)(L+K)\Bigr],\Label{Eq:LBG}
     \eea
where a term, e.g., $\Sl V_3 (L+K)J$ stands for $\sum_{i=1}^{n-2}\bigl[V_3^{\mu\nu\rho} (L^{\mu}(1,...,i)+K^{\mu}(1,...,i))J^{\rho}(i+1,...,n-1)\bigr]$ for short.
\end{itemize}

When we take the on-shell limit, the $n$-point amplitude $A(1,2,...,n)$ is given by
\bea
~~A(1,2,...,n)&=&\Bigl[{s_{12\dots n-1}} \epsilon_n\cdot J\big(1,2,\dots,n-1\big)\Bigr]\Big|_{s_{12\dots n-1}=k_n^2\to 0}\Label{Eq:On-ShellLimit1}\\
&=&\Bigl[{s_{12\dots n-1}} \big(\epsilon_n\cdot \W J(1,2,\dots,n-1)\nn
&&~~~~~~~~~~~~~+\epsilon_n\cdot K(1,2,\dots,n-1)
+\epsilon_n\cdot  L(1,2,\dots,n-1)\big)\Bigr]\Big|_{s_{12\dots n-1}=k_n^2\to 0}.\nonumber
\eea
The first term in the above equation precisely reproduces the decomposition (\ref{Eq:OnShellDec}) of on-shell amplitude with the expected BCJ numerators (\ref{Eq:BCJOnShell}). The second term, vanishes because of momentum conservation and the transversality condition $\epsilon_n\cdot k_n=0$. The last term cancels out due to \eqref{Eq:LID2}. Therefore, in the on-shell limit, the decomposition formula (\ref{Eq:GenForm1}) reproduces the expected decomposition formula (\ref{Eq:OnShellDec}) for $n$-point amplitudes.

In the following subsections, we prove the decomposition formula (\ref{Eq:GenForm1}) by Berends-Giele recursion (\ref{Eq:BerendsGiele}). We first show that the current $J^{\rho}(1,2,\dots,n-1)$ can be written as the form (\ref{Eq:GenForm1}) with the expected $K^{\rho}$ term (\ref{Eq:GenK}) and $L^{\rho}$ term (\ref{Eq:GenL}).  We further prove that the remaining term is just  $\W J^{\rho}(1,2,\dots,n-1)$ that satisfies the expansion formula (\ref{Eq:GenForm1-1}).


\subsection{General decomposition formula for Berends-Giele currents}

Supposing that the decomposition formula (\ref{Eq:GenForm1}) is satisfied by $J^{\rho}(1,...,m)$ ($m<n-1$), we now evaluate the YM current $J^{\rho}(1,...,n-1)$ according to Berends-Giele recursion (\ref{Eq:BerendsGiele}).

When the lower-point currents are expressed by the decomposition formula (\ref{Eq:GenForm1}), the first term in the Berends-Giele recursion (\ref{Eq:BerendsGiele}), which corresponds to the contribution of $3$-point vertex, is written as
\bea
T^{\rho}_A&\equiv&{1\over s_{12\dots n-1}}\Sl_{1\leq i<n-1} \W V_{3}^{\mu\nu\rho}\Bigl[\W J_{\mu}(1,...,i)+K_{\mu}(1,...,i)+L_{\mu}(1,...,i)\Bigr]\nn
&&~~~~~~~~~~~~~~~~~~~~~~~~\,\times\Bigl[\W J_{\nu}(i+1,...,n-1)+K_{\nu}(i+1,...,n-1)+L_{\nu}(i+1,...,n-1)\Bigr]\nn
&&~~~\,+{1\over s_{12\dots n-1}}\Sl_{1\leq i<n-1} \Bigl[J(1\dots i)\cdot J(i+1\dots n-1)\Bigr]k^{\rho}_{1,n-1},\Label{Eq:TA}
\eea
where we have expressed the $3$-point vertex $V_3$ by the effective one $\W V_3$ according to \eqref{Eq:CubicVertex1} and applied the identity (\ref{Eq:YMBGProperty-1}) for lower point Berends-Giele currents.
The above expression can be further arranged as
\bea
T^{\rho}_A&=&\,\,\,\,{1\over s_{12\dots n-1}}~~\Sl_{1\leq i<n-1}\,\W V_{3}^{\mu\nu\rho}\W J_{\mu}(1,...,i)\W J_{\nu}(i+1,...,n-1)\nn
&&+{1\over s_{12\dots n-1}}\biggl\{\Sl_{1\leq i <n-1} \W V_{3}^{\mu\nu\rho}J_{\mu}(1,...,i)\Bigl[K_{\nu}(i+1,...,n-1)+L_{\nu}(i+1,...,n-1)\Bigr]\nn
&&~~~~~~~~~~~\,+\Sl_{1\leq i<n-1} \W V_{3}^{\mu\nu\rho}\Big[K_{\mu}(1,...,i)+L_{\mu}(1,...,i)\Big]J_{\nu}(i+1,...,n-1)\nn
&&~~~~~~~~~~~\,-\Sl_{1\leq i<n-1} \W V_{3}^{\mu\nu\rho}\Big[K_{\mu}(1,...,i)+L_{\mu}(1,...,i)\Big]\Big[K_{\nu}(i+1,...,n-1)+L_{\nu}(i+1,...,n-1)\Big]\biggr\}\nn
&&+{1\over s_{12\dots n-1}}\,\,\,\Sl_{1\leq i<n-1}\,\Bigl[J(1\dots i)\cdot J(i+1\dots n-1)\Bigr]k^{\rho}_{1,n-1}.
\eea
When we reexpress the effective $3$-point vertices $\W V_{3}^{\mu\nu\rho}$ in the braces by $V_{3}^{\mu\nu\rho}$ according to \eqref{Eq:CubicVertex1}, the above expression turns into
\bea
T^{\rho}_A&=&~~\,{1\over s_{12\dots n-1}}~\Sl_{1\leq i<n-1}~\W V_{3}^{\mu\nu\rho}\W J_{\mu}(1,...,i)\W J_{\nu}(i+1,...,n-1)\nn
&&+{1\over s_{12\dots n-1}}~\Sl_{1\leq i<n-1}~\Bigl[\W J(1\dots i)\cdot \W J(i+1\dots n-1)\Bigr]k^{\rho}_{1,n-1}\nn
&&+{1\over s_{12\dots n-1}}\biggl\{\Sl_{1\leq i<n-1}V_{3}^{\mu\nu\rho}J_{\mu}(1,...,i)\Bigl[K_{\nu}(i+1,...,n-1)+L_{\nu}(i+1,...,n-1)\Bigr]\nn
&&~~~~~~~~~~~\,+\Sl_{1\leq i<n-1}V_{3}^{\mu\nu\rho}\Big[K_{\mu}(1,...,i)+L_{\mu}(1,...,i)\Big]J_{\nu}(i+1,...,n-1)\nn
&&~~~~~~~~~~~\,-\Sl_{1\leq i<n-1} V_{3}^{\mu\nu\rho}\Big[K_{\mu}(1,...,i)+L_{\mu}(1,...,i)\Big]\Big[K_{\nu}(i+1,...,n-1)+L_{\nu}(i+1,...,n-1)\Big]\biggr\}\nn
&&+{1\over s_{12\dots n-1}}\biggl\{\Sl_{1\leq i<n-2}\W J^{\rho}(1,...,i)\Bigl[k_{i+1,n-1}\cdot \bigl(K(i+1,...,n-1)+L(i+1,...,n-1)\bigr)\Bigr]\nn
&&~~~~~~~~~~~\,-\Sl_{2\leq j<n-1}\Bigl[k_{1,j}\cdot\big(K(1,...,j)+L(1,...,j)\big)\Bigr]\W J^{\rho}(j+1,...,n-1)\biggr\},
\eea
where the second term is just the $K^{\rho}(1,...,n-1)$ term (see \eqref{Eq:GenK}). We should note that the identity (\ref{Eq:YMBGProperty-1}) for a full Berends-Giele current $J^{\rho}$ is not satisfied by the object $\big[K^{\mu}(1,...,i)+L^{\mu}(1,...,i)\big]$ which only plays as a part of the current. According to the inductive assumption, the $L^{\mu}(1,...,j)$ term has the pattern (\ref{Eq:GenL}) thus satisfies the identity (\ref{Eq:LID1}), while the $K^{\mu}(1,...,j)$ term with the form (\ref{Eq:GenK}) survives when contracted with $(k_{1,j})_{\mu}$.
Therefore, the last term in the above expression can be simplified as
\bea
&&\Sl_{i=1}^{n-1}\Sl_{j=i+1}^{n-2}\W J^{\rho}(1,...,i)\Bigl[\W J(i+1,...,j)\cdot \W J(j+1,...,n-1)\Bigr]\nn
&&~~~~~~~~~~~~\,-\Sl_{i=1}^{j-1}\Sl_{j=2}^{n-2}\Bigl[\W J(1,...,i)\cdot \W J(i+1,...,j)\Bigr]\W J^{\rho}(j+1,...,n-1),
\eea
where $k_{1,n-1}\cdot k_{1,n-1}=s_{1...n-1}$ has been divided out. The $T^{\rho}_A$ in \eqref{Eq:TA} is then arranged as
\bea
T^{\rho}_A=T^{\rho}_{A1}+T^{\rho}_{A2}+K^{\rho}(1,...,n-1),
\eea
in which the explicit expressions of $T^{\rho}_{A1}$ and $T^{\rho}_{A2}$ are respectively given by
\bea
T^{\rho}_{A1}&\equiv&{1\over s_{12\dots n-1}}\Sl_{i=1}^{n-2}\W V_{3}^{\mu\nu\rho}\W J_{\mu}(1,...,i)\W J_{\nu}(i+1,...,n-1)\nn
&&+ {1\over s_{12\dots n-1}}\Sl_{1\leq i<j<n-1}\biggl\{\W J^{\rho}(1,...,i)\left[\W J(i+1,...,j)\cdot \W J(j+1,...,n-1)\right]\nn
&&~~~~~~~~~~~~~~~~~~~~~~~~~~~~~-\left[\W J(1,...,i)\cdot\W J(i+1,...,j)\right]\W J^{\rho}(j+1,...,n-1)\biggr\},\Label{Eq:TA1}
\eea
\bea
T^{\rho}_{A2}&\equiv&{1\over s_{12\dots n-1}}\Sl_{i=1}^{n-1}V_{3}^{\mu\nu\rho}J_{\mu}(1,...,i)\Bigl[K_{\nu}(i+1,...,n-1)+L_{\nu}(i+1,...,n-1)\Bigr]\nn
&&~~+{1\over s_{12\dots n-1}}\Sl_{i=1}^{n-1}V_{3}^{\mu\nu\rho}\Big[K_{\mu}(1,...,i)+L_{\mu}(1,...,i)\Big]J_{\nu}(i+1,...,n-1)\nn
&&-{1\over s_{12\dots n-1}}\Sl_{i=1}^{n-1}V_{3}^{\mu\nu\rho}\Big[K_{\mu}(1,...,i)+L_{\mu}(1,...,i)\Big]\Big[K_{\nu}(i+1,...,n-1)+L_{\nu}(i+1,...,n-1)\Big].\Label{Eq:TA2}
\eea

The second term  in the Berends-Giele recursion (\ref{Eq:BerendsGiele}), which corresponds to the contribution of $4$-point vertex, can be written as the following sum
\bea
T^{\rho}_B=T^{\rho}_{B1}+T^{\rho}_{B2},
\eea
where $T_{B1}$ and $T_{B2}$ are respectively defined by
\bea
T^{\rho}_{B1}&\equiv&\frac{1}{s_{1...n-1}}\Bigg[\Sl_{1\leq i<j\leq n-2} V_4^{\mu\nu\tau\rho}\W J_{\mu}\big(1,...,i\big)\W J_{\nu}\big(i+1,...,j\big)\W J_{\tau}\big(j+1,...,n-1\big)\Bigg],\Label{Eq:TB1}\\
T^{\rho}_{B2}&\equiv&\frac{1}{s_{1...n-1}}\Bigg\{\Sl_{1\leq i<j\leq n-2} V_4^{\mu\nu\tau\rho}\Bigl[J_{\mu}\big(1,...,i\big)J_{\nu}\big(i+1,...,j\big)J_{\tau}\big(j+1,...,n-1\big)\\
&&~~~~~~~~~~~~~~~~~~~~~~~~~~~~~~~~~~-\W J_{\mu}\big(1,...,i\big)\W J_{\nu}\big(i+1,...,j\big)\W J_{\tau}\big(j+1,...,n-1\big)\Bigr]\Bigg\}.\nonumber
\eea
Decomposing lower-point Berends-Giele currents according to (\ref{Eq:GenForm1}), one can verify the following expression of $T^{\rho}_{B2}$
\bea
T_{B2}&\equiv&~\,\,\,\frac{1}{s_{1...n-1}}\,\Sl_{1\leq i<j\leq n-2}\,\Bigl\{ V_4^{\mu\nu\tau\rho}\left[K_{(1,i)}+L_{(1,i)}\right]_{\mu}\,J_{\nu}\big(S_{i+1,j}\big)\,J_{\tau}\big(S_{j+1,n-1}\big)\nn
&&+V_4^{\mu\nu\tau\rho}J_{\mu}\big(S_{1,i}\big)\,\left[K_{(i+1,j)}+L_{(i+1,j)}\right]_{\nu}\,J_{\tau}\big(S_{j+1,n-1}\big)\nn
&&+V_4^{\mu\nu\tau\rho}J_{\mu}\big(S_{1,i}\big)\,J_{\nu}\big(S_{i+1,j}\big)\,\left[K_{(j+1,n-1)}+L_{(j+1,n-1)}\right]_{\tau}\nn
&&-V_4^{\mu\nu\tau\rho}\left[K_{(1,i)}+L_{(1,i)}\right]_{\mu}\,\left[K_{(i+1,j)}+L_{(i+1,j)}\right]_{\nu}\,J_{\tau}\big(S_{j+1,n-1}\big)\nn
&&-V_4^{\mu\nu\tau\rho}J_{\mu}\big(S_{1,i}\big)\,\left[K_{(i+1,j)}+L_{(i+1,j)}\right]_{\nu}\,\left[K_{(j+1,n-1)}+L_{(j+1,n-1)}\right]_{\tau}\nn
&&-V_4^{\mu\nu\tau\rho}\left[K_{(1,i)}+L_{(1,i)}\right]_{\mu}\,J_{\nu}\big(S_{i+1,j}\big)\,\left[K_{(j+1,n-1)}+L_{(j+1,n-1)}\right]_{\tau}\nn
&&+V_4^{\mu\nu\tau\rho}\left[K_{(1,i)}+L_{(1,i)}\right]_{\mu}\,\left[K_{(i+1,j)}+L_{(i+1,j)}\right]_{\nu}\,\left[K_{(j+1,n-1)}+L_{(j+1,n-1)}\right]_{\tau}\Bigr\}.\Label{Eq:TB2}
\eea
Here, we have respectively written the sequence $a,a+1,...,b$ as $S_{a,b}$ and the $K^{\mu}(a,a+1,...,b)$, $L^{\mu}(a,a+1,...,b)$ as $K^{\mu}_{(a,b)}$, $L^{\mu}_{(a,b)}$ for short.

The sum $T_{A2}+T_{B2}$ is just the expected Berends-Giele recursion expression (\ref{Eq:LBG}) of the $L^{\rho}(1,...,n-1)$ term. Hence the total current is finally given by
\bea
J^{\rho}(1,...,n-1)=T^{\rho}_{A1}+T^{\rho}_{B1}+L^{\rho}(1,...,n-1)+K^{\rho}(1,...,n-1),\Label{Eq:GenForm2}
\eea
with the expected $K$ term and $L$ term. In the remaining part of this section, we introduce an $(n-1)!$-expansion of the effective current and an expansion of the generalized strength tensor. By the use of these helpful expansions, we prove that $T^{\rho}_{A1}+T^{\rho}_{B1}=\W J^{\rho}(1,...,n-1)$, where $\W J^{\rho}(1,...,n-1)$ is the effective current that satisfies the expansion formula (\ref{Eq:GenForm1-1}).

\subsection{Expansions of effective currents and generalized strength tensors}
Before proving the expansion formula (\ref{Eq:GenForm1-1}) for $T^{\rho}_{A1}+T^{\rho}_{B1}$, we introduce an $(n-1)!$-expansion formula for effective currents and an expansion formula for a generalized strength tensor, which correspond to \eqref{Eq:2ptCurrentForm2} and \eqref{Eq:2ptGeneralizedF} in \secref{Sec:Examples}. These formulas will be helpful in the study of the expansion of $T^{\rho}_{A1}+T^{\rho}_{B1}$.

\subsubsection{$(n-1)!$-expansion formula of effective currents}

The effective current $\W J^{\rho}(1,2,\dots,n-1)$ that  satisfies  the expansion formula (\ref{Eq:GenForm1-1}) with $(n-2)!$ terms can also be expressed by the following  $(n-1)!$-expansion:
\bea\boxed{
\W J^{\rho}(1,2,\dots,n-1)=\Sl_{\pmb{\sigma}\in P(1,n-1)}N_B^{\rho}(\pmb{\sigma})\,\phi(1,2,...n-1|\pmb{\sigma})},\Label{Eq:GenForm2-1}
\eea
where $N_B^{\rho}(\pmb{\sigma})$ are the type-B numerators (\ref{Eq:Type-B}). For example, on can directly check that the three-point effective current $\W J^{\rho}(1,2,3)$ (see \eqref{Eq:3ptEFFJ}) satisfies the following expansion formula
\bea
\W J^{\rho}(1,2,3)&=&N_{B}^{\rho}(1,2,3)\phi(123|123)+N_{B}^{\rho}(1,3,2)\phi(123|132)+N_{B}^{\rho}(2,1,3)\phi(123|213)\nn
&&+N_{B}^{\rho}(2,3,1)\phi(123|231)+N_{B}^{\rho}(3,1,2)\phi(123|312)+N_{B}^{\rho}(3,2,1)\phi(123|321),\Label{Eq:2-1}
\eea
in which the type-B generalized numerators are explicitly given by \eqref{Eq:Type-BEG}. We now prove the general expansion formula (\ref{Eq:GenForm2-1}) by the following steps.
 \begin{figure}
\centering
\includegraphics[width=0.32\textwidth]{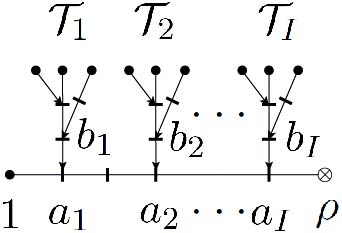}
\caption{The structure of a graph in the decomposition formula (\ref{Eq:GenForm1-1})}\label{Fig:(n-1)Graph}
\end{figure}

{\bf\emph{Step-1}}~~Reexpress the $(n-2)!$-expansion formula (\ref{Eq:GenForm1-1}) of the effective current $\W J^{\rho}$ by {\bf(i).} summing over all possible graphs $\mathcal{F}\in \mathcal{G}^A[\pmb{\sigma}]$ for the $\W N_A^{\rho}(1,\pmb{\sigma})$ in (\ref{Eq:GenForm1-1}),  {\bf(ii).} summing over all permutations $\pmb{\sigma}^{\mathcal{F}}$ (where the node $1$ is always the leftmost one) corresponding to a given graph $\mathcal{F}$. The effective current then becomes
\bea
\W J^{\rho}(1,2,\dots,n-1)=\Sl_{\mathcal{F}\in \mathcal{G}^A}C^{\rho}_{\mathcal{F}}\,\Sl_{\pmb{\sigma}^{\mathcal{F}}}\phi\big(1,2,...n-1|1,\pmb{\sigma}^{\mathcal{F}}\big),\Label{Eq:GenForm2-2}
\eea
which is just the off-shell extended version of \eqref{Eq:OnShellDec1}. The permutations $\pmb{\sigma}^{\mathcal{F}}$ are explicitly given as follows: (i). $1$ is the leftmost element, (ii). for two adjacent nodes $i$, $j$, we have $i\prec j$, if $i$ is nearer to $1$ than $j$, (iii). the relative orders of nodes on two branches which attach to a same node are given by shuffling the relative orders corresponding to both branches together.  Since each graph $\mathcal{F}$ contains a master chain $(-1)^{n-l}\,\big[\epsilon_1\cdot F_{i_1}\cdot F_{i_2}\cdot...\cdot F_{i_l}\big]^{\rho}$ as well as trees $\mathcal{T}_1$, ${\mathcal{T}_2}$, ..., ${\mathcal{T}_I}$ which are respectively planted at nodes $a_1,a_2,...,a_I\in \{1,i_1,...,i_l\}$ on the master chain,
the effective current (\ref{Eq:GenForm2-2}) can be formally written as
\bea
&&~~\W J^{\rho}(1,2,\dots,n-1)\Label{Eq:GenForm2-3}\\
&=&\Sl_{\text{master chains}}\,\Sl_{\substack{\text{tree structures}\\\text{$\mathcal{T}_1$... $\mathcal{T}_I$}}}(-1)^{n-l}\,\big(\epsilon_1\cdot F_{i_1}\cdot F_{i_2}\cdot...\cdot F_{i_l}\big)^{\rho}\,C^{\mathcal{T}_1}\,C^{\mathcal{T}_2}\,...\,C^{\mathcal{T}_I}\left[\,\Sl_{\pmb{\sigma}}\phi(1,2,...,n-1|1,\pmb{\sigma})\right],\nonumber
\eea
in which we have summed over all permutations $\pmb{\sigma}$ satisfying
\bea
\pmb{\sigma}\in\left[\big\{i_1,i_2,...,i_l\big\}\shuffle \mathcal{T}_1\big|_{b_1}\shuffle \mathcal{T}_2\big|_{b_2}...\shuffle \mathcal{T}_I\big|_{b_I}\right]\Big|_{a_j\prec b_j~(i=1,...,I)},
\eea
where $\mathcal{T}_i\big|_{b_i}$ denotes the permutations established by the tree structures $\mathcal{T}_i$ when $b_i\in \mathcal{T}_i$ is considered as the leftmost element in $\mathcal{T}_i$ (i.e. the node nearest to root $a_i$).

{\bf\emph{Step-2}}~~We rewrite the coefficient in \eqref{Eq:GenForm2-3} by the antisymmetry of strength tensors on the master chain
\bea
C_{\mathcal{F}}&=&(-1)^{n-l}\,\big(\epsilon_1\cdot F_{i_1}\cdot F_{i_2}\cdot...\cdot F_{i_l}\big)^{\rho}\,C^{\mathcal{T}_1}\,C^{\mathcal{T}_2}\,...\,C^{\mathcal{T}_I}\nn
&=&(-1)^{n}\,\big(F_{i_l}\cdot F_{i_{l-1}}\cdot...\cdot F_{i_1}\cdot \epsilon_1\big)^{\rho}\,C^{\mathcal{T}_1}\,C^{\mathcal{T}_2}\,...\,C^{\mathcal{T}_I},\Label{Eq:GenForm1-5}
\eea
which is the coefficient corresponding to a type-B graph $\mathcal{F}\in \mathcal{G}^B$, where no tree is planted at the off-shell node $\rho$.
On another hand, when the reflection relation (\ref{Eq:BSBGProperty-1}) and the generalized KK relation (\ref{Eq:BSBGProperty-3}) are applied,  the summation over $\pmb{\sigma}$ turns into
\bea
&&\Sl_{\pmb{\sigma}}\phi\bigl(1,2,...n-1\big|\pmb{\sigma}\in\bigl[\left\{1,i_1,i_2,...,i_l\right\}\shuffle \mathcal{T}_1\big|_{b_1}...\shuffle \mathcal{T}_I\big|_{b_I}\bigr]\big|_{a_j\prec b_j~(i=1,...,I)}\bigr)\nn
&=&(-1)^{n}\Sl_{\pmb{\sigma}}\phi\bigl(1,2,...n-1\big|\pmb{\sigma}\in\bigl[\left\{i_l,i_{l-1},...,i_1,1\right\}\shuffle (\mathcal{T}_1\big|_{b_1})^T...\shuffle (\mathcal{T}_I\big|_{b_I})^T\bigr]\big|_{a_j\succ b_j~(i=1,...,I)} \big)\nn
&=&(-1)^{l}\Sl_{\pmb{\sigma}}\phi\bigl(1,2,...n-1\big|\pmb{\sigma}\in\bigl[\left\{i_l,,i_{l-1},...,i_1,1\right\}\shuffle  \mathcal{T}_1\big|_{b_1}...\shuffle \mathcal{T}_I\big|_{b_I}\bigr]\big|_{a_j\prec b_j~(i=1,...,I)}\bigr).\Label{Eq:GenForm1-4}
\eea
On the second line of the above expression, $(\mathcal{T}_i\big|_{b_i})^T$ denotes the inverse permutations of $(\mathcal{T}_i\big|_{b_i})$ where $b_i$ is considered as the rightmost node in $\mathcal{T}_i$. On the third line, only the relative order of elements on the master chain is reversed. Permutations $\pmb{\sigma}$ on the third line of \eqref{Eq:GenForm1-4} are just those established by the type-B graph $\mathcal{F}$ whose contribution is (\ref{Eq:GenForm1-5}).
Hence \eqref{Eq:GenForm2-3} is further written as
\bea
\W J^{\rho}(1,2,\dots,n-1)&=&\Sl_{\mathcal{F}\in \mathcal{G}^B\Big/\Big\{\small\substack{\text{graphs with trees} \\ \text{planted at $\rho$ }}\Big\}}C^{\rho}_{\mathcal{F}}\,\Sl_{\pmb{\sigma}^{\mathcal{F}}} \phi\big(1,2,...n-1|\pmb{\sigma}^{\mathcal{F}}\big),\Label{Eq:GenForm2-2}
\eea
where we only summed over those type-B graphs $\mathcal{F}$ in which no tree is planted at the off-shell node $\rho$. For a given graph $\mathcal{F}\in \mathcal{G}^B$, permutations $\pmb{\sigma}^{\mathcal{F}}\in P(1,n-1)$ are summed over and the coefficient $C^{\rho}_{\mathcal{F}}$ is given by
\bea
C^{\rho}_{\mathcal{F}}
&=&(-1)^{n-l}\,\big(F_{i_l}\cdot F_{i_{l-1}}\cdot...\cdot F_{i_1}\cdot \epsilon_1\big)^{\rho}\,C^{\mathcal{T}_1}\,C^{\mathcal{T}_2}\,...\,C^{\mathcal{T}_I}.
\eea

{\bf\emph{Step-3}}~~The contribution of all type-B graphs where at least a tree is planted at the off-shell node $\rho$ is given by
\bea
\Sl_{\mathcal{F}\in \Big\{\substack{\text{graphs with trees} \\ \text{planted at $\rho$ }}\Big\}}C^{\rho}_{\mathcal{F}}\,\Sl_{\pmb{\sigma}^{\mathcal{F}}} \phi\big(1,2,...n-1|\pmb{\sigma}^{\mathcal{F}}\big).\Label{Eq:GenForm2-4}
\eea
Assuming that a tree $\mathcal{T}_0$ in a graph $\mathcal{F}$ is planted at the off-shell node $\rho$, all permutations $\pmb{\sigma}^{\mathcal{F}}$ allowed by this graph have the pattern
\bea
\mathcal{T}_0|_{b_0}\shuffle\left(\{i_{l},...,i_1\}\shuffle\mathcal{T}_{1}|_{b_1} \shuffle...\shuffle\mathcal{T}_{I}|_{b_I}\right).
\eea
Hence \eqref{Eq:GenForm2-4} is a vanishing object because of the generalized KK identity (\ref{Eq:BSBGProperty-4}). \emph{That is why we can introduce an arbitrariness $k_A$ into the coefficient $C_{\mathcal{F}}$ for graphs $\mathcal{F}\in \mathcal{G}^{B}$.}

{\bf\emph{Step-4}} When the vanishing expression (\ref{Eq:GenForm2-4}) is added and the momentum $k_A$ of the off-shell node is chosen as the total momentum of the substructure contracted to $\rho$, \eqref{Eq:GenForm2-2} becomes
\bea
\W J^{\rho}(1,2,\dots,n-1)&=&\Sl_{\mathcal{F}\in \mathcal{G}^B}C^{\rho}_{\mathcal{F}}\,\Sl_{\pmb{\sigma}^{\mathcal{F}}} \phi\big(1,2,...n-1|\pmb{\sigma}^{\mathcal{F}}\big),\Label{Eq:GenForm2-5}
\eea
which can be further arranged as \eqref{Eq:GenForm2-1} when we exchange the order of the two summations.

\subsubsection{Expansion of generalized strength tensors}

Generalized strength tensor $\W F^{\nu\rho}(1,...,n-1)$ is defined by
\bea
\W F^{\nu\rho}(1,...,n-1)\equiv 2 k_{1,n-1}^{\nu}\W J^{\rho}(1,...,n-1)-2 k_{1,n-1}^{\rho}\W J^{\nu}(1,...,n-1), \Label{Eq:GeneralizedF}
\eea
where $\W J^{\rho}(1,...,n-1)$ is the effective current (\ref{Eq:GenForm1-1}). This generalized strength tensor can be decomposed into two parts (i). a combination of BS currents and (ii). an extra term which is a sum of products of two lower-point effective currents. Particularly,
\bea
\W F^{\nu\rho}_{(1,n-1)}= \Sl_{\pmb{\sigma}\in P(1,n-1)}N^{\nu\rho}_{C}(\pmb{\sigma})\phi(1,...,n-1|\pmb{\sigma})+\Sl_{1\leq i< n-1}2\left[\W J^{\nu}_{(1,i)}\W J^{\rho}_{(i+1,n-1)}-\W J^{\rho}_{(i+1,n-1)}\W J^{\nu}_{(1,i)}\right],\Label{Eq:GeneralizedFDec1}
\eea
where we have respectively used $\W F^{\nu\rho}_{(1,n-1)}$ and $\W J^{\nu}_{(1,i)}$ to denote $\W F^{\nu\rho}{(1,...,n-1)}$ and $\W J^{\nu}{(1,...,i)}$ for short. The expansion coefficients $N^{\nu\rho}_{C}(\pmb{\sigma})$ are the type-C numerators which have already been defined by \eqref{Eq:TypeC}. The full proof of \eqref{Eq:GeneralizedFDec1} is complicated and we just provide a sketch of the proof in the appendix.



\subsection{The expansion of $T_{A1}+T_{B1}$}\label{sec:TA1TB1}

Having the expansion formulas (\ref{Eq:GenForm2-1}) and (\ref{Eq:GeneralizedFDec1}) in hand, let us prove that $T^{\rho}_{A1}+T^{\rho}_{B1}$ satisfies the decomposition formula (\ref{Eq:GenForm1-1}).

 To proceed, we insert the effective $3$-point vertex (\ref{Eq:EffCubicVertex}) and the $4$-point vertex (\ref{Eq:QuarticVertex}) into \eqref{Eq:TA1} and \eqref{Eq:TB1}, respectively. Then the $T^{\rho}_{A1}$ becomes
\bea
T^{\rho}_{A1}&=&{1\over s_{1\dots n-1}}\Sl_{1\leq i<n-1} \Bigl[\big(\W J_{(1,i)}\cdot \W F_{(i+1,n-1)}\big)^{\rho}-\W J_{(1,i)}^{\rho}\big(\W J_{(i+1,n-1)}\cdot 2k_{1,i}\bigr)\Bigr]\nn
&&+{1\over s_{1\dots n-1}}\Sl_{1\leq i<j<n-1} \Bigl[\W J_{(1,i)}^{\rho}\big(\W J_{(i+1,j)}\cdot \W J_{(j+1,n-1)}\big)-\big(\W J_{(1,i)}\cdot\W J_{(i+1,j)}\big)\W J_{(j+1,n-1)}^{\rho}\Bigr],\Label{Eq:TA1-1}
\eea
where the definition (\ref{Eq:GeneralizedF}) of generalized strength tensor has been used. The $T^{\rho}_{B1}$ term is given by
\bea
T^{\rho}_{B1}&=&{1\over s_{1\dots n-1}}\Sl_{1\leq i<j<n-1} \Big[2\big(\W J_{(1,i)}\cdot \W J_{(j+1,n-1)}\big)\W J^{\rho}_{(i+1,j)}\nn
&&~~~~~~~~~~~~~~~~~~~~~~~-\big(\W J_{(1,i)}\cdot \W J_{(i+1,j)}\big)\W J^{\rho}_{(j+1,n-1)}-\big(\W J_{(i+1,j)}\cdot \W J_{(j+1,n-1)}\big)\W J^{\rho}_{(1,i)}\Big].
\Label{Eq:GenForm1-2}
\eea
Hence the sum of $T^{\rho}_{A1}$ and $T^{\rho}_{B1}$ is presented as
\bea
T^{\rho}_{A1}+T^{\rho}_{B1}&=&{1\over s_{1\dots n-1}}\Sl_{1\leq i<n-1} \Bigl[\big(\W J_{(1,i)}\cdot \W F_{(i+1,n-1)}\big)^{\rho}-\W J_{(1,i)}^{\rho}\big(\W J_{(i+1,n-1)}\cdot 2k_{1,i}\bigr)\Bigr]\nn
&&+{1\over s_{1\dots n-1}}\Sl_{1\leq i<j<n-1} \Bigl[2\big(\W J_{(1,i)}\cdot \W J_{(j+1,n-1)}\big)\W J^{\rho}_{(i+1,j)}-2\big(\W J_{(1,i)}\cdot \W J_{(i+1,j)}\big)\W J^{\rho}_{(j+1,n-1)}\Bigr].\Label{Eq:TA1A2}
\eea
Expressing the  $\W J^{\rho}_{(1,i)}$, $\W J^{\rho}_{(i+1,n-1)}$ and $\W F_{(i+1,n-1)}^{\nu\rho}$ on the first line of \eqref{Eq:TA1A2}  by the expansion formulas (\ref{Eq:GenForm1-1}), (\ref{Eq:GenForm2-1}) and (\ref{Eq:GeneralizedFDec1}) respectively, we get
\bea
T_{A1}+T_{B1}&=&\frac{1}{s_{1...n-1}}\Sl_{i=1}^{n-2}\Sl_{\pmb{\alpha}\in P(2,i)}\Sl_{\pmb{\beta}\in P(i+1,n-1)}\bigl[N_A(1,\pmb{\alpha})\cdot N_C(\pmb{\beta})-N_A(1,\pmb{\alpha})N_{B}(\pmb{\beta})\cdot 2k_{1,i}\bigr]^{\rho}\nn
&&~~~~~~~~~~~~~~~~~~~~~~~~~~~~~~~~~~~~~~~~~~~~~~~~~~~\times\phi(1,2,...,i\big|1,\pmb{\alpha})\phi(i+1,...,n-1\big|\pmb{\beta}),
\eea
where the second line in \eqref{Eq:TA1A2} has been canceled with the corresponding contribution of the second term in \eqref{Eq:GeneralizedFDec1}.
According to \eqref{Eq:Num Relations}, the expression in the square brackets for given $\pmb{\alpha}\in P(2,i)$ and $\pmb{\beta}\in P(i+1,n-1)$ is just $N_A^{\rho}{(1,\pmb{\alpha},\pmb{\beta})}$. Thus, $T_{A1}+T_{B1}$ turns into
\bea
T_{A1}+T_{B1}&=&\frac{1}{s_{1...n-1}}\Sl_{i=1}^{n-2}\Sl_{\pmb{\alpha}\in P(2,i)}\Sl_{\pmb{\beta}\in P(i+1,n-1)}N_A^{\rho}{(1,\pmb{\alpha},\pmb{\beta})}\phi(1,2,...,i\big|1,\pmb{\alpha})\phi(i+1,...,n-1\big|\pmb{\beta})\\
&=&\frac{1}{s_{1...n-1}}\Sl_{i=1}^{n-2}\Sl_{\{l_2,...,l_i\}\subset \{2,...,n-1\}}\Sl_{\pmb{\alpha}\in P(l_2,l_i)}\Sl_{\pmb{\beta}\in P(l_{i+1},l_{n-1})}N_A^{\rho}{(1,\pmb{\alpha},\pmb{\beta})}\nn
&&~~~~~~\Big[\phi(1,2,...,i\big|1,\pmb{\alpha})\phi(i+1,...,n-1\big|\pmb{\beta}) -\phi(1,2,...,n-i-1\big|\pmb{\beta})\phi(n-i,...,n-1\big|1,\pmb{\alpha})\Big]. \nonumber
\eea
On the second line, we summed over all possible choices of order-$i$ subset $\{1,l_2,...,l_i\}$ while the elements in $\{1,...,n\}\setminus\{1,l_2,...,l_i\}$ are denoted as $l_{i+1},...,l_{n-1}$. In the boundary case $i=1$, we have $\pmb{\alpha}=\emptyset$ and $\pmb{\beta}\in P(2,n-1)$. The fact that the BS current $\phi(1,2,...,i\big|l_1,...,l_i)$ has to vanish if $\{1,2,...,i\}\setminus\{1,l_2,...,l_i\}\neq \emptyset$ has been considered. The above summation can be further arranged by collecting terms corresponding to a given permutation $\pmb{\sigma}$, then summing over all possible $\pmb{\sigma}\in P(2,n-1)$.
Thus $T_{A1}+T_{B1}$ becomes
\bea
T_{A1}+T_{B1}&=&\frac{1}{s_{1...n-1}}\Sl_{\pmb{\sigma}\in P(2,n-1)}N_A^{\rho}{(1,\pmb{\sigma})}\,\Sl_{i=1}^{n-2}\Sl_{\pmb{\sigma}\to\pmb{\sigma}_L,\pmb{\sigma}_R}\Big[\phi(1,2,...,i\big|1,\pmb{\sigma}_L)\phi(i+1,...,n-1\big|\pmb{\sigma}_R)\nn
&&~~~~~~~~~~~~~~~~~~~~~~~~~~-\phi(1,2,...,n-i-1\big|\pmb{\sigma}_R) \phi(n-i,...,n-1\big|1,\pmb{\sigma}_L)\Big]\nn
&=&\,\,\Sl_{\pmb{\sigma}\in P(2,n-1)}N_A^{\rho}{(1,\pmb{\sigma})}\,\phi(1,2,...,n-1\big|1,\pmb{\sigma}),
\eea
where $\pmb{\sigma}\to \pmb{\sigma}_L,\pmb{\sigma}_R$ stands for the division of $\pmb{\sigma}\to \pmb{\sigma}_L =\{\sigma_2,...,\sigma_i\},\pmb{\sigma}_R=\{\sigma_{i+1},...,\sigma_{n-1}\}$ (If $i=1$, then $\pmb{\sigma}_L=\emptyset$). In the above equation, we have applied the Berends-Giele recursion (\ref{Eq:BerendsGieleBS}) for BS currents. Hence the proof for $T^{\rho}_{A1}+T^{\rho}_{B1}=\W J^{\rho}$ where $\W J^{\rho}$ is the effective current satisfying \eqref{Eq:GenForm1-1}, has been completed.

\section{Off-shell extended BCJ numerators with Lie symmetries}\label{sec:NumLieSym}
\begin{figure}
\centering\includegraphics[width=0.28\textwidth]{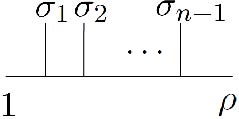}
\caption{ Half-ladder diagrams that characterize the numerators in DDM basis where $1$ and the off-shell line $n$ (with the Lorentz index $\rho$) are considered as the leftmost and the rightmost elements }\label{Fig:HalfLadder}
\end{figure}
\begin{figure}
\centering\includegraphics[width=0.95\textwidth]{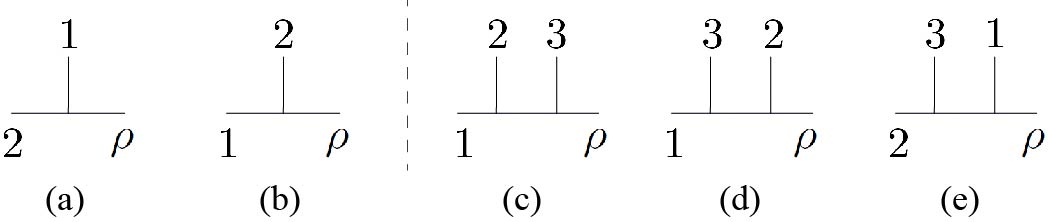}
\caption{Graphs (a) and (b) are the cubic diagrams corresponding to numerators $N^{\rho}(2,1)$ and $N^{\rho}(1,2)$ which are related with each other by antisymmetry. Numerators $N^{\rho}(1,2,3)$, $N_A^{\rho}(1,2,3)$ and $N_A^{\rho}(1,3,2)$ that are respectively characterized by (c), (d) and (e) are related via Jacobi identity.}\label{Fig:AntiSymJacobiEG}
\end{figure}
So far, the decomposition formula (\ref{Eq:GenForm1}) for Berends-Giele currents, whose on-shell limit reproduces the decomposition  (\ref{Eq:OnShellDec}) of YM amplitudes, have been proved. Since the leftmost and the rightmost elements in the off-shell extended BCJ numerators $N^{\rho}(1,\pmb{\sigma})\equiv N_A^{\rho}(1,\pmb{\sigma})$  in (\ref{Eq:GenForm1}) are respectively fixed as the element $1$ and the off-shell line (say the element $n$ with the Lorentz index $\rho$), these numerators correspond to the BCJ numerators in the DDM basis \cite{DelDuca:1999rs} that are characterized by half-ladder cubic diagrams (as shown in \figref{Fig:HalfLadder}). Other BCJ numerators can be expressed in terms of  $N_A^{\rho}(1,\pmb{\sigma})$ via  anti-symmetry and Jacobi identity. For example, the numerator $N^{\rho}(2,1)$ which corresponds to \figref{Fig:AntiSymJacobiEG} (a) can be expressed by $N^{\rho}(1,2)$ which corresponds to \figref{Fig:AntiSymJacobiEG} (b) via antisymmetry:
\bea
N^{\rho}(2,1)\equiv-N^{\rho}(1,2)=-N_A^{\rho}(1,2).
\eea
Moreover, having the numerators $N^{\rho}(1,2,3)\equiv N_A^{\rho}(1,2,3)$ and $N^{\rho}(1,3,2)\equiv N_A^{\rho}(1,3,2)$ which are characterized by \figref{Fig:AntiSymJacobiEG} (c) and (d) respectively, we can define the BCJ numerator $N^{\rho}(2,3,1)$ corresponding to \figref{Fig:AntiSymJacobiEG} (e) by considering the Jacobi identity and antisymmetry:
\bea
 N^{\rho}(2,3,1)\equiv -\left[N^{\rho}(1,2,3)-N^{\rho}(1,3,2)\right]=-\left[N_A^{\rho}(1,2,3)-N_A^{\rho}(1,3,2)\right].
\eea
 Although, these artificial constructions can be extended to more generic cases straightforwardly, the numerators constructed in this way do not have relabelling property. Specifically, different numerators corresponding to a same topology (for example $N^{\rho}(2,1)$ and $N^{\rho}(1,2)$) of cubic graph cannot be simply related with each other by relabelling the external lines. In this section, we symmetrize\footnote{Symmetrization of numerators was earlier proposed in \cite{Fu:2014pya}, where the symmetric numerators were obtained by averaging the full on-shell color-dressed YM amplitudes (or gravity amplitudes). The symmetrization process in this paper can be  considered as the generalization of \cite{Fu:2014pya} to off-shell color-ordered currents.} the off-shell extended numerators so that they have \emph{Lie symmetries}. Such numerators were first solved in earlier work \cite{Lee:2016tbd,Bridges:2019siz} for supersymmetric YM theory. As pointed in \cite{Mafra:2015vca,Lee:2016tbd,Bridges:2019siz}, numerators with Lie symmetries simultaneously have relabelling properties and satisfy the Jacobi identity and antisymmetry. The explicit relation between (the bosonic part of) numerators constructed in \cite{Lee:2016tbd} and in the current paper is investigated via two- and three-point examples.

\subsection{Off-shell extended numerators with Lie symmetries}
The symmetrization of numerators is achieved by the following two steps:
(i). We  symmetrize the  type-A numerators (\ref{Eq:Type-A}) by taking the average over all choices of the reference order while the first (i.e. the element $1$) and last elements (i.e. the Lorentz index $\rho$) are fixed. (ii). We construct the numerators with Lie symmetries by taking the average of distinct expressions of the full color-dressed Berends-Giele currents, and then apply the KK relation properly. These two steps are separately demonstrated in the following.

\subsubsection{Numerators with $(n-2)!$-relabelling property}
Although the type-A numerators defined in the previous sections are obtained by fixing the reference order $\mathsf{R}=\{\rho,n-1,...,2,1\}$, one can also define type-A numerators with a different choice of reference order $\mathsf{R}=\{\rho,\gamma_{n-2},...,\gamma_1,1\}$ (where $\pmb{\gamma}=\{\gamma_1,...,\gamma_{n-2}\}\in {P}(2,n-1)$). We use the notation  $N^{\mathsf{R},\,\rho}_{A}(1,\pmb{\sigma})$ to specify the type-A numerator with the reference order $\mathsf{R}$. Symmetric numerator $\bar {N}^{\rho}(1,\sigma_1,...,\sigma_{n-2})$ with the $(n-2)!$-relabelling property is introduced as follows
\bea
\bar {N}^{\rho}(1,\sigma_1,...,\sigma_{n-2})=\frac{1}{(n-2)!}\Sl_{\pmb{\gamma}\in {P}(2,n-1)}N_A^{\{\rho,\,\pmb{\gamma},\,1\},\,\rho}(1,\sigma_1,...,\sigma_{n-2}). \Label{Eq:NumRelabeling}
\eea
Apparently, all $\bar {N}^{\rho}(1,\sigma_1,...,\sigma_{n-2})$ can be obtained from $\bar {N}^{\rho}(1,2,...,n-1)$ by a simple relabelling $2\to \sigma_1$, $3\to\sigma_2$, ..., $n-1\to \sigma_{n-2}$. With this symmetrization, the Berends-Giele current $J^{\rho}(1,2,...,n-1)$, which has been shown to satisfy \eqref{Eq:GenForm1}, is reexpressed by
\bea
J^{\rho}(1,2,...,n-1)=\Sl_{\pmb{\sigma}\in P(2,n-1)}\bar {N}^{\rho}(1,\pmb{\sigma})\phi(1,2,...,n-1| 1,\pmb{\sigma})+ H^{\rho}(1,2,...,n-1).  \Label{Eq:GenExpansionRelabelling}
\eea
The $H^{\rho}(1,2,...,n-1)$ in the above expression is given by
\bea
H^{\rho}(1,2,...,n-1)&\equiv& K^{\rho}(1,2,...,n-1)+L^{\rho}(1,2,...,n-1)\nn
&&~~~~~~+\Sl_{\pmb{\sigma}\in P(2,n-1)}\Delta N^{\rho}(1,\pmb{\sigma})\phi(1,2,...,n-1| 1,\pmb{\sigma}),
\eea
where $K^{\rho}$ and $L^{\rho}$ were respectively defined by \eqref{Eq:GenK} and  \eqref{Eq:GenL}, while $\Delta N^{\rho}(1,\pmb{\sigma})$ for a given $\pmb{\sigma}$ is explicitly written as
\bea
\Delta N^{\rho}(1,\pmb{\sigma})\equiv \frac{1}{(n-2)!}\Sl_{\substack{\pmb{\gamma}\in P(2,n-1)\\ \pmb{\gamma}\neq\{n-1,...,3,2\}}}\left[{N}_A^{\{\rho,n-1,,...,2,1\},\,\rho}(1,\pmb{\sigma})-{N}_A^{\{\rho,\,\pmb{\gamma},1\},\,\rho}(1,\pmb{\sigma})\right].\Label{Eq:Hterm}
\eea
As pointed in \cite{Fu:2017uzt,Du:2017kpo},  numerators corresponding to all reference orders can reproduce the correct on-shell amplitude $A(1,2,...,n)$ via \eqref{Eq:OnShellDec}. Thus $\Delta N^{\rho}(1,\pmb{\sigma})$  has to vanish in the on-shell limit. As a consequence, the $H^{\rho}(1,2,...,n-1)$ must also vanish and \eqref{Eq:GenExpansionRelabelling} precisely gives a decomposition of amplitude $A(1,2,...,n)$ with BCJ numerators in DDM basis, when the on-shell limit is taken.

\paragraph{\bf \emph{Two-point case}}  The numerator $\bar {N}^{\rho}(1,2)$ is just given by  ${N}_A^{\{\rho,\,2,\,1\},\,\rho}(1,2)$ (see \eqref{Eq:2ptOffShellNum}), thus it is written as
\bea
\bar {N}^{\rho}(1,2)=\text{\figref{Fig:2ptGraphsA} (a)}+\text{\figref{Fig:2ptGraphsA} (b)}. \Label{2ptSymN}
\eea
Since the $L^{\rho}$ and $\Delta N^{\rho}$ in this case are both zero, the $H^{\rho}(1,2)$ term is
\bea
H^{\rho}(1,2)=K^{\rho}(1,2)={1\over s_{12}}(\epsilon_1\cdot\epsilon_2)k_{1,2}^{\rho}.  \Label{2ptHterm}
\eea
\paragraph{\bf \emph{Three-point case}} According to \eqref{Eq:NumRelabeling}, the numerator $\bar {N}^{\rho}(1,2,3)$ is defined by the following expression
\bea
\bar {N}^{\rho}(1,2,3)=\frac{1}{2}\left[{N}_A^{\{\rho,3,2,1\},\,\rho}(1,2,3)+{N}_A^{\{\rho,2,3,1\},\,\rho}(1,2,3)\right], \Label{Eq:NumRelabeling-1}
\eea
where ${N}_A^{\{\rho,3,2,1\},\,\rho}(1,2,3)$ was already given by the first line of \eqref{Eq:3ptOffShellNum}.
According to the graphic rule, the numerator ${N}_A^{\{\rho,2,3,1\},\,\rho}(1,2,3)$ with the reference order $\mathsf{R}=\{\rho,2,3,1\}$ is given by the sum of \figref{Fig:3ptGraphsA} (a), (b), (c), (e), (h) and (i)$(2\leftrightarrow3)$. Therefore, the $\bar {N}^{\rho}(1,2,3)$ is finally expressed as the following sum of graphs
\bea
\bar {N}^{\rho}(1,2,3)&=&\text{\figref{Fig:3ptGraphsA} (a)}+\text{\figref{Fig:3ptGraphsA} (b)}+\text{\figref{Fig:3ptGraphsA} (c)}+\text{\figref{Fig:3ptGraphsA} (e)}+\text{\figref{Fig:3ptGraphsA} (h)}\nn
&&~~~~~~+{1\over 2}\Big[\text{\figref{Fig:3ptGraphsA} (d)}+ \text{\figref{Fig:3ptGraphsA} (i)} (2\leftrightarrow3)\Big].   \Label{3ptSymN}
\eea
The numerator $\bar {N}^{\rho}(1,3,2)$ is straightforwardly obtained by exchanging $2$ and $3$ in $\bar {N}^{\rho}(1,2,3)$:
\bea
\bar {N}^{\rho}(1,3,2)&=&\text{\figref{Fig:3ptGraphsA} (f)}+\text{\figref{Fig:3ptGraphsA} (g)}+\text{\figref{Fig:3ptGraphsA} (h)}+\text{\figref{Fig:3ptGraphsA} (c)}+\text{\figref{Fig:3ptGraphsA} (e)}\nn
&&~~~~~~+{1\over 2}\Big[\text{\figref{Fig:3ptGraphsA} (i)}+ \text{\figref{Fig:3ptGraphsA} (d)} (2\leftrightarrow3)\Big].
\eea
In this example, $H^{\rho}(1,2,3)$ is given by
\bea
H^{\rho}(1,2,3)\equiv K^{\rho}(1,2,3)+L^{\rho}(1,2,3)+\Sl_{\pmb{\sigma}\in P\{2,3\}}\Delta N^{\rho}(1,\pmb{\sigma})\phi(1,2,3| 1,\pmb{\sigma}),  \Label{3ptHterm}
\eea
where $K^{\rho}(1,2,3)$ and $L^{\rho}(1,2,3)$ have already been presented by \eqref{Eq:3ptKterm} and \eqref{Eq:3ptLterm}, respectively. The $\Delta N^{\rho}(1,\pmb{\sigma})$ reads
\bea
\Delta N^{\rho}(1,\pmb{\sigma})&\equiv& \frac{1}{2}\left[{N}_A^{\{\rho,3,2,1\},\rho}(1,\pmb{\sigma})-{N}_A^{\{\rho,2,3,1\},\rho}(1,\pmb{\sigma})\right].
\eea
When the graphic expression of type-A numerators defined with different reference orders are considered, $\Delta N^{\rho}(1,\pmb{\sigma})$ ($\pmb{\sigma}\in P(2,3)$) can be written explicitly
\bea
\Delta N^{\rho}(1,2,3)&=&{1\over 2}\Big[\text{\figref{Fig:3ptGraphsA} (d)}- \text{\figref{Fig:3ptGraphsA} (i)} (2\leftrightarrow3)\Big]={1\over 2}(\epsilon_3\cdot\epsilon_2)(2k_2\cdot 2k_1)\epsilon_1^{\rho},\nn
\Delta N^{\rho}(1,3,2)&=&{1\over 2}\Big[\text{\figref{Fig:3ptGraphsA} (i)}-\text{\figref{Fig:3ptGraphsA} (d)}(2\leftrightarrow3) \Big]=-{1\over 2}(\epsilon_2\cdot\epsilon_3)(2k_3\cdot 2k_1)\epsilon_1^{\rho}.
\eea
 \begin{figure}
\centering
\includegraphics[width=0.9\textwidth]{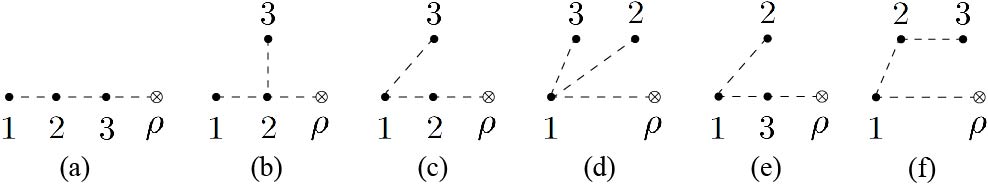}
\caption{Graphs without distinguishing line styles. In each graph, the dashed lines between nodes only reflect the relative orders between nodes but do not involve any kinematic information.}\label{Fig:DashedLineGraphs}
\end{figure}
\paragraph{\bf\emph{Comment on graphs contributing to $\bar {N}^{\rho}(1,\pmb{\sigma})$}:} In the three-point example, graphs contributing to the $\bar {N}^{\rho}(1,2,3)$ in \eqref{3ptSymN} are independent of the choice of reference order. In fact, they can be obtained by the following steps: (i) Construct all possible graphs (as shown by \figref{Fig:DashedLineGraphs}) corresponding to the permutation $1,2,3,\rho$ without distinguishing line styles (all lines in \figref{Fig:DashedLineGraphs} are dashed lines \cite{Hou:2018bwm,Du:2019vzf}, which only encode the relative orders between nodes and do not imply any kinematic information); (ii) For a given structure obtained in the previous step (e.g. \figref{Fig:DashedLineGraphs} (f)), draw all possible graphs (where the path between $1$ and $\rho$ is considered as the master chain) corresponding to distinct reference orders (e.g. for the structure \figref{Fig:DashedLineGraphs} (f), there are two graphs \figref{Fig:3ptGraphsA} (d) and \figref{Fig:3ptGraphsA} (i)($2\leftrightarrow3$) corresponding to the reference order $\{\rho,3,2,1\}$ and $\{\rho,2,3,1\}$) and then take the average of them. This observation is straightforwardly generalized to an arbitrary $\bar {N}^{\rho}(1,\pmb{\sigma})$.

\subsubsection{Numerators with Lie symmetries}
 \begin{figure}
\centering
\includegraphics[width=0.18\textwidth]{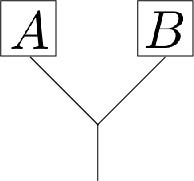}
\caption{A binary graph which describe the antisymmetric object $\bar {N}^{\rho}({[A,B]})$, where $A$ and $B$ are substructures (which may also be expressed by binary graphs) attached to a same cubic vertex.}\label{Fig:BinaryGraph}
\end{figure}
 Although, the numerators $\bar {N}^{\rho}(1,\pmb{\sigma})$ in \eqref{Eq:GenExpansionRelabelling} already has $(n-2)!$-relabeling property, they are still not the numerators which simultaneously satisfy the $(n-1)!$-relabeling property and the algebraic identities. To make up this shortcoming, we introduce a further symmetrization  by rewriting \eqref{Eq:GenExpansionRelabelling} as
\bea
J^{\rho}(1,2,...,n-1)={1\over n-1}\Sl_{i=1}^{n-1}\Sl_{\pmb{\sigma}\in P(1,...,\cancel{i},...n-1)}\bar {N}^{\rho}(i,\pmb{\sigma})\phi(1,2,...,n-1| i,\pmb{\sigma})+ H'^{\rho}(1,2,...,n-1),\Label{Eq:GenExpansionRelabelling1}
\eea
where $P(1,...,\cancel{i},...n-1)$ denotes the set of all permutations of elements in $\{1,2,...,i-1,i+1,...,n-1\}$, $\bar {N}^{\rho}(i,\pmb{\sigma})$ are the numerators with $(n-2)!$-relabeling property, in which $i$ plays as the first element. The $H'^{\rho}(1,2,...,n-1)$ in the above is defined by
\bea
H'^{\rho}(1,2,...,n-1)&=&H^{\rho}(1,2,...,n-1)+{1\over n-1}\Sl_{i=2}^{n-1}\Bigl[\Sl_{\pmb{\sigma}\in P(2,n-1)}\bar {N}^{\rho}(1,\pmb{\sigma})\phi(1,2,...,n-1| 1,\pmb{\sigma})\nn
&&~~~~~~~~-\Sl_{\pmb{\sigma}\in P(1,...,\cancel{i},...,n-1)}\bar {N}^{\rho}(i,\pmb{\sigma})\phi(1,2,...,n-1| i,\pmb{\sigma})\Bigr].\Label{Eq:HPrime}
\eea
The critical point is following: when contracting the last term (for a given $i$) in the square brackets with $\epsilon_n$ and taking the on-shell limit, we also get the color-ordered YM amplitude $A(1,2,...,n)$\footnote{This point can be clarified in the framework of CHY formula  \cite{Cachazo:2013hca,Cachazo:2013gna,Cachazo:2013iea, Cachazo:2014nsa,Cachazo:2014xea}: (i) Since the expansion of reduced Pfaffian in \cite{Du:2017kpo} is independent of the choice of reference orders, one can also take the average of the expansions of the reduced Pfaffian over all $(n-2)!$ possible choices of reference orders $\{\rho,\,\pmb{\gamma},\,1\}$. This corresponds to the on-shell limit of the first term of \eqref{Eq:GenExpansionRelabelling}. (ii). One can also expand the reduced Pfaffian in terms of the KK basis of Parke-Taylor factors where $i$ and $n$ play as the first and the last elements, then take the average over all choices of reference orders. After this step, the coefficient of each Parke-Taylor factor is just the on-shell limit of $\bar {N}^{\rho}(i,\pmb{\sigma})$, while the factor accompanied to this coefficient (i.e. the corresponding BS amplitude) is just $A(1,2,...,n|i,\pmb{\sigma},n)$ which is the on-shell limit of $\phi(1,2,...,n-1| i,\pmb{\sigma})$. Thus, if we average expansions of reduced Pfaffians over all the choices of $i$, the CHY formula also gives rise to the on-shell YM amplitude $A(1,2,...,n)$.}. As a result, the $H'^{\rho}(1,2,...,n-1)$ has to vanish in the on-shell limit. Supposing a permutation $\pmb{\sigma}$ in the first term of \eqref{Eq:GenExpansionRelabelling1} and the second term in the square brackets of \eqref{Eq:HPrime} can be written as $\pmb{\sigma}=\pmb{\sigma}_L,1,\pmb{\sigma}_R$, one can apply KK relation to express the corresponding BS current $\phi(1,2,...,n-1| i,\pmb{\sigma}_L,1,\pmb{\sigma}_R)$ as follows
\bea
\phi(1,2,...,n-1|i,\pmb{\sigma}_L,1,\pmb{\sigma}_R)=\Sl_{\pmb{\gamma}\in \left\{\pmb{\sigma}^T_L,i\right\}\,\shuffle\,\pmb{\sigma}_R}(-1)^{|\pmb{\sigma}_L|+1}\phi(1,2,...,n-1|1,\pmb{\gamma}).\Label{Eq:KKBS2}
\eea
 As pointed in \cite{DelDuca:1999rs,Bern:2011ia}, when the above expression is substituted into the first term of \eqref{Eq:GenExpansionRelabelling1} and the coefficients for a given $\pmb{\gamma}\in \left\{\pmb{\sigma}^T_L,i\right\}\,\shuffle\,\pmb{\sigma}_R$ are collected together, \eqref{Eq:GenExpansionRelabelling1} turns into
\bea
&&J^{\rho}(1,2,...,n-1)\nn
&=&{1\over n-1}\Sl_{\pmb{\gamma}\in P(2,n-1)}\bar {N}^{\rho}([[[1,\gamma_1],\gamma_2],...])\phi(1,2,...,n-1| 1,\pmb{\gamma})+ H'^{\rho}(1,2,...,n-1),\Label{Eq:GenExpansionRelabelling2}
\eea
where $H'^{\rho}(1,2,...,n-1)$ is given by
\bea
H'^{\rho}(1,2,...,n-1)&=&H^{\rho}(1,2,...,n-1)\nn
&&+{1\over n-1}\Sl_{\pmb{\sigma}\in P(2,n-1)}\left[\bar {N}^{\rho}(1,\pmb{\gamma})-\bar {N}^{\rho}([[[1,\gamma_1],\gamma_2],...])\right]\phi(1,2,...,n-1| 1,\pmb{\gamma}).\Label{Eq:HPrime2}
\eea
\begin{figure}
\centering\includegraphics[width=0.28\textwidth]{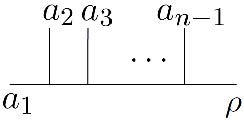}
\caption{ A half-ladder diagram where $a_1$ and the off-shell line $n$ (with the Lorentz index $\rho$) play as the two ends }\label{Fig:HalfLadder2}
\end{figure}
In \eqref{Eq:GenExpansionRelabelling2} and \eqref{Eq:HPrime2}, we have introduced  numerators $\bar {N}^{\rho}({[[[a_1,a_2],a_3],...,a_{n-1}]})$ with Lie symmetries which are defined by nested commutators $[[[\,,\,],\,]...,\,]$. For each $[A,B]$, $\bar {N}^{\rho}({[A,B]})$ is defined by
\bea
\bar {N}^{\rho}({[A,B]})=\bar {N}^{\rho}({A,B})-\bar {N}^{\rho}({B,A}),
\eea
which can be conveniently described by a binary graph (see \figref{Fig:BinaryGraph}) \cite{Mafra:2015syt,Lee:2016tbd,Bridges:2019siz,Frost:2020bmk,Mafra:2020qst}. As demonstrated in \cite{Mafra:2015syt,Lee:2016tbd,Bridges:2019siz,Frost:2020bmk,Mafra:2020qst},   $\bar {N}^{\rho}({[[[a_1,a_2],a_3],...,a_{n-1}]})$ with Lie symmetries are the symmetric objects corresponding to the half-ladder diagrams which simultaneously satisfy (i). relabeling property and (ii) the algebraic properties including Jacobi identity and antisymmetry. Therefore, it is reasonable to define numerators with Lie symmetries as
 \bea
 {N}_{\text{sym}}^{\rho}(a_1,a_2,...,a_{n-1})\equiv{1\over n-1}\bar {N}^{\rho}({[[[a_1,a_2],a_3],...,a_{n-1}]}),
 \eea
 which correspond to the half-ladder diagram \figref{Fig:HalfLadder2}. All other BCJ numerators can be generated by numerators of the form ${N}_{\text{sym}}^{\rho}(a_1,...)$, where $a_1$ and the off-shell line $n$ (with the Lorentz index $\rho$) are fixed as the first and the last elements. We now provide the explicit expressions of the ${N}_{\text{sym}}^{\rho}$ with two and three external particles, as examples.

\paragraph{\bf \emph{Two-point case}} BCJ numerator ${N}_{\text{sym}}^{\rho}(1,2)$ with Lie symmetry, which corresponds to the cubic graph \figref{Fig:AntiSymJacobiEG} (b),  is presented as
\bea
{N}_{\text{sym}}^{\rho}(1,2)={1\over 2}\left[\bar {N}^{\rho}(1,2)-\bar {N}^{\rho}(2,1)\right],\Label{Eq:LieSymmetricNum2pt}
\eea
where $\bar {N}(1,2)$ is given by \eqref{2ptSymN} and $\bar {N}(2,1)$ is obtained by exchanging $1$ and $2$ in $\bar {N}(1,2)$.
The antisymmetry of $\bar {N}(1,2)$ is naturally encoded in the construction (\ref{Eq:LieSymmetricNum2pt}). When the graphs are expressed explicitly, we get
\bea
{N}_{\text{sym}}^{\rho}(1,2)={1 \over 2}\bigl[\epsilon_1\cdot (F_2-\epsilon_2\cdot2k_1) -\epsilon_2\cdot(F_1-\epsilon_1\cdot2k_2)\bigr]^\rho.
\eea

\paragraph{\bf \emph{Three-point case}}  Three-point BCJ numerator ${N}_{\text{sym}}^{\rho}(1,2,3)$ with Lie symmetries, which corresponds to the cubic graph \figref{Fig:AntiSymJacobiEG} (c), reads
\bea
{N}_{\text{sym}}^{\rho}(1,2,3)={1\over 3}\left[\bar {N}^{\rho}(1,2,3)-\bar {N}^{\rho}(2,1,3)-\bar {N}^{\rho}(3,1,2)+\bar {N}^{\rho}(3,2,1)\right],\Label{Eq:LieSymmetricNum3pt}
\eea
where $\bar {N}^{\rho}(1,2,3)$ is given by \eqref{3ptSymN}, while other $\bar {N}^{\rho}(a,b,c)$ are obtained from  $\bar {N}^{\rho}(1,2,3)$ via the replacement $1\to a$, $2\to b$ and $3\to c$. Numerators of the form ${N}_{\text{sym}}^{\rho}(a,b,c)$ are defined by imposing the replacement $1\to a$, $2\to b$ and $3\to c$ on ${N}_{\text{sym}}^{\rho}(1,2,3)$. With this definition, the anti-symmetries are naturally encoded, while the Jacobi identity between numerators
\bea
{N}_{\text{sym}}^{\rho}(1,2,3)+{N}_{\text{sym}}^{\rho}(2,3,1)+{N}_{\text{sym}}^{\rho}(3,1,2)=0
\eea
is also satisfied.

\subsection{Comment on the numerators with Lie symmetries}

 In the papers \cite{Lee:2016tbd,Bridges:2019siz}, off-shell extended BCJ numerators with Lie symmetries in supersymmetric YM theory have been constructed by solving nonlinear equations, while supersymmetric version of the decomposition of Berends-Giele currents in terms of BS currents were suggested in \cite{Mafra:2016ltu, Broedel:2013tta}. Along this line, a decomposition of the Berends-Giele currents in terms of local BCJ numerators was introduced. Thus, results in \cite{Lee:2016tbd,Bridges:2019siz} have a strong relation with the constructions in the current paper. Particularly, the  $J^{\rho}(1,2,...,n-1)$ on the LHS of \eqref{Eq:GenExpansionRelabelling2} corresponds to the Berends-Giele current in Lorentz gauge \cite{Lee:2016tbd,Bridges:2019siz}, while the first term on the RHS of  \eqref{Eq:GenExpansionRelabelling2} corresponds to the Berends-Giele current in BCJ gauge \cite{Lee:2016tbd,Bridges:2019siz}. The $H'^{\rho}$ term, which gets contributions of the $K^{\rho}$, $L^{\rho}$ as well as terms which can be expressed via off-shell extended graphs (i.e. the last term in \eqref{3ptHterm} and the last term in \eqref{Eq:GenExpansionRelabelling2}), correspond to the gauge transformation term of the (3.10) in \cite{Lee:2016tbd} (where the $k_{12...p}\hat{H}_{[12...p-1,p]}$ correspond to the $K^{\rho}$ term in this paper). Furthermore, \eqref{Eq:GeneralizedFDec1} satisfied by generalized strength tensor correspond to the (2.14) in \cite{Lee:2016tbd}  where the extra terms with lower-point currents come from the nonlinear definition of the strength tensor ((2.4) in \cite{Lee:2016tbd}). In the following, we compare the explicit two-point and three-point constructions  (\ref{Eq:LieSymmetricNum2pt}) and (\ref{Eq:LieSymmetricNum3pt}) with those given in \cite{Lee:2016tbd}.

In the paper \cite{Lee:2016tbd}, the two-point numerator is given by the ${K}_{12}=A_{12}$ in eq. (3.12) (of \cite{Lee:2016tbd}), whose explicit expression is eq. (3.3) (of \cite{Lee:2016tbd}). Since we are only considering the bosonic part, terms with $\hat W$ do not appear, and the ${A}_{12}$ in \cite{Lee:2016tbd} can be written as is
\bea
{A}^{\rho}_{12}={1\over 4} \bar{N}^{\rho}([1,2])={1\over 2}{N}_{\text{sym}}^{\rho}(1,2),
\eea
Thus the ${A}_{12}$ in \cite{Lee:2016tbd} (upto a normalization factor ${1\over 2}$) has the same form with the numerator (\ref{Eq:LieSymmetricNum2pt}) that is given in the current paper.

The three-point case is more subtle. In the paper \cite{Lee:2016tbd}, three-point numerator with Lie symmetry is displayed by the second equation in (3.13). When we only consider the pure-YM sector and express the result in  \cite{Lee:2016tbd} in terms of graphs, we find that the three-point numerator  in \cite{Lee:2016tbd} can be rearranged into
\bea
A^{\rho}_{123}&=&\hat A^{\rho}_{[12,3]}-k^{\rho}_{123}\hat{H}_{[12,3]}\nn
&=&{1\over 4}{N}_{\text{sym}}^{\rho}(1,2,3)-{1\over 12}\big[\epsilon_3\cdot(\epsilon_1\epsilon_2-\epsilon_2\epsilon_1)\cdot k_{1,3}\big]k_{1,3}^{\rho},
\eea
where ${N}_{\text{sym}}^{\rho}(1,2,3)$ is the numerator (\ref{Eq:LieSymmetricNum3pt}) in the current paper. Apparently, the two numerators are not proportional to each other. Nevertheless, they only differ by a term which is proportional to the total momentum of the on-shell lines. This difference must vanish in the on-shell limit since $\epsilon_4\cdot k_4=-\epsilon_4\cdot k_{1,3}=0$. Therefore, the two constructions reproduce  the same amplitude at on-shell level, only upto a normalization factor $1\over 4$.

\section{Conclusions} \label{sec:Conclusions}

In this work, we studied the expansion of Berends-Giele current in YM theory. We proposed three types of off-shell extended graphs and the corresponding off-shell extended numerators. By the help of the Berends-Giele recursion, we showed that a Berends-Giele current in Feynman gauge could split into three parts: the effective current, a term proportional to the total momentum and a term expressed by lower point Berends-Giele currents (where some external lines are redefined). The last two parts vanish in the on-shell limit, while the effective current can be decomposed in terms of the Berends-Giele currents in BS theory. The coefficients for the BS currents are the type-A numerators which reproduce the same graphic interpretations with the on-shell numerators proposed in \cite{Du:2017kpo}.  BCJ numerators with Lie symmetries that satisfy both relabeling property and algebraic properties have also been obtained via a symmetrization procedure.

This work provides a connection between Feynman diagrams and the CHY formula via graphs, for the graphic expansion of YM amplitudes has been derived from the CHY formula.  In addition, the expansion formula may provide a hint for generalizing the discussions to loop levels. Specifically,
the expansion formula (\ref{Eq:GenForm1}) still holds if the current is attached to a vertex on loop. Moreover, if a 3-gluon vertex is attached by a tree level Berends-Giele current and two loop gluons,  it can also be expressed by the effective vertex which further induces a generalized strength tensor satisfying \eqref{Eq:GeneralizedFDec1}. We leave a systematic study of the loop level extensions to future work. Although we have discussed the relationship between constructions provided in \cite{Lee:2016tbd,Bridges:2019siz} and results presented in this paper, it is still worth studying the full connection between this work and other earlier progresses on  Berend-Giele currents/off-shell extended BCJ numerators, (see e.g., \cite{Mafra:2016ltu,Frost:2020bmk,Mafra:2015syt,Lee:2016tbd,Mizera:2018tdc,Bridges:2019siz}) in a systematical way.

%
\section*{Acknowledgments}
In the end of this paper, we would like to thank the organizers of Summer school on scattering amplitudes 2021 which was held in Hangzhou. We are grateful to the referee of this paper for a lot of constructive suggestions. This work is supported by NSFC under Grant No. 11875206, Jiangsu Ministry of Science and Technology under contract BK20170410.

\begin{figure}
\centering\includegraphics[width=1\textwidth]{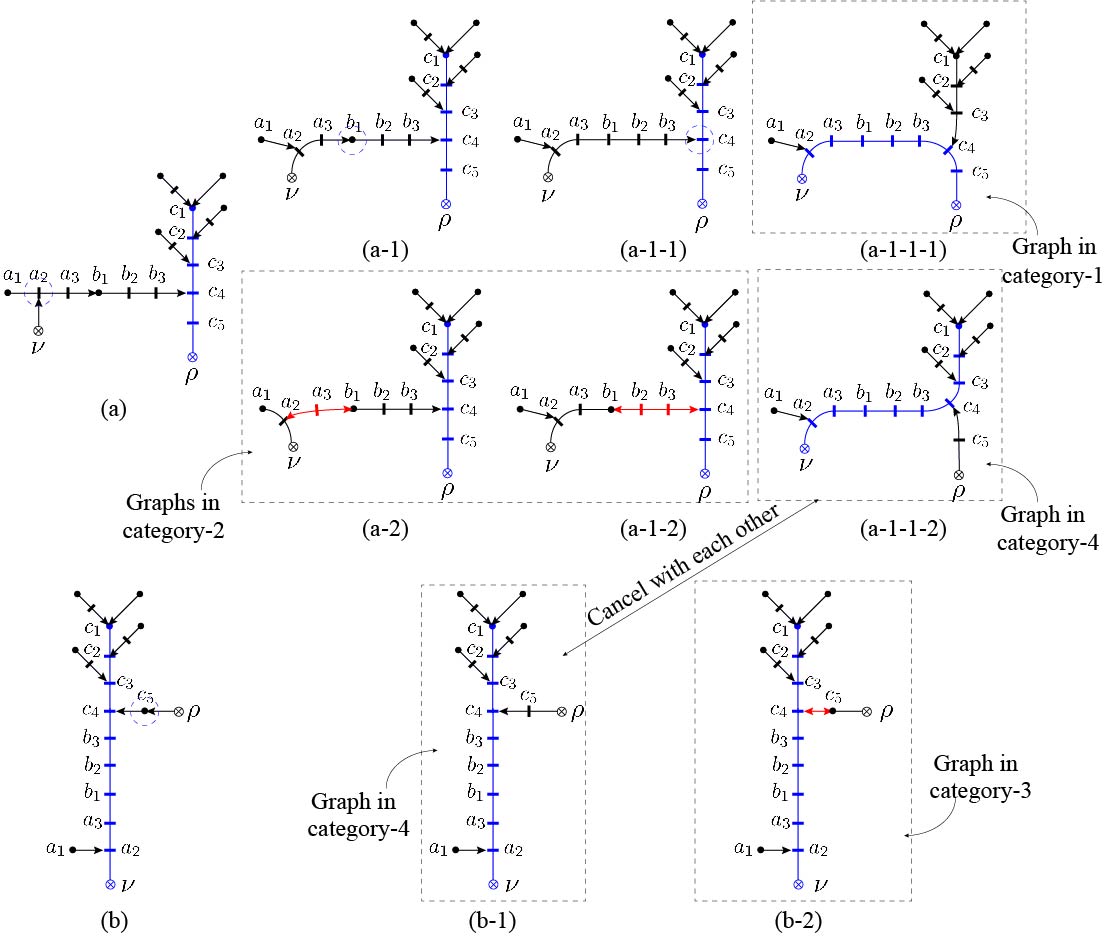}
\caption{ In these graphs, $c_1$ denotes the node $1$, which is the highest-weight node in the reference order. The master chain in each graph is colored blue while a chain  of the form $2k\cdot F\dots F\cdot2k$ is colored red. By transforming the nodes step by step, graph (a) can be decomposed into graphs in three categories: (a-1-1-1) in category-1, (a-2) and (a-1-2) in category-2, and (a-1-1-2) in category-4. Similarly, graph (b) is decomposed into (b-2) in category-3 and (a-1) in category-4. }\label{Fig:NewGraph1}
\end{figure}
%

%


%
\begin{figure}
\centering
\includegraphics[width=0.5\textwidth]{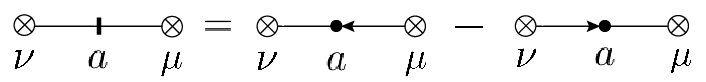}
\caption{ When $k_a^{\mu}$ is denoted by an arrow line pointing towards node $a$ and $\epsilon^{\nu}$ is denoted by a line with no arrow, the strength tensor $F_{a}^{\mu\nu}$ can always be decomposed into the difference of the two graphs on the RHS.}\label{Fig:NodeRelation1}
\end{figure}

\begin{figure}
\centering
\includegraphics[width=0.5\textwidth]{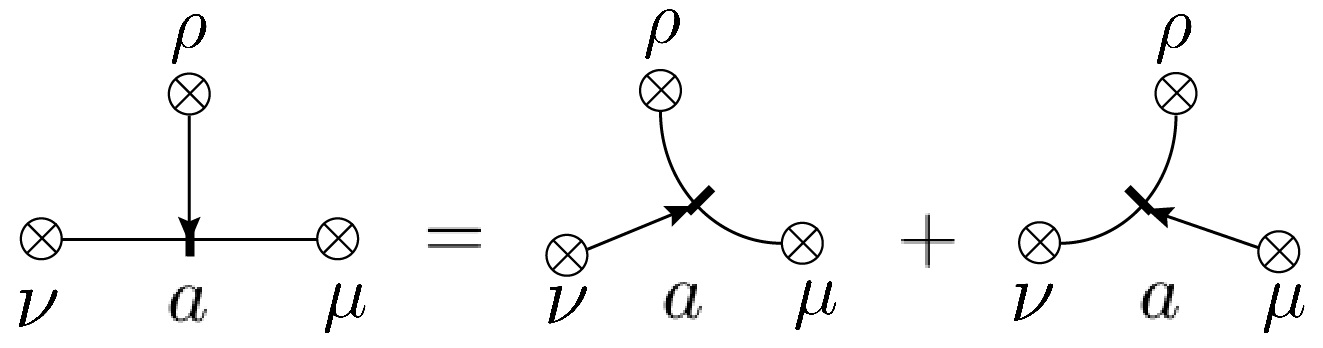}
\caption{ A graphic interpretation of the relation $F_{a}^{\mu\nu}2k_a^{\rho}=F_{a}^{\mu\rho}2k_a^{\nu}+F_{a}^{\rho\nu}2k_a^{\mu}$, which can be verified directly. }\label{Fig:NodeRelation2}
\end{figure}

\appendix

\section{A sketch of the proof of  \eqref{Eq:GeneralizedFDec1}} \label{sec:expression_F_decomposition}

Now we prove \eqref{Eq:GeneralizedFDec1}, via expanding the effective currents in \eqref{Eq:GeneralizedF} according to \eqref{Eq:GenForm2-1}. Since the graphs in  \eqref{Eq:GenForm2-1} and \eqref{Eq:GeneralizedFDec1}, which involve tree(s) planted at the outer Lorentz index, all cancel out, we neglect these graphs in the following discussions. The property (\ref{Eq:GeneralizedFDec1}) for the generalized strength tensor (\ref{Eq:GeneralizedF}) can be proved by the following steps\footnote{A full proof involves much more details, we just provide a rough sketch of the proof and claim that all details in the proof have been confirmed.}.

\paragraph{\bf\emph{Step-1: Manipulation of graphs}} When the effective currents in \eqref{Eq:GeneralizedF} are expanded according to \eqref{Eq:GenForm2-5}, $\W F^{\nu\rho}_{(1,n-1)}$ becomes
    \bea
    \W F^{\nu\rho}_{(1,n-1)}&=&\Sl_{i=1}^{n-1}\biggl[\,\Sl_{\mathcal{F}\in \mathcal{G}_1^B}k^{\nu}_iC^{\rho}_{\mathcal{F}}\Sl_{\pmb{\sigma}^{\mathcal{F}}}\phi\big(1,2,...n-1|\pmb{\sigma}^{\mathcal{F}}\big)-\Sl_{\mathcal{F}\in \mathcal{G}_2^B}k^{\rho}_iC^{\nu}_{\mathcal{F}}\Sl_{\pmb{\sigma}^{\mathcal{F}}}\phi\big(1,2,...n-1|\pmb{\sigma}^{\mathcal{F}}\big)\biggr],  \Label{Eq:GeneralizedFDec2}
    \eea
    where the two terms correspond to the two terms in the definition (\ref{Eq:GeneralizedF}). The $\mathcal{G}_1^B$ and $\mathcal{G}_2^B$ stand for the sets of type-B graphs with the outer Lorentz indices $\rho$ and $\nu$, respectively. Each graph in the first or the second term, associated with a $k_i^{\nu}$ or $k_i^{\rho}$ correspondingly, can be expressed by a graph with two Lorentz indices, as shown by \figref{Fig:NewGraph1} (a) or (b). To reduce these graphs, we start from {\bf\emph{a graph in the first term of \eqref{Eq:GeneralizedFDec2}}}   (e.g. \figref{Fig:NewGraph1} (a)) and find out the path from the node $i$ (in \figref{Fig:NewGraph1} (a) the node $a_2$) to the master chain. Along this path, we apply the relations \figref{Fig:NodeRelation1}\footnote{Graphs with all strength tensors expanded are mentioned as refined graphs in \cite{Hou:2018bwm,Du:2019vzf}.} or \figref{Fig:NodeRelation2} again and again, when we encounter the corresponding substructure on the LHS of \figref{Fig:NodeRelation1} or \figref{Fig:NodeRelation2}. This process terminates till we get a graph in which the path between $\nu$ and $\rho$ has the form $(F\cdot F\cdot...\cdot F)^{\nu\rho}$ (such as \figref{Fig:NewGraph1} (a-1-1-1)). For example, \figref{Fig:NewGraph1} (a) turns into \figref{Fig:NewGraph1} (a-1) and (a-2) while \figref{Fig:NewGraph1} (a-1) further turns into  \figref{Fig:NewGraph1} (a-1-1) and (a-1-2). The graph \figref{Fig:NewGraph1} (a-1-1) is finally reduced into \figref{Fig:NewGraph1} (a-1-1-1) and \figref{Fig:NewGraph1} (a-1-1-2). To summarize
    \bea
    \text{\figref{Fig:NewGraph1} (a)}=\text{\figref{Fig:NewGraph1} (a-1-1-1)}+[\text{\figref{Fig:NewGraph1} (a-2)}+\text{\figref{Fig:NewGraph1} (a-1-2)}]+\text{\figref{Fig:NewGraph1} (a-1-1-2)}.
    \eea
   The first term,  \figref{Fig:NewGraph1} (a-1-1-1), in the above equation is a standard type-C graph (with an incorrect sign $(-1)^{n-l'}$ ($l'=4$) which inherits from the original type-B graph \figref{Fig:NewGraph1} (a-1)), while the terms \figref{Fig:NewGraph1} (a-2) and {\figref{Fig:NewGraph1} (a-1-2)} are graphs which involve a chain of the form $2k\cdot F\cdot...\cdot F\cdot 2k$. The graph {\figref{Fig:NewGraph1} (a-1-1-2)} is a graph whose master chain is of the form $( F\cdot...\cdot F\cdot\epsilon)^{\nu}$ and the path from $\rho$ toward the master chain has the form $(F\cdot...\cdot F\cdot 2k)^{\rho}$ (the sign is also $(-1)^{n-l'}$ ($l'=4$)). On another hand,{\bf\emph{ any graph in the second term of \eqref{Eq:GeneralizedFDec2} can be reduced in the same way}}. Nevertheless, instead of producing a type-C graph, we just terminate the reduction process once the path from $\rho$ to the master chain becomes the form $(F\cdot...\cdot F\cdot 2k)^{\rho}$. Then a graph in the second term  of \eqref{Eq:GeneralizedFDec2} is in general given by the sum of two kinds of graphs: (i). graphs involving a chain of the form $2k\cdot F\cdot...\cdot F\cdot 2k$ and (ii). graphs involving a chain $(F\cdot...\cdot F\cdot 2k)^{\rho}$ towards the master chain. As an example
   \bea
    \text{\figref{Fig:NewGraph1} (b)}=\text{\figref{Fig:NewGraph1} (b-1)}+\text{\figref{Fig:NewGraph1} (b-2)}.
   \eea
   In general, we classify all graphs, which are obtained from the above reduction, into the following four categories.
  \begin{itemize}
\item \emph{Graphs in category-1:}  Type-C graphs (with incorrect sign $(-1)^{n-l'}$) such as \figref{Fig:NewGraph1} (a-1-1-1).
\item \emph{Graphs in category-2 and -3:} Graphs which contain a chain $2k\cdot F\cdot...\cdot F\cdot 2k$ and the master chain of the form $(F\cdot...\cdot F\cdot\epsilon)^{\rho}$ (or $(F\cdot...\cdot F\cdot\epsilon)^{\nu}$) are mentioned as graphs in  category-2 (or -3), such as  \figref{Fig:NewGraph1} (a-2), (a-1-2) (or (b-2)).
\item \emph{Graphs in category-4} Graphs where the master chain has the form $( F\cdot...\cdot F\cdot\epsilon)^{\nu}$ and the path from $\rho$ toward the master chain has the form $(F\cdot...\cdot F\cdot 2k)^{\rho}$, such as \figref{Fig:NewGraph1} (a-1-1-2) and (b-1).
\end{itemize}
A graph in the first term of \eqref{Eq:GeneralizedFDec2} is in general written as
\bea
\text{A graph in category-1}+\text{graphs in category-2}+\text{a graph in category-4},
\eea
while a graph in the second term of \eqref{Eq:GeneralizedFDec2} becomes
\bea
\text{A graph in category-3}+\text{a graph in category-4}.
\eea

\paragraph{\bf\emph{Step-2: Transformations of BS currents}} Accompanied by the manipulations of graphs,  the combination of BS currents corresponding to each graph can also be transformed in a proper way.
For any graph $\mathcal{F}\in \mathcal{G}_1^{B}$, the combination of BS currents is given by
\bea
\Sl_{\pmb{\sigma}^{\mathcal{F}}}\phi\big(1,2,...n-1|\pmb{\sigma}^{\mathcal{F}}\big).
\eea
In the above expression, $\pmb{\sigma}^{\mathcal{F}}\in \mathcal{F}|_{x_1}$ are permutations established by the type-B graph $\mathcal{F}$ where the node $x_1$, which is the one nearest to the Lorentz index $\rho$, is considered as the leftmost node in $\pmb{\sigma}^{\mathcal{F}}$. Assuming that \emph{$x_2\in \mathcal{F}$ is a node adjacent to $x_1$},  we have the following relation
\bea
\Sl_{\pmb{\sigma}\in{\mathcal{F}|_{x_1}}}\phi\big(1,2,...n-1|\pmb{\sigma}\big) =(-1)\Sl_{\pmb{\sigma}\in{\mathcal{F}|_{x_2}}}\phi\big(1,2,...n-1|\pmb{\sigma}\big).\Label{Eq:GeneralizedFDec3}
\eea
This relation is a consequence of the generalized KK relation (\ref{Eq:BSBGProperty-4}). Applying this relation, one can transform the combination of the BS currents which was defined according to a type-B graph  (such as \figref{Fig:NewGraph1} (a)) to a new combination based on a type-C graph (such as \figref{Fig:NewGraph1} (a-1-1-1)). In particular, supposing there are $l$ nodes between $\nu$ and $\rho$ and the nodes nearest to $\nu$ and $\rho$ are respectively $x$ and $y$, we must have
\bea
\Sl_{\pmb{\sigma}\in \mathcal{F}|_{y}}\phi\big(1,2,...n-1|\pmb{\sigma}\big)=(-1)^{l-1}\Sl_{\pmb{\sigma}\in \mathcal{F}|_{x}}\phi\big(1,2,...n-1|\pmb{\sigma}\big).\Label{Eq:GeneralizedFDec3-1}
\eea
In the case of \figref{Fig:NewGraph1} (a), $x=a_2$, $y=c_5$, $l=7$.

\paragraph{\bf\emph{Step-3: Contributions of graphs in different categories}} Having the relations between graphs which were discussed in step-1 and the relations (\ref{Eq:GeneralizedFDec3}), (\ref{Eq:GeneralizedFDec3-1}) between BS currents which were given in step-2,  we now collect the contribution of graphs of distinct categories.
\begin{itemize}
\item \emph{Category-1} The contribution of the graph \figref{Fig:NewGraph1} (a-1-1-1) is given by
\bea
&&(-1)^{n-l'}(F_{a_2}\cdot F_{a_3}\cdot F_{b_1}\cdot F_{b_3}\cdot F_{b_2})^{\nu}_{~\gamma}(F_{c_5}\cdot F_{c_4})^{\rho\gamma}(2k_{c_4}\cdot F_{c_3}\cdot F_{c_2}\cdot \epsilon)(...)\biggl[\,\Sl_{\pmb{\sigma}\in \mathcal{F}|_{c_5}}\phi\big(1,2,...n-1|\pmb{\sigma}\big)\biggr]\nn
&=&(F_{a_2}\cdot F_{a_3}\cdot F_{b_1}\cdot F_{b_3}\cdot F_{b_2}\cdot F_{c_4}\cdot F_{c_5})^{\nu\rho}(\epsilon\cdot F_{c_2}\cdot F_{c_3}\cdot 2k_{c_4})(...)\biggl[\,(-1)^{n-l+1}\Sl_{\pmb{\sigma}\in \mathcal{F}|_{a_2}}\phi\big(1,2,...n-1|\pmb{\sigma}\big)\biggr]\nn
&=&C^{\nu\rho}_{\mathcal{F}}\bigg[\,\Sl_{\pmb{\sigma}^{\mathcal{F}}}\phi\big(1,2,...n-1|\pmb{\sigma}\big)\biggr],
\eea
where $(...)$ denotes the contribution of other parts of this graph which was not changed. The first line in the above equation comes from the reduction of the original type-B graph \text{\figref{Fig:NewGraph1} (a)} in which there are $l'=4$ internal nodes (i.e. ${F}$'s) on the master chain. On the second line, the sign $(-1)^{n-l'}$ has been absorbed into the coefficient, while the relation (\ref{Eq:GeneralizedFDec3-1}) reproduces a sign $(-1)^{n-l+1}$ ($l=7$). The coefficient together with the sign on the second line precisely match with the $C^{\nu\rho}_{\mathcal{F}}$ where $\mathcal{F}$ is a standard type-C graph.
This example reveals the general pattern: any graph in category-1 contributes a type-C graph with the correct sign. Conversely, for any type-C graph (for example \figref{Fig:NewGraph1} (a-1-1-1)) in the first summation on the RHS of \eqref{Eq:GeneralizedFDec1}, one can reverse the above discussion and find out a unique type-B graph (e.g. \figref{Fig:NewGraph1} (a)) corresponding to it.  Therefore, {\bf\emph{we conclude that the  contribution of all graphs in category-1 gives rise to the first summation on the RHS of \eqref{Eq:GeneralizedFDec1} .}}

\item \emph{Category-2 and -3}
We claim\footnote{The proof in fact is
analogous to the intricate discussion in \cite{Hou:2018bwm}.} that the total contribution of all graphs in category-2 can be written as
\bea
\Sl_{\substack{\{1,...,n\}\\ \to \pmb{A}_1,\pmb{A}_2}}\Sl_{\mathcal{F}_1\in \mathcal{G}^{B}_{\pmb{A}_1}}C^{\rho}_{\mathcal{F}_1}\Sl_{\mathcal{F}_2\in \mathcal{G}^{B}_{\pmb{A}_2}}C^{\nu}_{\mathcal{F}_2}\left[\Sl_{a\in \mathcal{F}_1}\Sl_{b\in \mathcal{F}_2}f^{b}(-2k_a\cdot 2k_b)\Sl_{\pmb{\sigma}^{\mathcal{F}_1}}\Sl_{\pmb{\sigma}^{\mathcal{F}_2|_{b}}}\phi\Big(1,...,n-1\Big|\pmb{\sigma}^{\mathcal{F}_1}\shuffle\pmb{\sigma}^{\mathcal{F}_2|_{b}}\big|_{a\prec b}\Big)\right],\Label{Eq:GeneralizedFDec4}\nn
\eea
\begin{figure}
\centering
\includegraphics[width=0.5\textwidth]{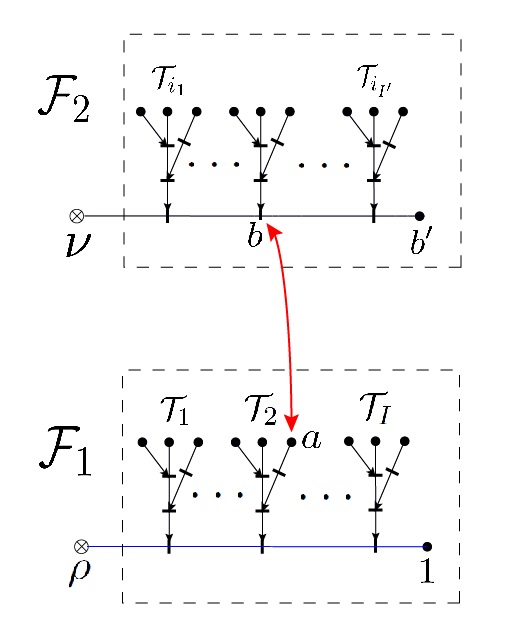}
\caption{ Suppose $b'$ is the highest-weight node in $\mathcal{F}_2$. Any $a\in\mathcal{F}_1$ and $b\in\mathcal{F}_2$ are connected through the red line (the $2k\cdot 2k$ line). The chain colored blue is the master chain of $\mathcal{F}_1$. This corresponds to a term in \eqref{Eq:GeneralizedFDec4}. }
\label{Fig:off-shellBCJ}
\end{figure}
where the first summation is taken over all possible splittings of $\{1,...,n-1\}\to \pmb{A}_1,\pmb{A}_2$ such that the highest-weight node, say $1$, belongs to the set $\pmb{A}_1$. The $\mathcal{G}^{B}_{\pmb{A}_1}$ and $\mathcal{G}^{B}_{\pmb{A}_2}$ denote the type-B graphs constructed by nodes in $\pmb{A}_1$ and $\pmb{A}_2$ with the reference order $\{n-1,...,1\}\setminus \pmb{A}_2$ and $\{n-1,...,1\}\setminus \pmb{A}_1$,  respectively.
Coefficients $C^{\rho}_{\mathcal{F}_1}$ and $C^{\nu}_{\mathcal{F}_2}$ are the coefficients corresponding to the type-B graphs. For given $\mathcal{F}_1$ and $\mathcal{F}_2$ in \figref{Fig:off-shellBCJ}, we pick out a pair of nodes $a\in \mathcal{F}_1$ and $b\in \mathcal{F}_2$ and connect these two nodes by a $(-2k_a\cdot 2k_b)$ line. We then sum over (i). all possible permutations established by $\mathcal{F}_1$ (where the node nearest to $\rho$ is considered as the leftmost one) and $\mathcal{F}_2$ where the node $b$ is considered as the leftmost one and (ii). all $\pmb{\sigma}^{\mathcal{F}_1}\shuffle \pmb{\sigma}^{\mathcal{F}_2|_b}$ s.t. $a\prec b$. Finally, we sum over all $a\in \mathcal{F}_1$ and $b\in \mathcal{F}_2$. The sign $f^b$ is determined as follows: if $b$ is the node nearest to $\rho$, $f^b=1$, (ii). for two adjacent nodes $x$ and $y$, $f^x=-f^y$. The expression in the square brackets has a similar form with the LHS of the so called \emph{graph-based BCJ relation} \cite{Hou:2018bwm}\footnote{They are not exact the same one since the LHS of graph-based BCJ relation is given by a combination of on-shell amplitudes in KK basis (i.e. the $(n-2)!$ basis), while the expression in the square brackets is a combination of $(n-1)!$ off-shell currents. }. We just present our observation on the expression in the square brackets and propose the following off-shell extended graph based BCJ relation:
\bea
&&\Sl_{a\in \mathcal{F}_1}\Sl_{b\in \mathcal{F}_2}f^{b}(-2k_a\cdot 2k_b)\Sl_{\pmb{\sigma}^{\mathcal{F}_1}}\Sl_{\pmb{\sigma}^{\mathcal{F}_2|_{b}}}\phi\Big(1,...,n-1\Big|\pmb{\sigma}^{\mathcal{F}_1}\shuffle\pmb{\sigma}^{\mathcal{F}_2|_{b}}\big|_{a\prec b}\Big)\nn
&=&-2f^{b_0}\Sl_{\pmb{\sigma}^{\mathcal{F}_1}}\Sl_{\pmb{\sigma}^{\mathcal{F}_2|_{b_0}}}\biggl[\phi\Big(1,...,i\Big|\pmb{\sigma}^{\mathcal{F}_1}\Big)\Big(i+1,...,n-1\Big|\pmb{\sigma}^{\mathcal{F}_2|_{b_0}}\Big)\nn
&&~~~~~~~~~~~~~~~~~~~~~~~~~~~~~~~~~~~~~~~-\phi\Big(1,...,n-1-i\Big|\pmb{\sigma}^{\mathcal{F}_2|_{b_0}}\Big)\Big(n-i,...,n-1\Big|\pmb{\sigma}^{\mathcal{F}_1}\Big)\biggr],\Label{Eq:OffShellGraphBCJ}
\eea
where $i$ is the number of on-shell nodes in the graph $\mathcal{F}_1$ (in other words, the elements in $\pmb{A}_1$) and $b_0$ is an arbitrarily chosen node in $\mathcal{F}_2$. When we multiply both sides by $s_{1...n-1}$ and take the on-shell limit $s_{1...n-1}=k_n^2\to 0$, we get an on-shell version of graph based BCJ relation where the RHS is zero. We leave the details of the proof of \eqref{Eq:OffShellGraphBCJ} in a future work and just insert \eqref{Eq:OffShellGraphBCJ} into \eqref{Eq:GeneralizedFDec4}. Since the node $1$ is always involved in $\mathcal{F}_1$ and the BS current satisfies  $\phi(\pmb{A}|\pmb{B})=0$ if $\pmb{A} \setminus \pmb{B}\neq \emptyset$, only the first term in \eqref{Eq:OffShellGraphBCJ} provides nonvanishing contribution when the nodes in $\mathcal{F}|_{1}$ are $1,2,...,i$. Therefore, (\ref{Eq:GeneralizedFDec4}) becomes
\bea
&&-2\Sl_{i=1}^{n-2}\biggl[\Sl_{\mathcal{F}_1\in \mathcal{G}^{B}_{\pmb{A}_1}}C^{\rho}_{\mathcal{F}_1}\Sl_{\pmb{\sigma}^{\mathcal{F}_1}}\phi\Big(1,...,i\Big|\pmb{\sigma}^{\mathcal{F}_1}\Big)\biggr]\biggl[\Sl_{\mathcal{F}_2\in \mathcal{G}^{B}_{\pmb{A}_2}}C^{\nu}_{\mathcal{F}_2}\Sl_{\pmb{\sigma}^{\mathcal{F}_2|_{b_0}}}\phi\Big(i+1,...,n-1\Big|\pmb{\sigma}^{\mathcal{F}_2}\Big)\biggr]\nn
&=&-\Sl_{i=1}^{n-2}2\W J^{\rho}_{(1,i)}\W J^{\nu}_{(i+1,n-1)},
\eea
where $\pmb{A}_1=\{1,...,i\}$, $\pmb{A}_2=\{i+1,...,n-1\}$ for a given $i$ on the first line, while the $b_0$ denotes the node nearest to the Lorentz index $\nu$. The above expression precisely agrees with \emph{ the second term in the second summation of \eqref{Eq:GeneralizedFDec1}}. Following a similar discussion, \emph{the contribution of all graphs in category-3 reproduces the first term in the second summation of \eqref{Eq:GeneralizedFDec1}.}
\item \emph{Category-4} The contributions of graphs in category-4 appear  in pairs (which correspond to the first and the second terms in \eqref{Eq:GeneralizedFDec2}) with opposite signs (e.g. \figref{Fig:NewGraph1} (a-1-1-2) and (b-1)), thus they all cancel out.
\end{itemize}
To sum up, the contribution of all graphs in \eqref{Eq:GeneralizedFDec2} matches with the RHS of \eqref{Eq:GeneralizedFDec1}, hence the proof of \eqref{Eq:GeneralizedFDec1} has been completed.

\section{The identity of Berends-Giele current}


Identities (\ref{Eq:YMBGProperty-2}) and (\ref{Eq:YMBGProperty-3}) are generalizations of (\ref{Eq:YMBGProperty-1}) which was proved in \cite{Berends:1987me}. The (\ref{Eq:YMBGProperty-3}) can always be obtained via replacing the $k_{1,n-1}$ in (\ref{Eq:YMBGProperty-2}) by another momentum, e.g., $k_{a_1,b_1}$. In this part, we briefly review the proof of (\ref{Eq:YMBGProperty-1}), based on which, the proof of  \eqref{Eq:YMBGProperty-2} is further given.

{\bf The proof of \eqref{Eq:YMBGProperty-1} which was given in \cite{Berends:1987me}}
~~When we expand the current $J^{\mu}(1,...,n-1)$ according to Berends-Giele recursion \eqref{Eq:BerendsGiele}, the LHS is given by summing all terms that have the following forms
\bea
V^{\mu\nu\rho}_3J_\mu(\pmb{A})J_\nu(\pmb{B})(k_{\pmb{A}}+k_{\pmb{B}})_\rho
&=&(s_{\pmb{A}}-s_{\pmb{B}})J(\pmb{A})\cdot J(\pmb{B})\nn
&&~~~+J(\pmb{A})\cdot k_{\pmb{B}}J(\pmb{B})\cdot k_{\pmb{B}}-J(\pmb{B})\cdot k_{\pmb{A}}J(\pmb{A})\cdot k_{\pmb{A}}, \Label{Eq:BGIdentity1} \\
V^{\mu\nu\tau\rho}_4J_\mu(\pmb{A})J_\nu(\pmb{B})J_\tau(\pmb{C})(k_{\pmb{A}}+k_{\pmb{B}}+k_{\pmb{C}})_\rho
&=&V^{\mu\nu\rho}_3\left[J_\rho(\pmb{A})J_\mu(\pmb{B})J_\nu(\pmb{C})-J_\mu(\pmb{A})J_\nu(\pmb{B})J_\rho(\pmb{C})\right],\Label{Eq:BGIdentity2}
\eea
where the $\pmb{A}$ and $\pmb{B}$ in \eqref{Eq:BGIdentity1} are supposed to be the ordered sets produced by a division $\{1,2,...,n-1\}\to \pmb{A},\pmb{B}$. For example, if $n=4$, $\pmb{A}$, $\pmb{B}$ can be $\{1\}$, $\{2,3\}$ or $\{1,2\}$, $\{3\}$.  Similarly, the $\pmb{A}$, $\pmb{B}$  and $\pmb{C}$ in \eqref{Eq:BGIdentity2} are produced by the division $\{1,2,...,n-1\}\to \pmb{A},\pmb{B},\pmb{C}$.  When all possible divisions are summed over and $V^{\mu\nu\rho}_3J_\mu(\pmb{B})J_\nu(\pmb{C})$ is expressed by subtracting the contribution of the four-point vertex term (i.e. the second term in \eqref{Eq:BerendsGiele}) from the current $J_\mu(\pmb{B},\pmb{C})$, as pointed in \cite{Berends:1987me}, the total contribution of all terms of the form \eqref{Eq:BGIdentity2} is reduced into
\bea
&&\sum^{n-2}_{m=1}(s_{m+1\ldots n-1}-s_{1\ldots m})J(1,\ldots, m)\cdot J(m+1,\ldots,n-1) \nn &&+k_1^2J(1)\cdot J(2,\dots,n-1)-k_{n-1}^2J(1,\ldots,n-2)\cdot J(n-1).
\Label{Eq:BGIdentity6}
\eea
where, the last two terms vanish due to the on-shell condition $k_1^2=k_{n-1}^2=0$. On another hand, the last two terms in \eqref{Eq:BGIdentity1} vanish according to our inductive assumption. When all divisions are summed over, the first term in \eqref{Eq:BGIdentity1} precisely cancel with the first term in \eqref{Eq:BGIdentity6}. Thus the identity (\ref{Eq:YMBGProperty-1}) is proven.

{\bf Proof of the identity~(\ref{Eq:YMBGProperty-2})}~~~Now we prove \eqref{Eq:YMBGProperty-2} via replacing some of the external lines in the above proof  by $k_{a_i,b_i}$. The starting point of \eqref{Eq:YMBGProperty-2} is given by
\bea
J(1,2)\cdot(k_1+k_2)=J(k_1,2)\cdot(k_1+k_2)=J(k_1,k_2)\cdot(k_1+k_2)=0,
\eea
which can be verified straightforwardly. The above relation can be directly generalized to
\bea
J(k_{1,n-2},n-1)\cdot k_{1,n-1}=J(k_1,k_{2,n-1})\cdot k_{1,n-1}=J(k_{1,i},k_{i+1,n-1})\cdot k_{1,n-1}=0,
\eea
for any $1\leq i<n-2$. Suppose that the identity~(\ref{Eq:YMBGProperty-2}) holds for all lower-point cases and all possible choices of the consecutive sequences $\{a_i,...,b_i\}$. The subcurrents $J(\pmb{A})$, $J(\pmb{B})$ and $J(\pmb{C})$ in \eqref{Eq:BGIdentity1} and \eqref{Eq:BGIdentity2} are then replaced by those involving external lines $k_{a_i,b_i}$. (i). All cancellations in the previous proof work, if the first element $1$ and the last element $n-1$ are not replaced by any $k_{a_i,b_i}$. (ii). If the element $1$ (or $n-1$) is replaced by $k_{a_1=1,b_1}$ (or $k_{a_I,b_I=n-1}$), the cancellation also works except the special terms: the term $-J(\pmb{B})\cdot k_{\pmb{A}}J(\pmb{A})\cdot k_{\pmb{A}}$ (or $J(\pmb{A})\cdot k_{\pmb{B}}J(\pmb{B})\cdot k_{\pmb{B}}$) for $\pmb{A}=\{a_1=1,b_1\}$ (or the term $\pmb{B}=\{a_I,b_I=n-1\}$) in \eqref{Eq:BGIdentity1}
must be replaced by $-\big(J(b_{1}+1,...,n-1)\cdot k_{1,b_1}\big) k^2_{1,b_1}$ (or $\big(J(1,...,a_I-1)\cdot k_{a_I,b_I}\big) k^2_{a_I,b_I}$) which does not vanish, while the term $k_1^2J(1)\cdot J(2,\dots,n-1)$ (or $-k_{n-1}^2J(1,\ldots,n-2)\cdot J(n-1)$) in (\ref{Eq:BGIdentity6}) is replaced by $k_{a_1,b_1}^2\big(k_{a_1,b_1}\cdot J(b_1+1,\dots,n-1)\big)$ (or $-k_{a_I,b_I}^2J(1,\ldots,a_I-1)\cdot k_{a_I,b_I}$). They also cancel in pair.

\bibliographystyle{JHEP}
\bibliography{FeynEYMExpansion}

\end{document}